\documentclass[twocolumn]{aastex631}

\shorttitle{The optical photometric variability of Herbig Ae/Be stars from {\em TESS}}
\shortauthors{Cody et al.}
\begin{document}

\title{The optical photometric variability of Herbig Ae/Be stars from {\em TESS}}

\author[0000-0002-3656-6706]{Ann Marie Cody}
\affiliation{SETI Institute, 339 Bernardo Ave., Suite 200, Mountain View, CA 94043, USA}
\correspondingauthor{Ann Marie Cody} \email{acody@seti.org}

\author{Lynne A. Hillenbrand}
\affiliation{Department of Astronomy, California Institute of Technology, Pasadena, CA 91125, USA}

\author{Shreya Chandragiri}
\affiliation{Santa Clara University, Santa Clara, CA 95053, USA}

\author[0000-0003-4022-6234]{Marvin Morgan}
\affiliation{Department of Physics, University of California, Santa Barbara, Santa Barbara, CA 93106, USA}



\begin{abstract}

We have carried out a photometric time domain study of 188 intermediate-mass young stars 
observed in Full Frame Image mode with the {\em TESS} satellite over the first 3.3 years of 
its mission. The majority of these targets are classified as Herbig Ae/Be stars (HAeBes). All 
were monitored at optical wavelengths for at least one 27-day {\em TESS} sector, with many 
having multiple sectors of data. From a custom aperture photometry pipeline, we produced 
light curves and analyzed the variability therein, as a function of stellar and circumstellar 
properties. Based on visual and statistical analysis, we find that $\sim$95\% of HAeBes are 
variable on timescales of 10 minutes to 1 month, with the most common light curve morphology 
being stochastic. Approximately 15\% of the set display quasi-periodic variability. In 
comparison to sets of low-mass T Tauri stars monitored with optical space telescopes, the 
Herbig Ae/Be stars display a much lower incidence of ``dipper" behaviors (quasi-periodic or 
aperiodic fading events), as well as periodic modulations. As posited by previous work, we 
conclude that magnetic starspots are rare on HAeBes, and that the inner circumstellar dust 
rims of these objects lie at substantially larger radii than for low-mass young stars. Beyond 
these differences, the accretion dynamics of young stars less than $\sim$7$M_\odot$ appear to 
be largely consistent based on their time domain properties from data streams of up to three 
months' duration. We do, however, find tentative evidence for a change in variability 
amplitude above this mass boundary, particularly for quasi-periodic behavior.

\end{abstract}



\section{Introduction} 
\label{sec:intro}

The pre-main sequence (PMS) stage of stellar evolution occupies a key point at the end of star formation, during which stars contract toward eventual main sequence equilibrium. T Tauri stars (masses $\lesssim$2~$M_\odot$) are the most common constituent of the young population, with the classical subtype (CTTS) hosting dusty circumstellar accretion disks. At the high-mass end ($M > 10M_\odot$), stars do not become optically visible until reaching the main sequence. The intermediate-mass regime is occupied by Herbig Ae/Be stars (HAeBes), with masses of approximately 2 to 10~$M_\odot$ \citep{1960ApJS....4..337H,1972ApJ...173..353S,1998ARA&A..36..233W,2023SSRv..219....7B}. Like their lower mass counterparts, HAeBes are characterized by spectroscopic signatures of accretion (e.g., hydrogen emission lines) as well as circumstellar dust emission, as signified by (near-)infrared excesses \citep{1992ApJ...397..613H, 2001A&A...365..476M}. They are often found in association with nebulosity, as originally defined by \cite{1960ApJS....4..337H}. That work considered HAeBes to comprise A and B spectral types, but the class is now often extended to include F spectral types as well \citep{1998A&A...330..145V}.

Of particular interest is how the structure and dynamics of young star-disk systems change as a function of mass. The transition from convective interiors (CTTSs) to radiative interiors (HAeBes) may correlate with other physical differences within and around these stars. For example, the magnetospheric accretion paradigm used to explain gas inflow for CTTSs \citep{1992ApJ...386..239C,1994ApJ...426..669H,2016ARA&A..54..135H} may break down for the HAeBes. Indeed, spectroscopic observations indicate a change in accretion rate trends between the Herbig Ae and Be regimes \citep{2011AJ....141...46D,2011A&A...529A..34M,2015MNRAS.453..976F}. Boundary layer accretion and potentially different inner disk structure has thus been proposed for stars more massive than about 4~$M_\odot$ \citep{2015ApJ...810....5C,2019AJ....157..159A,2020MNRAS.493..234W}. In general, inner disk edges lie at larger distances from higher mass stars, due to their higher luminosities and hence larger sublimation radii \citep{2002ApJ...579..694M,2005ApJ...624..832M}. Other differences, e.g., in dust processing, rim height, outer disk radii, and total dust mass, may exist between CTTSs and HAeBes \citep{2016JASS...33..119S,2019A&A...632A..53G,2021A&A...655A..73G}. 

One way to probe putative mass dependences in young accreting stars is via comparative investigations of young populations, through the lens of photometric variability. Thousands of CTTSs have been tallied \citep[e.g.,][and references therein]{2022AJ....163...25L,2023AJ....165...37L} and monitored from both the ground and space \citep[e.g.,][]{1993A&A...272..176B,1994AJ....108.1906H,2018AJ....156...71C,2021AJ....162..101V,2022AJ....163..212C}. Based on high-cadence, long-duration (sub-hour continuous cadence over months) space-based monitoring, distinct optical variability patterns have been identified on timescales from minutes to months and amplitudes from 1\% to over 100\% \citep{2014AJ....147...82C}. The patterns have been associated with different physical phenomena, including unsteady accretion, occultation by circumstellar dust, and cool or hotspots on the stellar surface. 

Likewise, time variations of HAeBes have been the subject of investigation for decades, albeit with smaller numbers of targets and typically less-intensive ground-based monitoring campaigns \citep[e.g.,][]{1999AJ....118.1043H,2002A&A...384.1038E,2013ARep...57...89B,2022ARep...66..236E}. Early photometric monitoring revealed variability at many different timescales, and lower amplitude among the Herbig Be stars, compared with the Ae subclass \citep{1984A&AS...55..109F,1991A&AS...89..319B, 1998A&A...330..145V}.  Additionally, some objects exhibit a tendency toward pronounced fading (up to several magnitudes) on timescales of days to months \citep{1994ASPC...62...23T,2003ApJ...594L..47D,2004PASJ...56S.183U}. These have been dubbed ``UXors" after prototype UX~Ori. In an optical photometric monitoring study of 230 HAeBes, \citet{1999AJ....118.1043H} observed irregular variations on day to year timescales, which they also categorized as part of the UXor class. Notably, they did not detect periodic variations consistent with hot or cool spots on the stellar surface. Since then, periodic variations {\em have} been noted, although primarily associated with pulsational instabilities \citep{1998ApJ...507L.141M}. The question of whether the radiative HAeBes can support large-scale magnetic fields and associated stellar surface spots remains largely unanswered, though progress has been made in making and explaining magnetic field detections in a handful of objects \citep{2019MNRAS.489..886J,2021mfob.book.....H}. The physical origin of irregular variability also remains the subject of speculation, whether tied to circumstellar occultation \citep{2018A&A...620A.128V} or changes in accretion luminosity \citep{1999AJ....118.1043H}. 

A path forward in understanding the drivers of Herbig Ae/Be star variability and its connections to (circum)stellar physics is through high-cadence space-based photometric monitoring. With the advent of NASA Transiting Exoplanet Survey Satellite ({\em TESS}) mission \citep{2015JATIS...1a4003R,2019ESS.....410001R}, nearly all-sky optical photometric coverage is now available on timescales from minutes to months and, in many cases, years. In parallel, data from Gaia \citep{2016A&A...595A...1G} and other missions have made it possible to identify hundreds of new HAeBe candidates based on their photometric properties alone \citep{2020A&A...638A..21V}. 

Using observations from {\em TESS}, we have embarked on a photometric monitoring study of 188 HAeBes on timescales of minutes to years. In this paper, we define a sample with uniformly derived stellar properties (\S\ref{sec:sample}) and create custom light curves for the included stars from Full Frame Images of available {\em TESS} sectors (\S\ref{sec:lcs}). Based on those light curves, we then assess HAeBe variability properties via various statistical metrics (\S\ref{sec:varanalysis}). Finally, we attempt to correlate these variability properties with circumstellar parameters (\S\ref{sec:vartocircumstellar}). We present our key findings in the context of previous works in \S\ref{sec:discussion}; concluding remarks are provided in \S\ref{sec:conclusions}.

\section{Stellar sample}
\label{sec:sample}

In recent decades, imaging and spectroscopic studies have been undertaken to catalog HAeBes across the sky \citep{1992ApJ...397..613H,1995AJ....109.2146T}. 
The canonical list of \citet{1994A&AS..104..315T} serves as a starting point for building a photometric sample, with the caveat that it contains some post-main-sequence star contaminants. These classical Be stars can be difficult to distinguish from HAeBes, as they also display infrared excesses and line emission due to expelled circumstellar material \citep{2013A&ARv..21...69R}. High-resolution spectroscopic confirmation \citep[e.g.,][]{2003AJ....126.2971V,2004AJ....127.1682H,2010A&A...517A..67C} has played a crucial role in disentangling Herbig Be stars from these more evolved systems. 
% Chen IR study: \cite{2016NewA...44....1C}

Within the past few years, a flurry of studies have been undertaken to mine WISE, 2MASS, and now the Gaia dataset for the analysis of Herbig Ae/Be star properties and identification of additional candidates. \cite{2006Ap&SS.305...11Z} assembled a set of 206 HAeBes, later adding in additional stars in the F spectral type regime \citep{2022ApJ...936..151Z}. \citet{2018A&A...620A.128V,2020A&A...638A..21V} developed a larger catalog based on machine learning to identify over 1300 potential new Herbig Ae/Be 
stars. While the contamination fraction of classical Be stars here is likely low, spectroscopic confirmation for the majority of new candidates has not yet been reported in the literature. 

\begin{figure}
\centering
\includegraphics[width=0.52\textwidth]{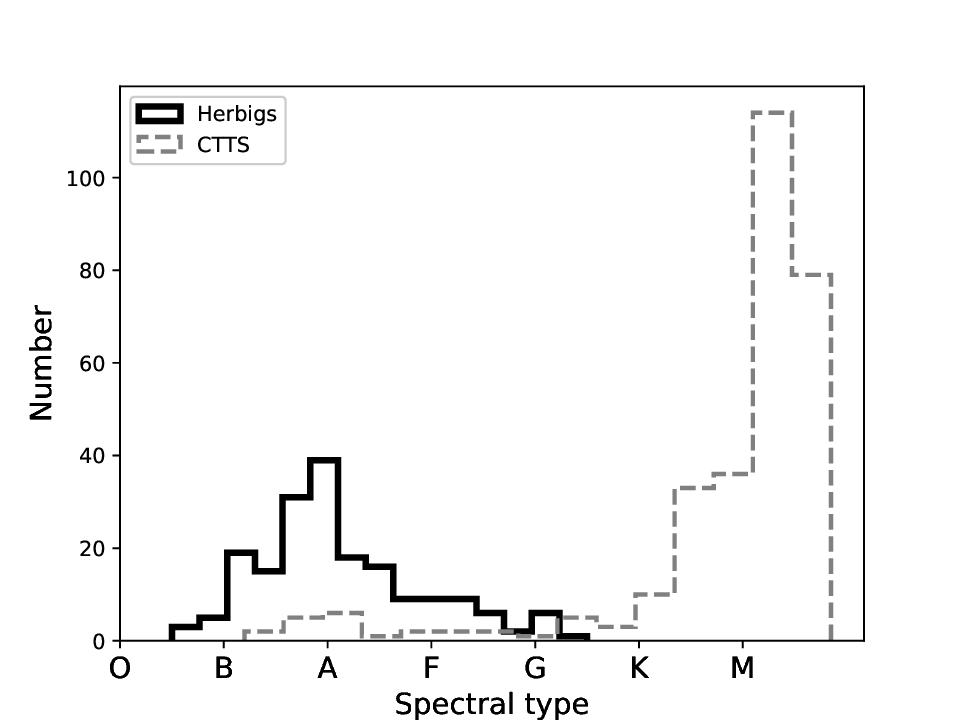}
\caption{Spectral type distribution for the photometric sample employed in this work (Herbig Ae/Be stars), as well as those from previous works on lower mass stars (``CTTS"). Spectral types are labeled at the zero subtype (i.e., ``A" is A0), and adopted from \cite{2021AA...650A.182G}. If a range of spectral types is given, we take the subtype at the middle. As seen in the histogram, late B and early A stars comprise the most common types here, while earlier samples were dominated by late K and M types.}
\label{fig:SpTdist}
\end{figure}

In terms of probing HAeBe variability properties, it is advantageous to select as large as possible a sample that is both clean and homogeneous. For our photometric monitoring study, we therefore adopted the catalog presented by \citet{2021AA...650A.182G}, dubbed the ``HArchiBe." The HArchiBe draws confirmed young stars from the above-cited compilations and also includes uniformly derived sets of stellar and disk parameters. From its spectral type list, we discarded stars listed as being K or M. A few G stars are present in the HArchiBe, and we do retain those for the purpose of continuity with previous low-mass samples. The spectral type distribution of our adopted set, after removal of stars with missing or compromised {\em TESS} photometric data (as described in \S\ref{sec:lcs}), is shown in Figure~\ref{fig:SpTdist}.

We provide these spectral types, along with crossed-matched stellar identifiers and {\em TESS} magnitudes for our photometric sample, in Table~1. With the 21\arcsec\ size of {\em TESS} pixels, blending is an issue of consideration. The multiplicity of HAeBes was investigated recently by \citet{2023AJ....165..135T}, and the fraction of stars with companions was reported to be over 50\%. In Table~1, we note which stars are either known multiples or photometric blends in {\em TESS} imaging, as indicated by the TESS Input Catalog \citep[TIC;][]{2021arXiv210804778P}, visual examination of higher spatial resolution images from the Digitized Sky Survey (DSS), and/or literature reports of spectroscopic multiplicity. %CHeck also tables 1 and 2 of Vioque et al. 2018 for spectroastrometric binaries. 
We consider an object to be blended if a neighboring star within the same {\em TESS} pixel is brighter or no more than two {\em TESS} magnitudes fainter than the intended target. Blended objects were only thrown out if we were unable to selected a suitable photometric aperture (\S\ref{sec:lcs}). 

\startlongtable
\begin{deluxetable*}{c c c c c c}
\tablecaption{Intermediate-mass young stars with {\em TESS} photometry \label{tab:sample}}
\tablehead{
\colhead{TIC id} & \colhead{2MASS id} & \colhead{Other name}& \colhead{TESS mag} & \colhead{Spectral type} & \colhead{Blend?}
}
\startdata
4206238 & J05233100-0104237 & HD 290380 & 9.9 & F5-F8 & N \\ 
8806247 & J16271510-4839268 & Hen 3-1191 & 12.28 & B1 & N \\ 
11199521 & J05380526-0115216 & HD 290764 & 9.51 & A5-A7 & N \\ 
11402323 & J05410229-0243006 & HD 37806 & 7.86 & B8 & N \\ 
12161661 & J06065848-0555066 & PDS 124 & 11.94 & B8-B9 & N \\ 
12876844 & J22531560+6208450 & IL Cep & 8.53 & B2 & Y \\ 
13002932 & J20202825+4121514 & V1685 Cyg & 9.59 & B2-B3 & N \\ 
13417240 & J20213607+4047561 & VOS 321 & 12.27 & A8 & N \\ 
14740791 & J20253882+4019008 & VOS 398 & 12.95 & G & N \\ 
15898682 & J20292692+4140439 & MWC 1021 & 9.02 & B6-B7 & N \\ 
17130641 & J20324553+4039366 & V1478 Cyg & 9.72 & B5-B6 & Y \\ 
24344701 & J05160047-0948353 & HD 34282 & 9.74 & A0-A1 & N \\ 
30243973 & J08554594-4425140 & SAO 220669 & 8.22 & B4 & N \\ 
31638754 & J09060000-4718581 & PDS 286 & 10.1 & B0 & N \\ 
35485982 & J06484168-1648056 & PDS 24 & 12.82 & B9 & N \\ 
45124112 & J06074953+1839264 & LkHa 208 & 10.99 & A9-F0 & N \\ 
46749954 & J17125876-3214335 & HD 155448 & 8.73 & B8 & N \\ 
50656450 & J05294805-0023434 & HD 290500 & 10.75 & A2-A3 & N \\ 
50745417 & J05320993-0249467 & RY Ori & 10.83 & F4-F9 & N \\ 
53898232 & J06542787-2502158 & PDS 25 & 12.69 & A3-A6 & N \\ 
54003409 & J01343776-1540348 & HD 9672 & 5.6 & A1-A2 & N \\ 
54185990 & J05240118+2457370 & HD 35187 & 8.22 & B9-A0 & Y \\ 
63705801 & J21212751+5059475 & HD 235495 & 9.46 & B9 & N \\ 
66918331 & J05534254-1024006 & V1818 Ori & 9.53 & B8 & N \\ 
67671603 & J16065795-2743094 & HD 144432 & 7.81 & A5-A7 & N \\ 
72553036 & J06133726-0625017 & PDS 126 & 11.3 & A5-A7 & N \\ 
72780071 & J05431188-0459499 & HD 38120 & 9.06 & B8 & N \\ 
72781767 & J05430057-0218454 & HD 38087 & 8.05 & B5-B6 & N \\ 
73816033 & J05302753+2519571 & HD 36112 & 7.98 & A9-F0 & N \\ 
74632969 & J05355845+2444542 & CQ Tau & 9.86 & F2-F3 & N \\ 
75242497 & J05393051+2622269 & RR Tau & 11.71 & B9 & N \\ 
75242566 & J05393048+2619552 & HD 245906 & 10.0 & A5-A7 & N \\ 
81297869 & J08114457-4405087 & HD 68695 & 9.76 & B9-A0 & N \\ 
85100457 & J04584626+2950370 & HD 31648 & 7.55 & A5-A6 & N \\ 
89189882 & J05585578+1639573 & HD 249879 & 10.64 & B8 & N \\ 
95586779 & J07092789-1212160 & VOS 465 & 12.79 & A7 & N \\ 
96680681 & J04554582+3033043 & AB Aur & 6.94 & A1-A2 & N \\ 
97638779 & J06010465+1654407 & VOS 131 & 11.86 & F6-G4 & N \\ 
97807159 & J06015998+1630567 & HD 250550 & 9.37 & B8 & N \\ 
101186168 & J17332078-3234256 & VOS 15 & 13.11 & A6 & N \\ 
107406463 & J07241754-2616053 & GSC 6546-3156 & 13.8 & B9-A0 & N \\ 
107544865 & J07250495-2545496 & PDS 133 & 13.71 & B7-B8 & N \\ 
110990131 & J19264025+2353508 & PX Vul & 10.57 & F3-F5 & N \\ 
120751782 & J06495854-0738522 & PDS 130 & 12.65 & B8-B9 & N \\ 
121137860 & J16590677-4242083 & V921 Sco & 9.24 & B0-B1 & N \\ 
125641610 & J07034316-1133062 & Z CMa & 8.36 & A4-A6 & N \\ 
125644224 & J07040669-1126084 & HU CMa & 11.42 & B7 & N \\ 
136712091 & J21111906+4738476 & VOS 22 & 11.8 & B8 & N \\ 
138670679 & J21200323+4657466 & VOS 522 & 12.53 & F1 & N \\ 
138941394 & J05330479+0228096 & HD 288012 & 9.64 & A2-A3 & Y \\ 
139019368 & J21220592+4651406 & VOS 336 & 12.03 & F6-G4 & N \\ 
139715411 & J07283676-2157493 & HD 59319 & 8.37 & B7-B8 & N \\ 
149542947 & J05444639-6544079 & HD 39014 & 4.14 & A5-A6 & N \\ 
163268072 & J17310584-3508292 & HD 319896 & 10.14 & B4-B5 & N \\ 
166896955 & J13574395-3958471 & PDS 69 & 10.0 & B4-B5 & Y \\ 
168627773 & J07083879-0419048 & PDS 241 & 11.48 & B1 & Y \\ 
178486401 & J07143123-1129271 & VOS 703 & 12.26 & F1-F6 & N \\ 
181142391 & J08421727-4044096 & V388 Vel & 14.33 & B9 & Y \\ 
183658281 & J08231185-3907015 & PDS 277 & 9.57 & F3-F5 & N \\ 
189283710 & J20384588+4207047 & VOS 1331 & 12.51 & B0.5 & N \\ 
189286260 & J20391272+4220546 & VOS 111 & 13.29 & A0 & N \\ 
190255749 & J08511143-4138500 & VOS 4 & 12.31 & A0 & N \\ 
199757113 & J05274279-0819386 & HD 35929 & 7.68 & A9-F0 & N \\ 
199967382 & J05441880+0008403 & V351 Ori & 8.63 & A7-A9 & N \\ 
200511734 & J05301903+1120199 & HD 244314 & 9.87 & A2-A3 & N \\ 
200517741 & J05315724+1117414 & HD 244604 & 9.23 & A2-A3 & N \\ 
207044437 & J06310692+1026049 & VY Mon & 12.74 & B6-B7 & N \\ 
207045303 & J06310363+1001133 & PDS 129 & 11.42 & F2-F3 & N \\ 
213103713 & J05042998-0347142 & UX Ori & 10.01 & A3-A6 & N \\ 
220322651 & J06405118+0944461 & HBC 222 & 11.47 & F3-F6 & N \\ 
229340364 & J06330519+1019199 & HD 259431 & 8.31 & B7 & N \\ 
254045303 & J19014081-3652337 & TY CrA & 8.65 & B8 & N \\ 
254387071 & J15404638-4229536 & HD 139614 & 8.03 & A7-A9 & N \\ 
255191560 & J15564188-4219232 & HD 142527 & 7.57 & F2-F3 & N \\ 
261539498 & J05081303+4214558 & VOS 84 & 12.85 & A8 & N \\ 
264460023 & J05244279+0143482 & V346 Ori & 9.9 & A7-A9 & N \\ 
265290711 & J02464395+5921168 & VOS 2047 & 12.52 & B9.5 & N \\ 
265464119 & J23584164+6626126 & LkHa 259 & 13.02 & A7-A9 & N \\ 
266773096 & J04054937+5128348 & VOS 200 & 12.13 & A0 & N \\ 
269112596 & J23172558+6050436 & MWC 1080 & 9.8 & B0 & N \\ 
272142683 & J23422673+6337387 & VOS 42 & 12.21 & B & N \\ 
272956728 & J11223166-5322114 & HD 98922 & 6.65 & B8-B9 & N \\ 
272995440 & J04301623+4852099 & VOS 898 & 12.25 & A0 & N \\ 
276661825 & J05370245-0137213 & HD 290770 & 9.2 & B8 & N \\ 
280773136 & J11332542-7011412 & HD 100546 & 6.66 & A0-A1 & N \\ 
283096205 & J06554005-0309506 & PDS 229N & 12.8 & B7 & Y \\ 
284214915 & J00365177+6329300 & VOS 70 & 11.94 & A7 & N \\ 
285920558 & J07591156-5022468 & GSC 8143-1225 & 12.18 & F1-F4 & N \\ 
286322252 & J02590513+6054041 & VOS 588 & 12.37 & B8 & N \\ 
287757971 & J05365925-0609164 & HD 37258 & 9.53 & A0-A1 & N \\ 
287840183 & J05371326-0635005 & BF Ori & 10.07 & A2-A3 & N \\ 
288924907 & J20590591+4805368 & VOS 69 & 12.26 & A3 & N \\ 
290064548 & J21035423+5015101 & LkHa 324 & 11.39 & B7-B8 & N \\ 
293537745 & J15580967-5351349 & Hen 3-1121S & 10.47 & O-B0 & Y \\ 
297154074 & J20480478+4347258 & V2019 Cyg & 10.46 & B8 & N \\ 
299797908 & J13524285-6332492 & Hen 3-938 & 11.36 & O & N \\ 
302824830 & J05321413+1703292 & HD 36408 & 6.08 & B6-B7 & Y \\ 
305447083 & J11034056-5925590 & HD 96042 & 8.41 & B0 & N \\ 
307606851 & J15154894-3708558 & HD 135344 & 7.71 & F0-F2 & N \\ 
309449806 & J06071539+2957550 & GSC 1876-0892 & 12.71 & B4 & Y \\ 
311826580 & J00385008+5954101 & VOS 32 & 13.82 & G0 & N \\ 
312493387 & J20140898+4029502 & VOS 250 & 12.46 & A3 & N \\ 
317789227 & J11311339-6350120 & VOS 290 & 10.96 & B2 & N \\ 
319939254 & J11374699-6342562 & VOS 1336 & 12.88 & B2.5-B9.5 & Y \\ 
320547523 & J11394445-6010278 & HD 101412 & 9.07 & B9-A0 & N \\ 
322312832 & J21425018+6606351 & V361 Cep & 9.53 & B4 & N \\ 
332856261 & J05374708-0642301 & HD 37357 & 7.38 & B9 & N \\ 
332856968 & J05380931-0649166 & V1787 Ori & 12.0 & A3-A4 & N \\ 
332914217 & J05385862-0716457 & V599 Ori & 11.66 & A4-A6 & N \\ 
332970982 & J05401176-0942110 & V350 Ori & 11.4 & A1-A3 & N \\ 
335623526 & J22114281+6036433 & VOS 48 & 13.15 & G0 & N \\ 
336518731 & J14185145-6124377 & VOS 1245 & 13.55 & A0 & Y \\ 
339989392 & J16544485-3653185 & AK Sco & 8.79 & F7-G5 & N \\ 
341984965 & J22111311+5253341 & VOS 475 & 13.02 & B9.5 & Y \\ 
349354945 & J03291977+3124572 & BD+30 549 & 9.9 & B8 & N \\ 
355320922 & J08442365-5956578 & GSC 8581-2002 & 10.55 & B9 & N \\ 
355676575 & J19251419+2240382 & VOS 481 & 12.52 & F6-G4 & N \\ 
356608297 & J20510271+4349318 & HBC 705 & 12.34 & B2 & N \\ 
357040880 & J12000511-7811346 & HD 104237 & 6.35 & A5-A7 & N \\ 
358568109 & J05505477+2014476 & UCAC4 552-019438 & 12.33 & B7-B8 & N \\ 
358901633 & J23101132+6555587 & VOS 35 & 12.53 & F & N \\ 
362789897 & J19422811+2306515 & VOS 105 & 13.48 & F5 & N \\ 
365776906 & J05273833+1125389 & CO Ori & 10.84 & F3-F6 & N \\ 
367036276 & J20473745+4347249 & V1977 Cyg & 10.22 & B8 & N \\ 
367420555 & J22165406+7003450 & BO Cep & 10.96 & F3-F5 & N \\ 
369496705 & J12484238-5954350 & Hen 3-823 & 10.08 & B4 & N \\ 
372070676 & J22014287+6944364 & BH Cep & 10.68 & F1-F4 & N \\ 
377617067 & J23122637+6058129 & VOS 1026 & 11.87 & B0.5 & N \\ 
377893466 & J23160757+6001437 & VOS 672 & 12.42 & B1.5-B8 & N \\ 
385894679 & J19255874+2112313 & WW Vul & 10.12 & A3-A4 & N \\ 
386040374 & J07255610-1410435 & HD 58647 & 6.71 & B8 & N \\ 
387813185 & J21013691+6809476 & HD 200775 & 6.77 & B4 & Y \\ 
388143588 & J22383182+5550052 & MWC 655 & 8.86 & B2-B3 & N \\ 
388935225 & J05362543-0642577 & V380 Ori & 9.78 & B9 & Y \\ 
390384783 & J12174750-5943590 & GSC 8645-1401 & 11.98 & A9-F2 & N \\ 
395172881 & J21400313+4802106 & VOS 759 & 11.71 & A8 & N \\ 
396284614 & J19111124+1547155 & HD 179218 & 7.32 & B9 & N \\ 
396300742 & J00093755+5813106 & MQ Cas & 12.97 & B9 & N \\ 
401214336 & J21380845+5726476 & GSC 3975-0579 & 11.26 & A0-A1 & N \\ 
408081161 & J19555428+1932354 & VOS 199 & 12.7 & F1 & N \\ 
409323949 & J07193593-1739180 & PDS 27 & 11.79 & B4-B5 & N \\ 
409406992 & J16130668-5023199 & WRAY 15-1435 & 11.67 & B2-B3 & N \\ 
410736126 & J03283262+5113543 & VOS 934 & 11.94 & A2 & N \\ 
410951921 & J16160533-5046035 & VOS 911 & 12.03 & B2-B4 & N \\ 
412645923 & J21590612+6040533 & VOS 473 & 12.19 & A0 & N \\ 
420018149 & J00431825+6154402 & V594 Cas & 9.73 & B7-B8 & N \\ 
420307432 & J22585908+5918024 & VOS 1126 & 12.59 & B2 & N \\ 
427352999 & J05341416-0536542 & HBC 442 & 9.65 & F8-G1 & N \\ 
427373765 & J05344698-0534145 & HD 36917 & 6.1 & B8 & N \\ 
427395300 & J05350983-0527532 & HD 36982 & 8.17 & B2 & N \\ 
427395473 & J05353136-0533088 & NV Ori & 9.37 & F0-F2 & N \\ 
427396230 & J05355043-0528349 & T Ori & 10.87 & A0-A1 & N \\ 
427577547 & J03390056+2941457 & PDS 4 & 10.3 & B9-A0 & N \\ 
427938329 & J21523408+4713436 & AS 477 & 9.7 & B8-B9 & N \\ 
428476066 & J00313068+6158509 & VX Cas & 11.19 & B9 & N \\ 
430006517 & J21351915+5736382 & VOS 121 & 12.75 & B9.5 & N \\ 
434633379 & J23022850+6316309 & VOS 274 & 12.54 & A3 & N \\ 
434972921 & J23050746+6215362 & V374 Cep & 9.58 & B3-B5 & N \\ 
436098606 & J05350960+1001515 & HD 245185 & 9.79 & B9 & N \\ 
437994564 & J06184553+1516522 & MWC 137 & 10.23 & B1 & N \\ 
441152806 & J02440733-1351312 & HD 17081 & 4.38 & B7 & N \\ 
441744490 & J09424032-5615341 & PDS 297 & 11.85 & B8 & N \\ 
444356276 & J00231367+6159318 & VOS 787 & 12.9 & B1.5 & N \\ 
445062554 & J06045913-1629039 & HD 41511 & 4.18 & B9-A0 & N \\ 
449260863 & J05194140+0538428 & HD 34700 & 8.61 & F8-G1 & N \\ 
452035938 & J11330559-5419285 & HD 100453 & 7.5 & A9-F0 & N \\ 
452248316 & J19471649+2719558 & VOS 104 & 12.28 & A2 & Y \\ 
454215115 & J15032380-6322588 & DG Cir & 11.72 & A2-A3 & N \\ 
454274408 & J15045606-6307526 & HD 132947 & 8.91 & B8-B9 & N \\ 
454291762 & J11080329-7739174 & HD 97048 & 8.07 & B8 & N \\ 
455573060 & J21541877+4712096 & LkHa 257 & 12.51 & A1-A2 & N \\ 
456030854 & J22411697+6302377 & VOS 56 & 13.18 & A4 & Y \\ 
457231768 & J05270547+0025075 & HD 290409 & 9.88 & B9 & N \\ 
459571845 & J10520871-5612066 & PDS 322 & 11.82 & B2 & N \\ 
459699513 & J10532726-5825245 & HD 94509 & 9.15 & B7 & N \\ 
459990291 & J05240804+0227468 & HD 287823 & 9.58 & A4-A6 & N \\ 
460571412 & J10330498-6019513 & HD 305298 & 9.46 & O & N \\ 
462436722 & J10043028-5839521 & HD 87643 & 8.04 & B4 & N \\ 
462878642 & J10100032-5702073 & PDS 37 & 11.51 & B2-B3 & N \\ 
463597276 & J17244469-3843514 & MWC 878 & 9.62 & B0 & N \\ 
465196254 & J17340462-3923413 & HD 323771 & 10.72 & B5-B6 & N \\ 
465790701 & J10572427-6253132 & PDS 324 & 13.62 & B2-B3 & N \\ 
467356861 & J22213319+7340270 & SV Cep & 10.03 & A3-A4 & N \\ 
469534433 & J21361422+5721309 & AS 470 & 11.49 & A4-A7 & Y \\ 
642035133 & J03493638+3858556 & XY Per A & 9.12 & B9 & Y \\ 
709004859 & J06033705-1453025 & PDS 22 & 10.72 & B9 & Y \\ 
816058534 & J08550867-4327596 & HD 76534 & 8.2 & B2 & Y \\ 
1139218134 & J15542182-5519443 & HD 141926 & 8.21 & B0 & Y \\
\enddata
\tablecomments{The sample of stars observed with {\em TESS} as part of this work. Alternate names and spectral types are taken from \cite{2021AA...650A.182G}, whereas magnitudes, TIC ids, and 2MASS ids are based on cross matching with the TESS Input Catalog. The final column indicates whether a target is deemed to be blended, as explained in the text.}
\end{deluxetable*}

\section{{\em TESS} observations and light curve production}
\label{sec:lcs}

The {\em TESS} mission has been underway since mid-2018, with new regions of sky (``sectors") observed approximately every 27 days. Many stars have been observed in multiple sectors, as camera fields overlap and {\em TESS's} extended mission campaigns have also revisited sky regions. During the primary mission (2018-2020; sectors 1-26), so-called Full Frame Images (FFIs) were acquired every 30 minutes during each sector. For the first extended mission (2020-2022; sectors 27-55), the FFI cadence was raised to once every 10 minutes. While {\em TESS} has continued with further extended missions, for the purposes of this work, we only consider FFI data taken through sector 45 (i.e., from 25 July 2018 through 2 December 2021).

From the initial list of 318 HArchiBe members, we checked which stars were observed by {\em TESS} in any of the sectors 1--45 and removed 39 objects with no data. For the remaining stars, we downloaded FFI data in the form of 11$\times$11 target pixel files (TPFs), as produced by the \texttt{astrocut} Python package. Many of the targets did not have publicly available light curves, so we ran several custom pipelines to create our own. 

The first pipeline consisted of circular aperture photometry with radii of one, two, three, and four pixels. The aperture was centered on the known coordinates of each star, with no shifts. Upon summing the enclosed flux, we subtracted background, as determined from the median flux level of the TPF after an iterative five-sigma cut to remove point sources. 

The second pipeline employed the \texttt{lightkurve} package to measure fluxes using non-circular aperture masks. We again created several versions, included 1-pixel and 2$\times$2-pixel masks centered on the target, as well as asymmetric masks defined by all pixels exceeding an eight or 12-sigma threshold relative to the median flux level. Upon summing the flux within these apertures, the background was estimated based on the median flux of pixels exterior to the mask, minus the fifth percentile of values over those regions. Background subtraction corrected most of the artificial time variability due to, e.g., scattered Earth light, but did not necessarily remove such trends completely. To further correct systematic effects, we applied the \texttt{lightcurve} RegressionCorrector function to subtract off up to four orders of principal component vectors associated with the background.  For cases where this was unsuccessful in reducing pronounced systematic trends, we instead applied \texttt{lightkurve's} pixel-level decorrelation corrector function, again with up to four principle components.

For stars included in the public TESS-SPOC reduction \citep{2020RNAAS...4..201C}, we considered both the simple aperture photometry (SAP) and pre-search data conditioning (PDC) products. The latter includes a prescription for removing common trends among neighboring stars, and thus typically displays a reduction in artificial systematics, similar to the \texttt{lightkurve} RegressionCorrector and PLDCorrector methods.  

Finally, for one star (TIC~460571412) for which we were unable to mitigate systematics with the above pipelines, we created a light curve using the \texttt{eleanor} package \citep{2019PASP..131i4502F}. Upon creation of all light curves for each star and in each sector, we selected the option with the lowest random and systematic noise levels on visual comparison. 
Most HAeBes are relatively bright, but some of the more distant ones fall outside of {\em TESS's} optimal range of magnitude 7-16. If a light curve was consistent with background noise and no star was visible in the TPF, we removed it from consideration. Likewise, if there was a clear blend and the aperture could not be centered, we also discarded the object. In some cases, a light curve version had large (multi-day) gaps, such that the variability pattern could not be inferred; we removed these sectors from the set. 

These procedures resulted in a set of 188 Herbig Ae/Be stars with a viable light curve in at least one {\em TESS} sector. This includes a total of 452 final light curves, all of which are displayed in the Appendix. Those that were not derived directly from the TESS-SPOC pipeline (``SAP" and ``PDC" apertures in Table 2) are made publicly available at MAST as a High Level Science Product via \dataset[10.17909/fkzx-2z29]{\doi{10.17909/fkzx-2z29}}. The corresponding list of observed TIC identification numbers, sectors, and chosen photometric method, is provided in Table~2. For each light curve, we retained data quality flags from the originating {\em TESS} TPF; all flagged points are included in the light curve files at MAST, but non-zero (i.e., bad) quality points are omitted from the light curve plots here.

\startlongtable
\begin{deluxetable*}{c c c c c c c c}
\tablecaption{Properties of the {\em TESS} Herbig light curve sample \label{tab:lcs}}
\tablehead{
\colhead{TIC id} & \colhead{Sector} & \colhead{Phot ap} & \colhead{Vartype} & \colhead{Amplitude} & \colhead{Timescale [d]} & \colhead{Q} & \colhead{M} \\
}
\startdata
4206238 & 6 & sap & APD & 0.0367 & 17.42 & 1.0 & 0.76 \\ 
4206238 & 32 & sap & APD & 0.0303 & 26.92 & 1.0 & 1.07 \\ 
8806247 & 12 & 1a & SC & 0.035 & 24.29 & 1.0 & 0.14 \\ 
8806247 & 39 & 1a & SC & 0.0369 & 19.54 & 1.0 & 0.08 \\ 
11199521 & 6 & sap & SC & 1.1557 & 16.6 & 1.0 & -0.12 \\ 
11199521 & 32 & sap & N & 0.021 & 7.94 & 0.88 & 0.07 \\ 
11402323 & 6 & 3a & S & 0.0344 & 3.62 & 0.74 & 0.02 \\ 
12161661 & 6 & sap & SC & 0.0432 & 19.75 & 1.0 & 0.04 \\ 
12161661 & 33 & sap & SC & 0.0381 & 31.39 & 1.0 & -0.26 \\ 
12876844 & 24 & 3a & EB & 0.0417 & 7.14 & 0.08 & 1.44 \\ 
13002932 & 14 & 3a & S & 0.1245 & 26.54 & 1.0 & 0.35 \\ 
13002932 & 15 & 3a & S & 0.0885 & 24.52 & 1.0 & -0.25 \\ 
13002932 & 41 & 3a & S & 0.0455 & 5.33 & 1.0 & 0.04 \\ 
13417240 & 41 & 2m & L & 0.0856 & 25.17 & 1.0 & 0.58 \\ 
14740791 & 14 & 1a & EB & 0.3339 & 0.29 & 0.13 & 0.19 \\ 
14740791 & 15 & 1a & EB & 0.3808 & 0.29 & 0.1 & 0.26 \\ 
14740791 & 41 & 1a & EB & 0.2986 & 0.29 & 0.06 & 0.58 \\ 
15898682 & 14 & 8m & S & 0.0244 & 14.83 & 1.0 & 0.38 \\ 
15898682 & 15 & 8m & S & 0.0253 & 16.96 & 1.0 & -0.26 \\ 
15898682 & 41 & 8m & S & 0.0311 & 23.49 & 1.0 & 0.45 \\ 
17130641 & 14 & 8m & SC & 0.0076 & 22.96 & 1.0 & -0.45 \\ 
17130641 & 15 & 8m & SC & 0.0061 & 6.33 & 1.0 & -0.19 \\ 
17130641 & 41 & 8m & SC & 0.0102 & 12.63 & 1.0 & 0.29 \\ 
24344701 & 5 & 2a & L & 0.1298 & 7.17 & 1.0 & 0.3 \\ 
24344701 & 32 & 2a & APDC & 0.1435 & 18.22 & 1.0 & 0.54 \\ 
30243973 & 8 & pdc & MP & 0.0125 & 1.45 & 0.43 & 0.46 \\ 
30243973 & 9 & pdc & MP & 0.0166 & 1.46 & 0.69 & -0.28 \\ 
30243973 & 35 & pdc & MP & 0.0132 & 1.47 & 0.38 & -0.21 \\ 
30243973 & 36 & pdc & MP & 0.0112 & 1.46 & 0.39 & 0.34 \\ 
31638754 & 9 & 3a & SC & 0.0167 & 7.81 & 0.38 & 0.02 \\ 
31638754 & 35 & 3a & SC & 0.0197 & 19.81 & 0.93 & 0.2 \\ 
31638754 & 36 & 3a & SC & 0.0179 & 3.51 & 0.86 & -0.19 \\ 
35485982 & 6 & 2m & S & 0.0176 & 5.54 & 1.0 & 0.26 \\ 
35485982 & 7 & 2m & S & 0.0104 & 2.19 & 0.8 & 0.06 \\ 
35485982 & 33 & 2m & APDC & 0.0232 & 9.47 & 1.0 & 0.47 \\ 
45124112 & 43 & 8m & QPSC & 0.0068 & 0.55 & 0.82 & -0.06 \\ 
45124112 & 44 & 8m & QPSC & 0.0048 & 0.55 & 0.83 & 0.21 \\ 
46749954 & 12 & 8m & QPS & 0.0068 & 5.1 & 0.47 & -0.07 \\ 
46749954 & 39 & 8m & QPS & 0.0086 & 5.44 & 0.64 & 0.27 \\ 
50656450 & 6 & pdc & MP & 0.0044 & 10.87 & 0.96 & 0.08 \\ 
50656450 & 32 & pdc & MP & 0.0103 & 1.61 & 1.0 & 0.19 \\ 
50745417 & 6 & 3a & S & 0.6238 & 22.75 & 1.0 & 0.35 \\ 
50745417 & 32 & 3a & S & 0.3894 & 21.29 & 1.0 & 0.0 \\ 
53898232 & 6 & 2a & APDC & 0.6157 & 16.63 & 1.0 & 0.38 \\ 
53898232 & 7 & 2a & APDC & 0.7859 & 15.17 & 1.0 & 0.77 \\ 
53898232 & 33 & 2a & APDC & 0.5644 & 30.56 & 1.0 & 0.81 \\ 
54003409 & 3 & pdc & N & 0.0001 & 3.33 & 0.86 & 0.11 \\ 
54185990 & 43 & sap & APDC & 0.0569 & 12.19 & 1.0 & 0.67 \\ 
54185990 & 44 & sap & SC & 0.0476 & 24.09 & 1.0 & -0.38 \\ 
54185990 & 45 & sap & LC & 0.0277 & 28.25 & 1.0 & -0.12 \\ 
63705801 & 15 & 3a & S & 0.0342 & 20.79 & 1.0 & 0.54 \\ 
63705801 & 16 & 3a & S & 0.0255 & 4.17 & 1.0 & 0.48 \\ 
66918331 & 6 & 2a & L & 0.2724 & 31.0 & 1.0 & -0.99 \\ 
66918331 & 33 & 2a & L & 2.7715 & 0.35 & 1.0 & -1.14 \\ 
67671603 & 12 & sap & QPS & 0.0152 & 1.12 & 0.74 & 0.12 \\ 
72553036 & 6 & pdc & S & 0.0083 & 3.96 & 0.92 & 0.1 \\ 
72780071 & 6 & 3a & QPS & 0.0091 & 2.36 & 0.67 & -0.19 \\ 
72781767 & 6 & sap & MPC & 0.0118 & 2.36 & 0.69 & -0.01 \\ 
73816033 & 43 & 3a & MP & 0.0264 & 18.71 & 1.0 & -0.04 \\ 
73816033 & 44 & 3a & APDC & 0.0512 & 17.94 & 0.99 & 0.96 \\ 
73816033 & 45 & 3a & MP & 0.0176 & 11.72 & 1.0 & 0.09 \\ 
74632969 & 43 & 2a & S & 1.0786 & 8.25 & 0.98 & -0.24 \\ 
74632969 & 44 & 2a & S & 0.6052 & 8.65 & 0.89 & 0.12 \\ 
74632969 & 45 & 2a & S & 0.7232 & 6.67 & 0.88 & -0.64 \\ 
75242497 & 43 & 2a & LC & 1.2832 & 7.64 & 1.0 & -1.1 \\ 
75242497 & 44 & 2a & SC & 1.0612 & 19.02 & 1.0 & -0.21 \\ 
75242497 & 45 & 2a & SC & 0.7765 & 8.1 & 1.0 & -0.26 \\ 
75242566 & 43 & 8m & QPS & 0.0107 & 2.08 & 0.58 & 0.1 \\ 
75242566 & 44 & 8m & QPS & 0.0107 & 2.1 & 0.44 & 0.19 \\ 
75242566 & 45 & 8m & QPS & 0.0125 & 2.38 & 0.64 & 0.02 \\ 
81297869 & 7 & 8m & MP & 0.0057 & 17.9 & 0.98 & 0.08 \\ 
81297869 & 8 & 8m & MP & 0.0071 & 0.06 & 0.8 & 0.16 \\ 
81297869 & 34 & 8m & MP & 0.0145 & 25.31 & 0.99 & 0.06 \\ 
81297869 & 35 & 8m & MP & 0.0171 & 33.82 & 0.9 & 0.04 \\ 
85100457 & 19 & 4a & S & 0.035 & 5.17 & 1.0 & 0.56 \\ 
85100457 & 43 & 4a & S & 0.0304 & 23.47 & 1.0 & -0.19 \\ 
85100457 & 44 & 4a & S & 0.0406 & 4.31 & 1.0 & 0.1 \\ 
89189882 & 6 & 2m & QPS & 0.0108 & 0.75 & 0.61 & 0.06 \\ 
89189882 & 33 & 2m & QPS & 0.01 & 0.75 & 0.57 & -0.22 \\ 
89189882 & 43 & 2m & QPS & 0.0112 & 0.76 & 0.47 & -0.09 \\ 
89189882 & 45 & 2m & QPS & 0.0118 & 0.75 & 0.56 & -0.01 \\ 
95586779 & 7 & 1a & EB & 0.0384 & 2.32 & 0.63 & 0.36 \\ 
95586779 & 33 & 1a & EB & 0.0595 & 2.32 & 0.37 & 0.56 \\ 
96680681 & 19 & 4a & S & 0.0354 & 13.56 & 1.0 & -0.36 \\ 
96680681 & 43 & 4a & S & 0.017 & 18.8 & 1.0 & -0.18 \\ 
96680681 & 44 & 4a & S & 0.025 & 20.13 & 1.0 & 0.21 \\ 
97638779 & 6 & 2a & QPS & 0.1455 & 1.63 & 0.73 & 0.54 \\ 
97638779 & 33 & 2a & S & 0.1565 & 13.75 & 1.0 & -0.03 \\ 
97638779 & 43 & 2a & QPS & 0.3268 & 1.63 & 0.71 & 0.46 \\ 
97638779 & 44 & 2a & QPS & 0.2225 & 1.63 & 0.34 & 0.07 \\ 
97638779 & 45 & 2a & QPS & 0.1204 & 1.63 & 0.19 & 0.2 \\ 
97807159 & 6 & 3a & S & 0.0593 & 12.67 & 0.815 & 0.19 \\ 
97807159 & 33 & 3a & B & 0.0408 & 9.91 & 0.85 & -0.33 \\ 
97807159 & 43 & 3a & S & 0.0574 & 8.51 & 0.91 & -0.01 \\ 
97807159 & 44 & 3a & B & 0.0562 & 6.97 & 1.0 & -0.3 \\ 
97807159 & 45 & 3a & B & 0.0488 & 5.1 & 0.6 & -0.54 \\ 
101186168 & 12 & 2m & APDC & 0.0495 & 20.79 & 1.0 & 0.78 \\ 
101186168 & 39 & 2m & APDC & 0.0516 & 16.01 & 1.0 & 0.27 \\ 
107406463 & 7 & 1m & QPSC & 0.011 & 6.41 & 0.27 & 0.12 \\ 
107406463 & 34 & 1m & QPSC & 0.0123 & 11.9 & 0.61 & 0.11 \\ 
107544865 & 7 & 1a & S & 0.5542 & 23.79 & 1.0 & 0.12 \\ 
107544865 & 34 & 1a & S & 0.3246 & 29.36 & 1.0 & -0.31 \\ 
110990131 & 40 & 8m & S & 0.0179 & 32.0 & 1.0 & 0.17 \\ 
110990131 & 41 & 8m & QPDC & 0.1187 & 5.92 & 0.36 & 0.34 \\ 
120751782 & 6 & 2a & LC & 0.096 & 10.33 & 1.0 & -0.55 \\ 
120751782 & 33 & 2a & NC & 0.0477 & 18.96 & 1.0 & 0.06 \\ 
121137860 & 12 & 2m & QPSC & 0.0417 & 1.89 & 0.6 & 0.31 \\ 
121137860 & 39 & 2m & QPSC & 0.1048 & 1.88 & 0.65 & 0.38 \\ 
125641610 & 7 & 2a & S & 0.0872 & 9.37 & 0.83 & -0.36 \\ 
125641610 & 33 & 2a & L & 0.2231 & 8.13 & 1.0 & -0.27 \\ 
125644224 & 7 & 1m & S & 0.0271 & 6.4 & 1.0 & -0.11 \\ 
125644224 & 33 & 1m & S & 0.0382 & 6.27 & 1.0 & -0.12 \\ 
136712091 & 15 & 2m & LC & 0.1735 & 33.29 & 1.0 & 0.72 \\ 
136712091 & 16 & 2m & SC & 0.0657 & 23.04 & 1.0 & 0.27 \\ 
138670679 & 15 & 2m & QPSC & 0.0124 & 2.08 & 0.71 & -0.06 \\ 
138670679 & 16 & 2m & QPSC & 0.0212 & 19.37 & 1.0 & 0.66 \\ 
138941394 & 6 & sap & APDC & 0.0834 & 16.71 & 1.0 & 0.34 \\ 
138941394 & 32 & sap & APDC & 0.0626 & 12.93 & 1.0 & 0.33 \\ 
139019368 & 16 & 2m & SC & 0.0575 & 21.06 & 1.0 & 0.05 \\ 
139715411 & 7 & pdc & P & 0.0083 & 0.73 & 0.05 & -0.03 \\ 
139715411 & 34 & pdc & P & 0.0134 & 0.73 & 0.02 & 0.12 \\ 
149542947 & 2 & 8m & MP & 0.0015 & 0.11 & 0.65 & 0.15 \\ 
149542947 & 3 & 8m & MP & 0.0014 & 0.11 & 0.34 & 0.04 \\ 
149542947 & 5 & 8m & MP & 0.0014 & 0.11 & 0.53 & 0.03 \\ 
149542947 & 6 & 8m & MP & 0.0013 & 0.11 & 0.48 & 0.01 \\ 
149542947 & 9 & 8m & MP & 0.0016 & 0.11 & 0.39 & 0.05 \\ 
149542947 & 11 & 8m & MP & 0.0013 & 0.11 & 0.27 & 0.01 \\ 
149542947 & 12 & 8m & MP & 0.0013 & 0.11 & 0.45 & 0.05 \\ 
149542947 & 13 & 8m & MP & 0.0014 & 0.11 & 0.47 & 0.02 \\ 
149542947 & 28 & 8m & MP & 0.0016 & 0.11 & 0.72 & 0.09 \\ 
149542947 & 29 & 8m & MP & 0.0014 & 0.11 & 0.66 & -0.03 \\ 
149542947 & 30 & 8m & MP & 0.0015 & 0.11 & 0.69 & -0.01 \\ 
149542947 & 31 & 8m & MP & 0.0014 & 0.11 & 0.61 & -0.01 \\ 
149542947 & 32 & 8m & MP & 0.0017 & 0.11 & 0.78 & -0.11 \\ 
149542947 & 33 & 8m & MP & 0.0014 & 0.11 & 0.65 & 0.03 \\ 
149542947 & 35 & 8m & MP & 0.0015 & 0.11 & 0.79 & 0.04 \\ 
149542947 & 36 & 8m & MP & 0.0016 & 0.11 & 0.59 & 0.0 \\ 
149542947 & 38 & 8m & MP & 0.0015 & 0.11 & 0.66 & -0.04 \\ 
149542947 & 39 & 8m & MP & 0.0014 & 0.11 & 0.6 & -0.05 \\ 
163268072 & 12 & 8m & MP & 0.0084 & 0.66 & 0.57 & -0.2 \\ 
163268072 & 39 & 8m & MP & 0.0119 & 0.66 & 0.83 & -0.13 \\ 
166896955 & 11 & 3a & S & 0.0349 & 5.12 & 1.0 & 0.31 \\ 
166896955 & 38 & 3a & S & 0.0471 & 28.53 & 1.0 & -0.59 \\ 
168627773 & 7 & 8m & EB & 0.0163 & 3.57 & 0.06 & 1.36 \\ 
168627773 & 33 & 8m & EB & 0.0199 & 3.57 & 0.16 & 1.41 \\ 
178486401 & 7 & 2m & APDC & 0.0404 & 23.33 & 1.0 & 0.77 \\ 
178486401 & 33 & 2m & APDC & 0.0235 & 8.97 & 0.84 & 0.45 \\ 
181142391 & 9 & 8m & QPSC & 0.0146 & 6.41 & 0.23 & 0.28 \\ 
181142391 & 35 & 8m & QPSC & 0.0132 & 6.41 & 0.52 & -0.16 \\ 
183658281 & 8 & pdc & APDC & 0.006 & 3.54 & 1.0 & 0.78 \\ 
183658281 & 34 & pdc & QPSC & 0.002 & 6.1 & 0.63 & 0.01 \\ 
183658281 & 35 & pdc & QPSC & 0.0073 & 4.81 & 0.52 & 0.23 \\ 
189283710 & 15 & 1a & S & 0.1034 & 8.06 & 1.0 & 0.19 \\ 
189283710 & 41 & 1a & S & 0.1108 & 14.67 & 1.0 & -0.31 \\ 
189286260 & 15 & 2m & LC & 0.1295 & 21.83 & 1.0 & 0.84 \\ 
189286260 & 41 & 2m & SC & 0.0315 & 16.51 & 1.0 & 0.28 \\ 
190255749 & 8 & 1m & APDC & 0.0883 & 3.96 & 0.85 & 0.41 \\ 
190255749 & 9 & 1m & SC & 0.1125 & 23.19 & 1.0 & -0.2 \\ 
199757113 & 6 & pdc & P & 0.0039 & 1.37 & 0.09 & 0.08 \\ 
199757113 & 32 & pdc & P & 0.003 & 1.37 & 0.2 & -0.03 \\ 
199967382 & 6 & 3a & APDC & 0.1017 & 14.44 & 1.0 & 0.8 \\ 
200511734 & 6 & sap & S & 0.0236 & 16.21 & 1.0 & -0.7 \\ 
200511734 & 32 & sap & S & 0.0195 & 0.02 & 0.64 & -0.04 \\ 
200517741 & 6 & pdc & QPSC & 0.0047 & 3.29 & 0.69 & -0.09 \\ 
200517741 & 32 & pdc & QPSC & 0.0037 & 3.05 & 0.79 & 0.18 \\ 
207044437 & 6 & 2a & L & 0.072 & 20.42 & 1.0 & 0.1 \\ 
207044437 & 33 & 2a & L & 0.1602 & 28.75 & 1.0 & -0.36 \\ 
207045303 & 6 & pdc & N & 0.0027 & 11.5 & 1.0 & -0.05 \\ 
207045303 & 33 & pdc & N & 0.0056 & 8.88 & 1.0 & 0.02 \\ 
213103713 & 5 & 3a & SC & 1.1527 & 12.75 & 1.0 & -0.22 \\ 
213103713 & 32 & 3a & APDC & 0.7898 & 11.17 & 1.0 & 0.0 \\ 
220322651 & 6 & pdc & QPSC & 0.0198 & 6.58 & 0.55 & 0.12 \\ 
220322651 & 33 & pdc & QPSC & 0.0266 & 10.87 & 0.19 & -0.12 \\ 
229340364 & 6 & 3a & S & 0.0445 & 2.96 & 1.0 & -0.09 \\ 
229340364 & 33 & 3a & BC & 0.0219 & 3.57 & 0.82 & 0.12 \\ 
254045303 & 13 & sap & EB & 0.1282 & 2.87 & 0.14 & 1.24 \\ 
254387071 & 12 & sap & MP & 0.0042 & 0.05 & 0.35 & 0.1 \\ 
254387071 & 38 & sap & MP & 0.0054 & 0.05 & 0.78 & 0.04 \\ 
255191560 & 12 & sap & L & 0.0187 & 10.5 & 1.0 & -0.49 \\ 
261539498 & 19 & 2a & APD & 0.1095 & 13.25 & 1.0 & 0.44 \\ 
264460023 & 6 & pdc & MP & 0.0068 & 5.73 & 0.98 & 0.05 \\ 
264460023 & 32 & pdc & MP & 0.0059 & 0.03 & 0.8 & 0.02 \\ 
265290711 & 18 & 8m & EB & 0.0074 & 0.32 & 0.56 & 0.65 \\ 
265464119 & 17 & 2a & S & 0.0621 & 10.19 & 1.0 & -0.08 \\ 
265464119 & 18 & 2a & S & 0.0653 & 11.79 & 1.0 & -0.36 \\ 
265464119 & 24 & 2a & S & 0.0606 & 10.75 & 0.91 & -0.2 \\ 
266773096 & 19 & 2a & S & 0.1174 & 32.46 & 0.79 & -0.13 \\ 
269112596 & 17 & 3a & QPS & 0.1348 & 1.44 & 0.74 & -0.16 \\ 
269112596 & 24 & 3a & QPS & 0.1505 & 1.44 & 0.72 & 0.12 \\ 
272142683 & 17 & 1a & L & 1.6777 & 11.08 & 1.0 & -0.91 \\ 
272142683 & 18 & 1a & LC & 0.5175 & 6.04 & 1.0 & 0.05 \\ 
272142683 & 24 & 1a & S & 0.3147 & 16.42 & 1.0 & 0.0 \\ 
272956728 & 10 & 3a & QPSC & 0.0189 & 5.95 & 0.63 & 0.24 \\ 
272956728 & 37 & 3a & SC & 0.0296 & 16.03 & 1.0 & -0.21 \\ 
272995440 & 19 & 8m & SC & 0.0103 & 5.83 & 1.0 & -0.22 \\ 
276661825 & 6 & sap & B & 0.0132 & 20.81 & 1.0 & -0.25 \\ 
276661825 & 32 & sap & B & 0.0214 & 15.06 & 1.0 & -0.07 \\ 
280773136 & 11 & 3a & B & 0.0141 & 15.17 & 1.0 & -0.34 \\ 
280773136 & 37 & 3a & APD & 0.0623 & 24.61 & 1.0 & 1.17 \\ 
280773136 & 38 & 3a & APD & 0.0631 & 2.33 & 1.0 & 0.31 \\ 
283096205 & 6 & 2m & S & 0.0164 & 7.17 & 1.0 & -0.44 \\ 
283096205 & 33 & 2m & S & 0.0125 & 5.11 & 1.0 & 0.15 \\ 
284214915 & 17 & 2m & APDC & 0.0434 & 8.33 & 0.56 & 0.51 \\ 
284214915 & 18 & 2m & APDC & 0.1889 & 9.79 & 1.0 & 0.9 \\ 
285920558 & 7 & sap & APD & 0.3656 & 15.65 & 1.0 & 0.76 \\ 
285920558 & 8 & sap & APD & 0.187 & 5.04 & 1.0 & 0.28 \\ 
285920558 & 9 & sap & APD & 0.3248 & 16.06 & 1.0 & 0.61 \\ 
285920558 & 34 & sap & APD & 0.2403 & 9.99 & 1.0 & 0.28 \\ 
285920558 & 35 & sap & APD & 0.3486 & 14.67 & 1.0 & 0.27 \\ 
285920558 & 36 & sap & APD & 0.4845 & 7.64 & 0.86 & 0.48 \\ 
286322252 & 19 & 1m & S & 0.0205 & 8.08 & 1.0 & 0.06 \\ 
287757971 & 6 & 3a & APDC & 0.0498 & 6.37 & 1.0 & 0.33 \\ 
287840183 & 6 & pdc & QPDC & 0.1403 & 4.46 & 0.68 & 0.71 \\ 
287840183 & 32 & pdc & QPDC & 0.2515 & 3.91 & 0.66 & 0.32 \\ 
288924907 & 15 & sap & APDC & 0.4023 & 24.67 & 1.0 & -0.18 \\ 
288924907 & 16 & sap & LC & 0.7486 & 14.5 & 1.0 & -0.2 \\ 
290064548 & 15 & sap & SC & 0.018 & 33.83 & 1.0 & -0.13 \\ 
290064548 & 16 & sap & SC & 0.0204 & 30.37 & 1.0 & 0.12 \\ 
293537745 & 12 & 3a & QPS & 0.0257 & 4.39 & 0.17 & -0.2 \\ 
293537745 & 39 & 2a & QPS & 0.034 & 4.46 & 0.32 & 0.01 \\ 
297154074 & 15 & 2a & S & 0.0835 & 5.67 & 1.0 & 0.2 \\ 
297154074 & 41 & 2a & S & 0.0585 & 5.68 & 0.81 & -0.19 \\ 
299797908 & 11 & 1m & QPS & 0.0323 & 1.69 & 0.6 & 0.28 \\ 
299797908 & 38 & 1m & QPS & 0.0184 & 1.8 & 0.79 & -0.34 \\ 
302824830 & 6 & 8m & QPSC & 0.0005 & 2.98 & 0.52 & 0.43 \\ 
302824830 & 43 & 8m & P & 0.0004 & 1.19 & 0.61 & -0.0 \\ 
302824830 & 44 & 8m & P & 0.0003 & 1.19 & 0.68 & 0.01 \\ 
302824830 & 45 & 8m & PC & 0.0005 & 1.19 & 0.83 & 0.08 \\ 
305447083 & 10 & 8m & MP & 0.0099 & 0.49 & 0.36 & -0.17 \\ 
305447083 & 11 & 8m & MP & 0.0101 & 0.49 & 0.52 & -0.04 \\ 
307606851 & 11 & 3a & LC & 0.03 & 17.88 & 1.0 & 0.53 \\ 
307606851 & 38 & 3a & APDC & 0.0071 & 10.35 & 1.0 & 0.1 \\ 
309449806 & 44 & 1a & EBC & 0.0289 & 0.39 & 0.49 & 0.02 \\ 
311826580 & 17 & pld & QPSC & 0.0936 & 5.91 & 0.62 & -0.24 \\ 
311826580 & 18 & pld & LC & 0.2096 & 10.92 & 1.0 & -0.57 \\ 
311826580 & 24 & pld & LC & 0.1666 & 11.33 & 1.0 & 0.2 \\ 
312493387 & 15 & 2m & APDC & 0.0204 & 11.19 & 1.0 & 0.74 \\ 
312493387 & 41 & 2m & APD & 0.0443 & 11.78 & 1.0 & 0.52 \\ 
317789227 & 10 & 2a & QPS & 0.035 & 0.16 & 0.06 & -0.09 \\ 
317789227 & 11 & 2a & S & 0.0473 & 22.54 & 1.0 & -0.12 \\ 
317789227 & 37 & 2a & QPS & 0.0245 & 0.16 & 0.76 & -0.12 \\ 
317789227 & 38 & 2a & SC & 0.0349 & 6.41 & 1.0 & 0.3 \\ 
319939254 & 10 & 1a & QPSC & 0.0509 & 6.41 & 0.69 & -0.05 \\ 
319939254 & 11 & 1a & SC & 0.0381 & 6.58 & 0.75 & 0.2 \\ 
319939254 & 37 & 1a & LC & 0.0457 & 24.77 & 1.0 & 0.42 \\ 
319939254 & 38 & 1a & LC & 0.0364 & 22.06 & 1.0 & -0.27 \\ 
320547523 & 10 & pdc & N & 0.0011 & 8.17 & 1.0 & -0.11 \\ 
320547523 & 11 & pdc & N & 0.0019 & 4.75 & 1.0 & -0.31 \\ 
320547523 & 37 & pdc & N & 0.0022 & 7.12 & 1.0 & -0.0 \\ 
320547523 & 38 & pdc & N & 0.0019 & 11.81 & 1.0 & -0.13 \\ 
322312832 & 16 & pdc & MP & 0.0069 & 5.54 & 0.9 & 0.12 \\ 
322312832 & 17 & pdc & MP & 0.0069 & 13.25 & 0.94 & 0.01 \\ 
322312832 & 18 & pdc & MP & 0.007 & 0.32 & 0.84 & -0.06 \\ 
322312832 & 24 & pdc & MP & 0.0055 & 5.88 & 1.0 & 0.26 \\ 
332856261 & 6 & pdc & S & 0.0101 & 17.79 & 1.0 & -0.05 \\ 
332856261 & 32 & pdc & N & 0.0082 & 10.24 & 1.0 & -0.14 \\ 
332856968 & 6 & 2a & LC & 0.0262 & 20.83 & 1.0 & 0.18 \\ 
332856968 & 32 & 2a & N & 0.0138 & 6.36 & 1.0 & 0.03 \\ 
332914217 & 6 & 3a & LC & 0.3852 & 18.12 & 1.0 & 0.77 \\ 
332914217 & 32 & 3a & MP & 0.0624 & 29.21 & 0.98 & -0.13 \\ 
332970982 & 6 & 2a & L & 0.9485 & 16.25 & 1.0 & 0.15 \\ 
332970982 & 32 & 2a & S & 0.6188 & 12.3 & 1.0 & -0.55 \\ 
335623526 & 16 & 1a & APD & 0.3295 & 2.91 & 0.84 & 1.02 \\ 
335623526 & 17 & 1a & APD & 0.1548 & 8.54 & 0.92 & 0.85 \\ 
335623526 & 24 & 1a & APDC & 0.2592 & 14.4 & 1.0 & 0.51 \\ 
336518731 & 11 & 1m & P & 0.0086 & 0.24 & 0.08 & 0.17 \\ 
336518731 & 12 & 1m & P & 0.0145 & 0.24 & 0.13 & 0.28 \\ 
336518731 & 38 & 1m & P & 0.0083 & 0.24 & 0.12 & -0.04 \\ 
339989392 & 12 & sap & APD & 0.3583 & 12.58 & 1.0 & 0.32 \\ 
339989392 & 39 & sap & APD & 0.5934 & 18.53 & 1.0 & 0.52 \\ 
341984965 & 16 & 2m & P & 0.0044 & 0.09 & 0.06 & 0.08 \\ 
341984965 & 17 & 2m & P & 0.0024 & 0.09 & 0.4 & -0.02 \\ 
349354945 & 42 & pld & N & 0.0158 & 26.28 & 1.0 & -0.1 \\ 
349354945 & 43 & pld & N & 0.0341 & 24.71 & 1.0 & -0.1 \\ 
349354945 & 44 & pld & N & 0.0055 & 25.86 & 1.0 & 0.05 \\ 
355320922 & 8 & 12m & MP & 0.0017 & 4.48 & 0.99 & -0.1 \\ 
355320922 & 9 & 12m & MP & 0.0016 & 17.0 & 0.95 & 0.16 \\ 
355320922 & 10 & 12m & MP & 0.0015 & 14.12 & 0.94 & 0.16 \\ 
355320922 & 35 & 12m & MP & 0.0023 & 10.82 & 0.87 & 0.05 \\ 
355320922 & 37 & 12m & MP & 0.0022 & 0.06 & 0.63 & -0.01 \\ 
355676575 & 14 & 1m & SC & 0.0191 & 18.98 & 1.0 & 0.07 \\ 
355676575 & 40 & 1m & SC & 0.0366 & 13.25 & 1.0 & -0.06 \\ 
355676575 & 41 & 1m & S & 0.0152 & 5.1 & 0.81 & -0.13 \\ 
356608297 & 15 & 8m & MPC & 0.0331 & 29.0 & 0.94 & 0.32 \\ 
357040880 & 11 & 3a & APDC & 0.0685 & 5.56 & 0.82 & 0.39 \\ 
357040880 & 12 & 3a & BC & 0.0508 & 15.62 & 0.35 & -0.46 \\ 
357040880 & 38 & 3a & SC & 0.0405 & 13.89 & 0.43 & 0.17 \\ 
357040880 & 39 & 3a & BC & 0.0485 & 14.71 & 0.33 & -0.28 \\ 
358568109 & 44 & 2a & QPSC & 0.0791 & 6.41 & 0.47 & -0.1 \\ 
358568109 & 45 & 2a & S & 0.064 & 3.92 & 0.94 & 0.06 \\ 
358901633 & 17 & 1a & QPS & 0.0487 & 2.32 & 0.26 & 0.08 \\ 
358901633 & 18 & 1a & APD & 0.3326 & 18.77 & 1.0 & 1.02 \\ 
358901633 & 24 & 1a & QPSC & 0.2347 & 2.27 & 0.76 & 0.63 \\ 
362789897 & 40 & 2m & APDC & 0.042 & 9.0 & 1.0 & 0.18 \\ 
362789897 & 41 & 2m & APDC & 0.0309 & 4.08 & 1.0 & 0.64 \\ 
365776906 & 6 & 3a & L & 1.3834 & 13.92 & 1.0 & -0.25 \\ 
365776906 & 32 & 3a & L & 0.7687 & 14.89 & 1.0 & 0.14 \\ 
367036276 & 15 & 2a & S & 0.1165 & 15.42 & 1.0 & 0.0 \\ 
367036276 & 41 & 2a & S & 0.0932 & 10.67 & 1.0 & 0.38 \\ 
367420555 & 17 & sap & QPD & 0.1083 & 10.0 & 0.34 & 1.26 \\ 
367420555 & 18 & sap & QPD & 0.1188 & 11.36 & 0.23 & 1.24 \\ 
367420555 & 24 & sap & QPD & 0.0539 & 10.87 & 0.41 & 1.13 \\ 
367420555 & 25 & sap & QPD & 0.0807 & 10.42 & 0.5 & 1.18 \\ 
369496705 & 11 & 2m & MPC & 0.0482 & 6.21 & 0.99 & -0.2 \\ 
369496705 & 37 & 2m & MPC & 0.0402 & 10.47 & 0.96 & -0.12 \\ 
369496705 & 38 & 2m & MPC & 0.0765 & 6.81 & 0.89 & -0.53 \\ 
372070676 & 17 & 2a & APD & 0.319 & 8.75 & 0.85 & 0.87 \\ 
372070676 & 18 & 2a & S & 0.0967 & 9.62 & 1.0 & 0.75 \\ 
372070676 & 24 & 2a & SC & 0.2533 & 6.52 & 1.0 & 0.54 \\ 
377617067 & 17 & 2m & QPSC & 0.0043 & 4.33 & 0.86 & 0.08 \\ 
377617067 & 24 & 2m & QPSC & 0.0042 & 14.0 & 1.0 & -0.06 \\ 
377893466 & 17 & 2m & MPC & 0.0075 & 0.59 & 0.71 & 0.42 \\ 
377893466 & 24 & 2m & MPC & 0.0069 & 11.42 & 0.95 & -0.03 \\ 
385894679 & 14 & sap & APD & 0.1369 & 21.79 & 1.0 & 0.48 \\ 
386040374 & 7 & 12m & QPS & 0.0007 & 0.52 & 0.82 & -0.06 \\ 
386040374 & 34 & 12m & QPS & 0.0012 & 1.04 & 0.56 & 0.02 \\ 
387813185 & 16 & 4a & S & 0.0159 & 4.33 & 1.0 & -0.01 \\ 
387813185 & 18 & 4a & S & 0.0237 & 20.58 & 1.0 & -0.29 \\ 
387813185 & 24 & 4a & S & 0.0227 & 4.31 & 0.83 & 0.15 \\ 
387813185 & 25 & 4a & S & 0.0183 & 34.88 & 1.0 & -0.03 \\ 
388143588 & 16 & 2m & MPC & 0.0556 & 1.06 & 0.75 & 0.08 \\ 
388143588 & 17 & 2m & MPC & 0.0568 & 21.85 & 0.93 & -0.08 \\ 
388935225 & 32 & pdc & B & 0.0547 & 4.54 & 0.73 & -0.35 \\ 
390384783 & 10 & 2a & APDC & 0.0968 & 17.1 & 1.0 & 0.5 \\ 
390384783 & 11 & sap & SC & 0.0505 & 26.67 & 1.0 & 0.77 \\ 
390384783 & 37 & 2a & APDC & 0.2239 & 3.74 & 1.0 & 1.09 \\ 
390384783 & 38 & 2a & QPSC & 0.1849 & 3.79 & 0.81 & 0.19 \\ 
395172881 & 15 & 2m & MP & 0.0134 & 0.04 & 0.36 & 0.1 \\ 
395172881 & 16 & 2m & MP & 0.012 & 0.04 & 0.28 & 0.05 \\ 
396284614 & 40 & 8m & BC & 0.0029 & 4.68 & 1.0 & -0.31 \\ 
396300742 & 17 & 2a & SC & 1.3896 & 22.96 & 1.0 & -0.09 \\ 
396300742 & 18 & 2a & SC & 1.8148 & 15.98 & 1.0 & -1.14 \\ 
396300742 & 24 & 2a & SC & 1.3655 & 5.79 & 1.0 & -0.46 \\ 
401214336 & 16 & pdc & MPC & 0.0026 & 1.71 & 0.7 & 0.06 \\ 
401214336 & 17 & pdc & P & 0.0037 & 1.7 & 0.38 & -0.04 \\ 
408081161 & 41 & 2m & APDC & 0.0321 & 6.06 & 1.0 & 0.61 \\ 
409323949 & 7 & 2a & QPS & 0.884 & 1.91 & 0.64 & -0.18 \\ 
409323949 & 33 & 2a & QPS & 0.2566 & 1.88 & 0.32 & -0.12 \\ 
409323949 & 34 & 2a & QPS & 0.204 & 1.89 & 0.25 & 0.14 \\ 
409406992 & 12 & pld & EB & 0.0439 & 2.36 & 0.06 & 1.1 \\ 
409406992 & 39 & pld & EB & 0.052 & 2.36 & 0.06 & 1.27 \\ 
410736126 & 18 & 8m & N & 0.0038 & 3.68 & 0.78 & 0.35 \\ 
410951921 & 12 & 1a & PC & 0.0051 & 0.56 & 0.69 & 0.06 \\ 
410951921 & 39 & 1a & PC & 0.0058 & 0.56 & 0.46 & 0.03 \\ 
412645923 & 16 & sap & S & 0.0113 & 16.83 & 1.0 & -0.1 \\ 
412645923 & 17 & sap & S & 0.0114 & 19.96 & 1.0 & -0.05 \\ 
420018149 & 17 & sap & S & 0.1546 & 32.96 & 1.0 & -0.03 \\ 
420018149 & 18 & sap & S & 0.1584 & 18.69 & 1.0 & -0.41 \\ 
420018149 & 24 & sap & S & 0.0966 & 11.83 & 1.0 & -0.22 \\ 
420307432 & 17 & 1m & QPSC & 0.0087 & 16.44 & 0.95 & -0.24 \\ 
420307432 & 24 & 1m & QPSC & 0.0096 & 25.77 & 0.92 & -0.22 \\ 
427352999 & 6 & 4a & APDC & 0.0652 & 11.29 & 1.0 & 0.1 \\ 
427352999 & 32 & 4a & SC & 0.1069 & 17.36 & 0.92 & 0.72 \\ 
427373765 & 6 & 3a & S & 0.005 & 17.94 & 1.0 & -0.33 \\ 
427373765 & 32 & 3a & S & 0.0061 & 11.62 & 1.0 & 0.09 \\ 
427395300 & 32 & pdc & MP & 0.002 & 0.74 & 0.77 & 0.03 \\ 
427395473 & 6 & 3a & QPD & 0.0317 & 3.91 & 0.66 & 0.42 \\ 
427395473 & 32 & 3a & APD & 0.0497 & 3.79 & 0.62 & 0.32 \\ 
427396230 & 6 & 2a & L & 0.0783 & 32.29 & 1.0 & -0.32 \\ 
427396230 & 32 & 2a & SC & 0.6767 & 16.48 & 1.0 & -0.04 \\ 
427577547 & 42 & 8m & MP & 0.0126 & 5.07 & 0.94 & -0.04 \\ 
427577547 & 43 & 8m & MP & 0.0129 & 9.14 & 0.73 & 0.01 \\ 
427577547 & 44 & 8m & MP & 0.0147 & 8.48 & 0.94 & -0.03 \\ 
427938329 & 16 & 8m & S & 0.041 & 20.4 & 0.89 & -0.03 \\ 
428476066 & 17 & sap & APD & 0.292 & 7.25 & 1.0 & 1.23 \\ 
428476066 & 18 & sap & APD & 0.2169 & 6.87 & 1.0 & 0.3 \\ 
428476066 & 24 & sap & APD & 0.3425 & 10.08 & 1.0 & 0.72 \\ 
430006517 & 16 & 1a & APDC & 0.1225 & 9.5 & 1.0 & 0.67 \\ 
430006517 & 17 & 1a & APDC & 0.222 & 7.35 & 0.54 & 0.58 \\ 
434633379 & 17 & 2m & S & 0.0346 & 21.92 & 1.0 & 0.18 \\ 
434633379 & 18 & 2m & S & 0.0376 & 10.42 & 1.0 & -0.22 \\ 
434633379 & 24 & 2m & S & 0.0279 & 22.1 & 1.0 & -0.06 \\ 
434972921 & 17 & 8m & QPS & 0.0075 & 0.08 & 0.84 & -0.04 \\ 
434972921 & 18 & 8m & QPSC & 0.0092 & 17.1 & 0.88 & 0.09 \\ 
434972921 & 24 & 8m & QPSC & 0.0103 & 3.0 & 0.95 & -0.06 \\ 
436098606 & 6 & 2a & APDC & 0.0218 & 11.65 & 1.0 & 0.45 \\ 
437994564 & 6 & 2a & QPS & 0.0468 & 1.94 & 0.54 & -0.26 \\ 
437994564 & 33 & 2a & QPS & 0.0446 & 1.94 & 0.74 & -0.05 \\ 
437994564 & 43 & 2a & QPS & 0.0589 & 1.94 & 0.67 & -0.25 \\ 
437994564 & 44 & 2a & QPS & 0.065 & 1.92 & 0.63 & -0.25 \\ 
437994564 & 45 & 2a & QPS & 0.0384 & 1.95 & 0.61 & -0.26 \\ 
441152806 & 4 & sap & QPS & 0.0028 & 2.58 & 0.67 & 0.28 \\ 
441152806 & 31 & sap & QPS & 0.0035 & 2.55 & 0.51 & -0.17 \\ 
441744490 & 9 & pld & QPS & 0.0098 & 0.9 & 0.65 & 0.16 \\ 
441744490 & 10 & pld & QPS & 0.0222 & 0.89 & 0.54 & -0.04 \\ 
441744490 & 36 & pld & QPS & 0.0347 & 0.9 & 0.79 & -0.29 \\ 
444356276 & 17 & 2m & QPSC & 0.0164 & 7.83 & 0.94 & -0.42 \\ 
444356276 & 18 & 2m & QPSC & 0.0233 & 14.21 & 0.98 & 0.03 \\ 
444356276 & 24 & 2m & QPSC & 0.0147 & 21.37 & 0.9 & -0.26 \\ 
445062554 & 6 & pdc & QPSC & 0.0076 & 3.01 & 0.66 & 0.01 \\ 
445062554 & 33 & pdc & QPSC & 0.01 & 2.81 & 0.59 & -0.26 \\ 
449260863 & 32 & pdc & QPS & 0.0025 & 4.54 & 0.72 & 0.05 \\ 
452035938 & 10 & sap & MP & 0.0071 & 0.55 & 0.58 & 0.06 \\ 
452035938 & 37 & sap & MP & 0.0059 & 0.55 & 0.54 & -0.14 \\ 
452248316 & 41 & 2m & S & 0.0829 & 26.05 & 1.0 & -0.21 \\ 
454215115 & 12 & 1a & B & 0.0997 & 4.72 & 0.74 & -0.31 \\ 
454215115 & 38 & 1a & S & 0.1796 & 13.07 & 0.97 & -0.06 \\ 
454215115 & 39 & 1a & B & 0.1119 & 7.58 & 0.67 & -0.31 \\ 
454274408 & 12 & 8m & QPSC & 0.0016 & 13.38 & 0.86 & 0.34 \\ 
454274408 & 38 & 8m & N & 0.0026 & 26.27 & 1.0 & 0.21 \\ 
454274408 & 39 & 8m & QPSC & 0.0032 & 6.25 & 0.61 & 0.4 \\ 
454291762 & 11 & 3a & S & 0.0357 & 27.25 & 1.0 & 0.51 \\ 
454291762 & 12 & 3a & S & 0.0256 & 14.79 & 1.0 & 0.39 \\ 
454291762 & 38 & 3a & S & 0.0125 & 3.47 & 0.68 & 0.43 \\ 
454291762 & 39 & 3a & S & 0.02 & 6.25 & 0.61 & 0.15 \\ 
455573060 & 16 & 1a & N & 0.0138 & 3.29 & 1.0 & -0.01 \\ 
456030854 & 18 & 1a & S & 0.4398 & 11.73 & 1.0 & -0.26 \\ 
456030854 & 24 & 1a & S & 0.7016 & 20.67 & 1.0 & 0.72 \\ 
457231768 & 6 & sap & APDC & 0.0542 & 17.96 & 1.0 & 0.45 \\ 
457231768 & 32 & sap & APDC & 0.0489 & 13.68 & 0.86 & 0.66 \\ 
459571845 & 36 & 1m & EB & 0.0077 & 7.65 & 0.13 & 0.71 \\ 
459571845 & 37 & 1m & EB & 0.0055 & 7.65 & 0.26 & 0.6 \\ 
459699513 & 10 & 8m & QPS & 0.0037 & 1.52 & 0.43 & 0.06 \\ 
459699513 & 36 & 8m & QPS & 0.0032 & 1.51 & 0.54 & -0.04 \\ 
459699513 & 37 & 8m & QPS & 0.0035 & 1.52 & 0.27 & -0.11 \\ 
459990291 & 6 & pdc & MP & 0.0014 & 4.9 & 1.0 & -0.02 \\ 
459990291 & 32 & pdc & MP & 0.0031 & 9.58 & 1.0 & -0.02 \\ 
460571412 & 10 & eleanor & N & 0.0025 & 11.71 & 1.0 & 0.14 \\ 
460571412 & 11 & eleanor & N & 0.0066 & 2.29 & 1.0 & 0.31 \\ 
460571412 & 36 & eleanor & N & 0.0049 & 13.2 & 1.0 & 0.03 \\ 
460571412 & 37 & eleanor & N & 0.0062 & 4.08 & 1.0 & -0.08 \\ 
462436722 & 9 & 3a & S & 0.0728 & 13.58 & 1.0 & 0.05 \\ 
462436722 & 10 & 3a & S & 0.0819 & 18.31 & 1.0 & -0.25 \\ 
462436722 & 36 & 3a & S & 0.0815 & 4.72 & 0.79 & -0.5 \\ 
462436722 & 37 & 3a & S & 0.0691 & 21.44 & 1.0 & 0.14 \\ 
462878642 & 9 & 2a & S & 0.0993 & 7.67 & 0.88 & 0.08 \\ 
462878642 & 10 & 2a & S & 0.1394 & 24.29 & 1.0 & 0.3 \\ 
462878642 & 36 & 2a & S & 0.159 & 5.18 & 0.93 & -0.48 \\ 
462878642 & 37 & 2a & S & 0.1033 & 7.83 & 0.9 & -0.25 \\ 
463597276 & 12 & 8m & MPC & 0.0073 & 6.54 & 0.89 & 0.27 \\ 
463597276 & 39 & 8m & QMP & 0.0094 & 5.81 & 0.88 & -0.12 \\ 
465196254 & 12 & 2a & S & 0.1041 & 8.25 & 1.0 & -0.28 \\ 
465196254 & 39 & 2a & S & 0.1456 & 13.47 & 1.0 & 0.14 \\ 
465790701 & 11 & 1a & EB & 0.0883 & 6.8 & 0.4 & 0.55 \\ 
465790701 & 37 & 1a & EB & 0.0487 & 6.8 & 0.18 & 0.62 \\ 
465790701 & 38 & 1a & EB & 0.049 & 6.8 & 0.33 & 0.46 \\ 
467356861 & 17 & 3a & S & 0.0653 & 17.5 & 1.0 & 0.22 \\ 
467356861 & 18 & 3a & S & 0.071 & 5.95 & 0.81 & 0.37 \\ 
467356861 & 19 & 3a & APDC & 0.0705 & 17.83 & 1.0 & 0.45 \\ 
467356861 & 24 & 3a & APDC & 0.0744 & 25.54 & 1.0 & 0.49 \\ 
467356861 & 25 & 3a & L & 0.0997 & 21.71 & 1.0 & 0.36 \\ 
469534433 & 16 & 2a & P & 0.0559 & 1.11 & 0.06 & 0.16 \\ 
469534433 & 17 & 2a & P & 0.0486 & 1.11 & 0.1 & 0.08 \\ 
642035133 & 18 & 4a & SC & 0.2452 & 19.38 & 1.0 & 0.41 \\ 
709004859 & 6 & 2a & SC & 0.0044 & 21.25 & 1.0 & 0.26 \\ 
709004859 & 33 & 2a & N & 0.006 & 25.42 & 1.0 & 0.09 \\ 
816058534 & 8 & pdc & MPC & 0.0105 & 0.38 & 0.48 & 0.06 \\ 
816058534 & 9 & pdc & MPC & 0.0109 & 0.39 & 0.72 & 0.12 \\ 
816058534 & 35 & pdc & MPC & 0.0108 & 1.47 & 0.89 & -0.01 \\ 
816058534 & 36 & pdc & MPC & 0.0101 & 0.39 & 0.83 & 0.02 \\ 
1139218134 & 12 & sap & QPS & 0.0445 & 0.9 & 0.44 & 0.38 \\ 
1139218134 & 39 & sap & QPS & 0.0404 & 0.91 & 0.42 & 0.29 \\ 
\enddata
\tablecomments{We provide here photometric data and variability metrics for each star and light curve derived from {\em TESS} FFI images in sectors 1--45. The information is listed separately for each sector, as shown in the second column. The ``phot ap" code refers to the selected photometric method and aperture, and is defined as follows: ``1m" and ``2m" are 1-pixel and 2$\times$2-pixel masks, respectively, centered on the target. ``8m" and ``12m" also employ masks, as determined by the collection of pixels eight and 12 standard deviations above the median flux level. Codes with ``a" in them are circular apertures, with the number in front denoting the aperture radius. ``SAP," ``PDC," and ``PLD" are as described in the text (\S\ref{sec:lcs}). Variability types (column ``vartype") are encoded following the descriptions in \S\ref{subsec:varclustering}. Finally, the listed peak-to-peak normalized flux amplitudes, timescales, and Q/M values are derived as described in \S\ref{sec:varanalysis}.
}
\end{deluxetable*}

\section{Variability analysis}
\label{sec:varanalysis}

Upon production and selection of {\em TESS} light curves for each target, we assessed variability behavior on a sector-by-sector basis. In general, a light curve was considered to be variable if its standard deviation exceeded the median photometric uncertainty, as determined by the chosen photometric pipeline. 

\subsection{Variability categories}
\label{subsec:varclustering}

The first step in variability analysis was visual classification, following the same comprehensive light curve morphology categories developed for CTTSs in our previous work \citep{2014AJ....147...82C}. 

In brief, we placed each light curve into one of nine categories according to the dominant behavior, or otherwise designated it as non-variable (``N" in Table~2). A light curve was considered periodic (``P") if it was comprised of a repeating trend with little change in shape or amplitude. If, instead, there were modulations in amplitude that repeated on longer timescales (i.e., a ``beating" pattern), then we considered a light curve to be multiperiodic (``MP"). Light curves that only displayed a single dominant repetition timescale, but had significant variations in shape or amplitude were designated quasi-periodic symmetric (``QPS"). When a quasi-periodic pattern was present but predominantly as fading events, then we classified the behavior as quasi-periodic dipping (``QPD"). Likewise, if fading events were present but no pattern was discernable, we labeled it as aperiodic dipper (``APD"). Perfectly periodic light curves with fading events were exclusively classified as eclipsing binary systems (``EB"). For light curves dominated by brightening events (whether quasi-periodic or aperiodic), we used the burster classificaton (``B"). Variable behavior that favored neither bursting nor dipping but showed significant variations on timescales less than the 27-day sector duration was labeled stochastic (``S"). On the other hand, if a light curve displayed undulations but the dominant feature was a trend on timescales {\em longer} than the full sector, then we classified it as a long-timescale variable (``L").

\begin{figure*}
\centering
\includegraphics[width=0.95\textwidth]{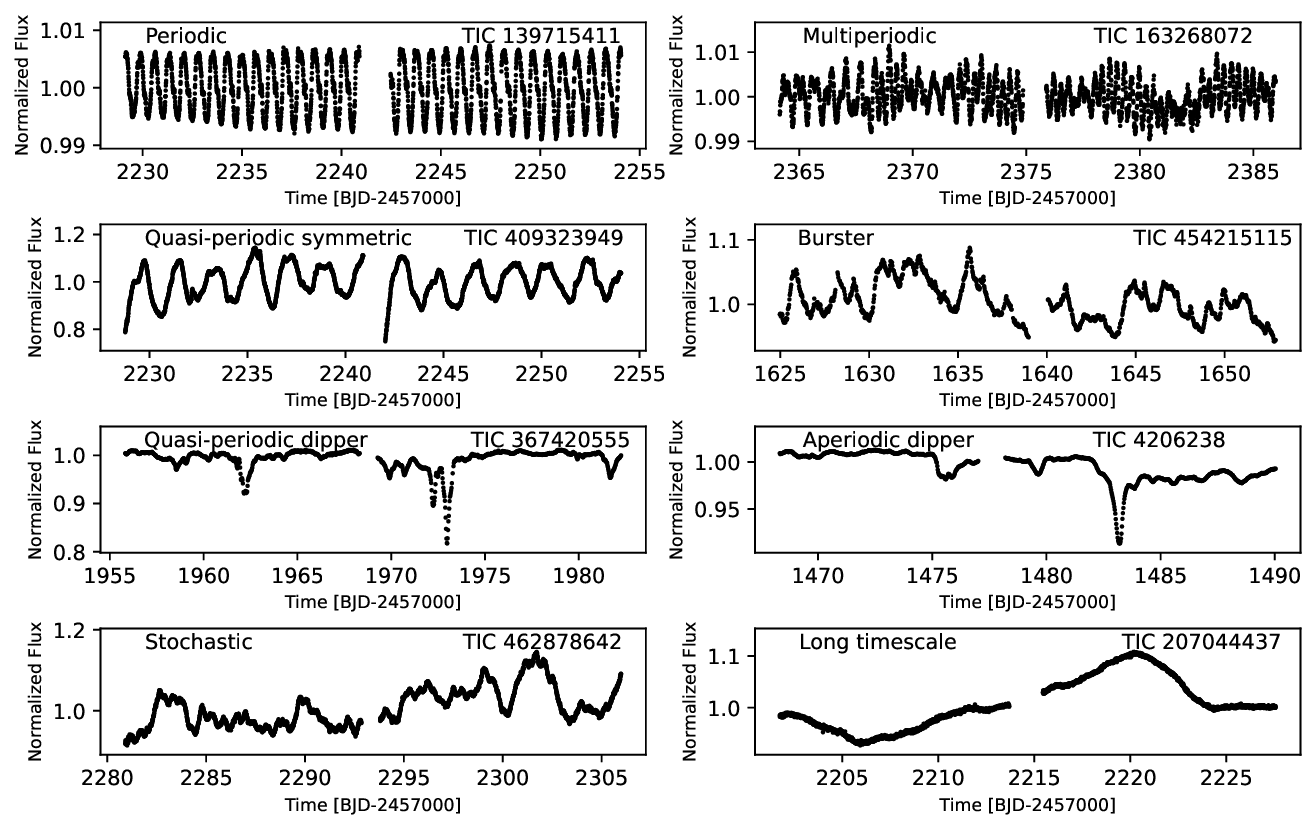}
\caption{Representative examples of light curve morphology types from our Herbig Ae/Be sample observed by {\em TESS}. We omit eclipsing binaries and non-variable objects.}
\label{fig:lcexamples}
\end{figure*}

Since light curve behavior may change over time, we classified light curves separately in each sector. In a number of cases, the variability type was difficult to select, particularly when a longer-term trend or noise-like features were present. When we were not confident in the classification, we still chose a type but appended a ``C" for ``candidate" to the end of the label. Variability morphology types are provided in Table~2 under the ``vartype" column. These same classifications are used throughout the rest of this paper, e.g., to color code the plots in Figures~3--9 and 11. Typical examples of each type are also provided in Figure~\ref{fig:lcexamples}.

Taking the set of light curves that have well determined variability behavior (i.e., no ``C" on the end in Table~2), we can obtain an approximate distribution of morphology types. For stars that were observed multiple times, we randomly selected a single sector for each, tabulating the variability type therein. We find the most common variability type is stochastic, comprising roughly 32\% of the sample. The next most common type are the quasi-periodic variables, at $\sim$15\%. Periodic and multi-periodic behavior makes up $\sim$5\% and $\sim$13\% of the sample, respectively, while long-timescale variables and aperiodic dippers are present at levels of $\sim$6\% and $\sim$7\%. Bursters make up $\sim$3\% of the sample, while quasi-periodic dippers are the most rare at $\sim$2\%.

\subsection{Statistical metrics}

To evaluate the behavior of the HAeBes in our sample on a more quantitative basis, we calculated several statistical metrics on each light curve. Since the distribution of available sectors varies across the stellar sample, we calculated the metrics independently for each sector. 

The first metric is a version of peak-to-peak amplitude, which involves first determining the fifth and 95th percentile of normalized flux values. The difference between these two numbers is then calculated, and normalized by the median flux value. 

Next, we calculated the metrics ``$Q$" and ``$M$" as described in \citet{2014AJ....147...82C} and \citet{2018AJ....156...71C}, again on a sector-by-sector basis. In brief, $M$ is a measure of a light curve's tendency toward either fading ($M>0.25$) or brightening ($M<-0.25$) events. It involves taking the difference of the median flux value and the mean of the tenth and 90th percentiles. This value is then normalized by the photometric noise level. $M$ values close to zero indicate relatively symmetric light curves, for which neither fading nor brightening dominates. 

The $Q$ metric, on the other hand, provides a measure of a light curve's degree of periodicity. Periodic signals are drawn from an autocorrelation function and verified with a Fourier transform periodogram. Once a periodicity is found, the light curve is phased up, and the phased pattern subtracted from the original time series. $Q$ is then the ratio of the residual variance to the original variance, after both have had the variance due to photometric noise subtracted off. This metric was designed such that perfectly periodic light curves attain $Q=0$ while those with little periodicity have $Q$ closer to 1 (i.e., subtracting off a phased pattern does not reduce variance much). In previous work with space-based monitoring facilities \citep[e.g.,][]{2014AJ....147...82C}, we established an empirical cut-off between periodic and quasi-periodic at $Q=0.15$. Upon cross-checking with visually-based variability classifications, this boundary value appears to remain suitable for the {\em TESS} data analyzed here. Likewise, the value $Q=0.8$ separates well quasi-periodic and aperiodic behavior. Light curves for which no periodicity was detected are automatically assigned $Q=1$. Additionally, if a long-timescale trend was suspected to be present, we fit it using a 15-day window filter and subtracted the result before computing $Q$ and $M$. 

\begin{figure*}
\centering
\includegraphics[width=0.85\textwidth]{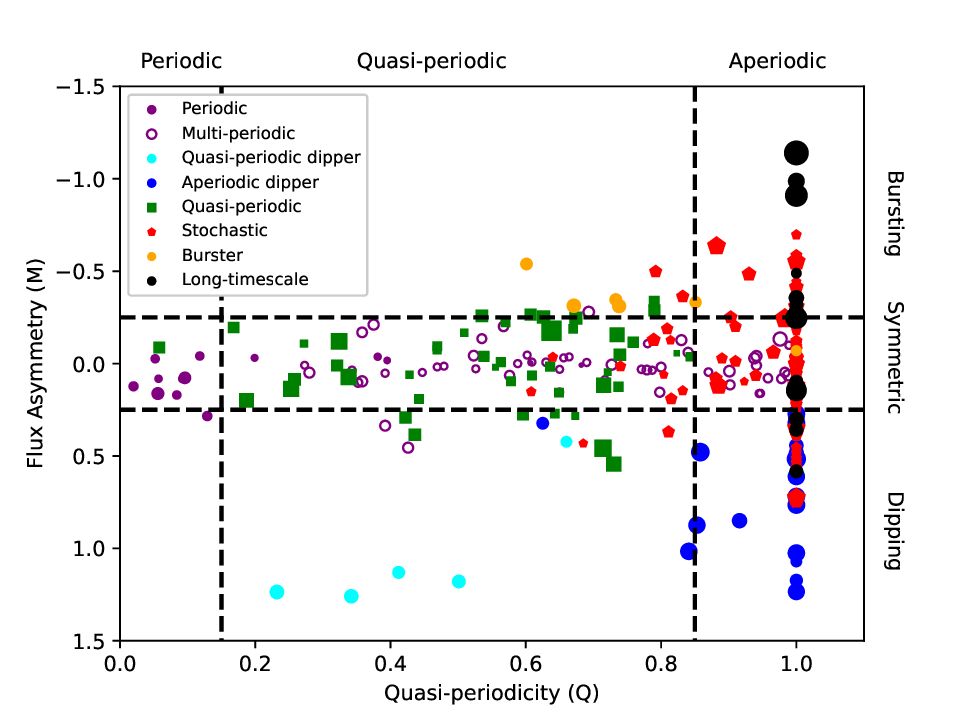}
\caption{Q-M diagram for the portion of the Herbig Ae/Be star sample with more certain variable types, as shown by the color coding. Point size represents a scaling of variability amplitude to the one third power. Data for objects with more than one sector of data appear as multiple points in this plot. The pile-up of points at $Q=1$ illustrates the sequence of objects with no discernible period.}
\label{fig:QMdiagram}
\end{figure*}

For lower mass T~Tauri stars, the diagnostic $Q$-$M$ diagram has proven useful in demonstrating correlation between visual light curve morphology classifications and statistical behavior, for both space and ground-based time series photometry \citep{2014AJ....147...82C,2022AJ....163..263H}. The corresponding plot for our HAeBe sample is displayed in Figure~\ref{fig:QMdiagram}, along with our previously established boundaries between different variability morphologies. In general, there is good correspondence between a light curve's position on this diagram and its visually assigned variability type (as indicated by color on the diagram). However, the parameter space is not nearly as well populated as is typical for lower-mass YSOs. In particular, we find a dearth of ``dipper" and ``burster" variables. Potential reasons for this finding are provided in \S6.

\subsection{Timescales}
\label{subsec:timescales}

Timescale is another metric of interest for young variable stars, and it is easiest to quantify in the case of a clear periodicity. For light curves that are periodic or quasi-periodic (i.e., $Q<0.8$), we directly adopt the dominant timescale that was identified in the autocorrelation function and periodogram analysis as part of computing $Q$. For light curves that do not show clear periods, there may still exist a characteristic timescale of stochasticity. In \citet{2014AJ....147...82C} we developed a way to estimate this timescale, by first identifying all peaks differing by an amplitude threshold. For each such threshold, we calculate the median timescale separating the peaks. We then plot the sequence of timescales against changing amplitude threshold. The largest amplitude difference is similar to the full amplitude of the light curve. We adopt the characteristic timescale as being the one that corresponds to an amplitude difference that is 70\% of this. For aperiodic light curves, the resulting timescales are presented in Table~2, alongside the aforementioned peak-to-peak amplitudes and $Q$-$M$ metrics. We assess these measures of variability in the context of (circum-)stellar properties in \S\ref{sec:vartocircumstellar}.

\begin{figure*}
\centering
\epsscale{0.9}
\plotone{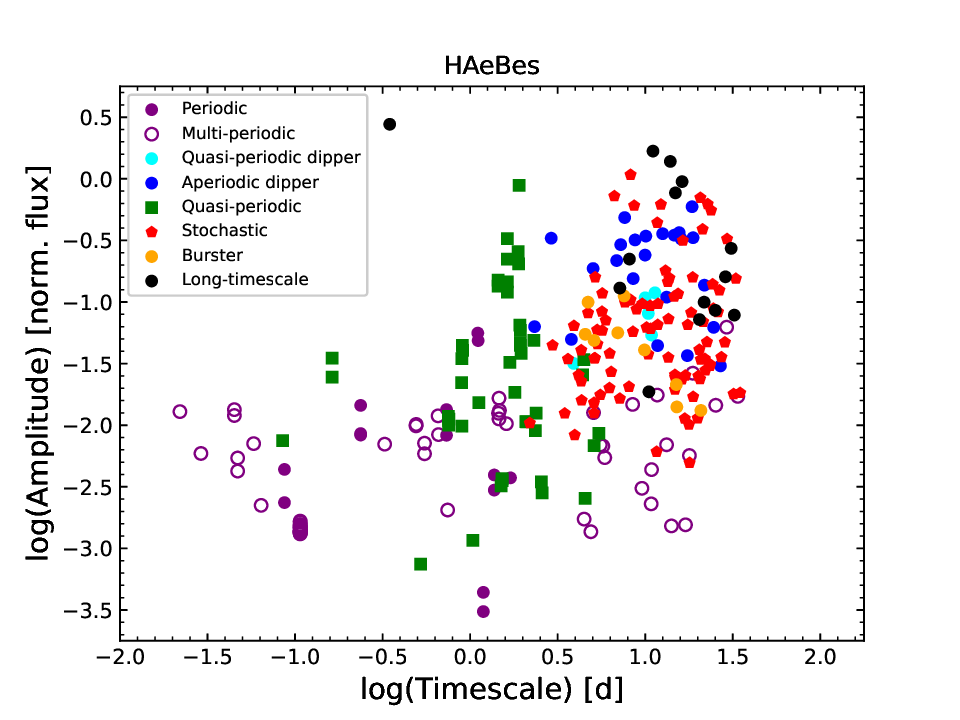}
\plotone{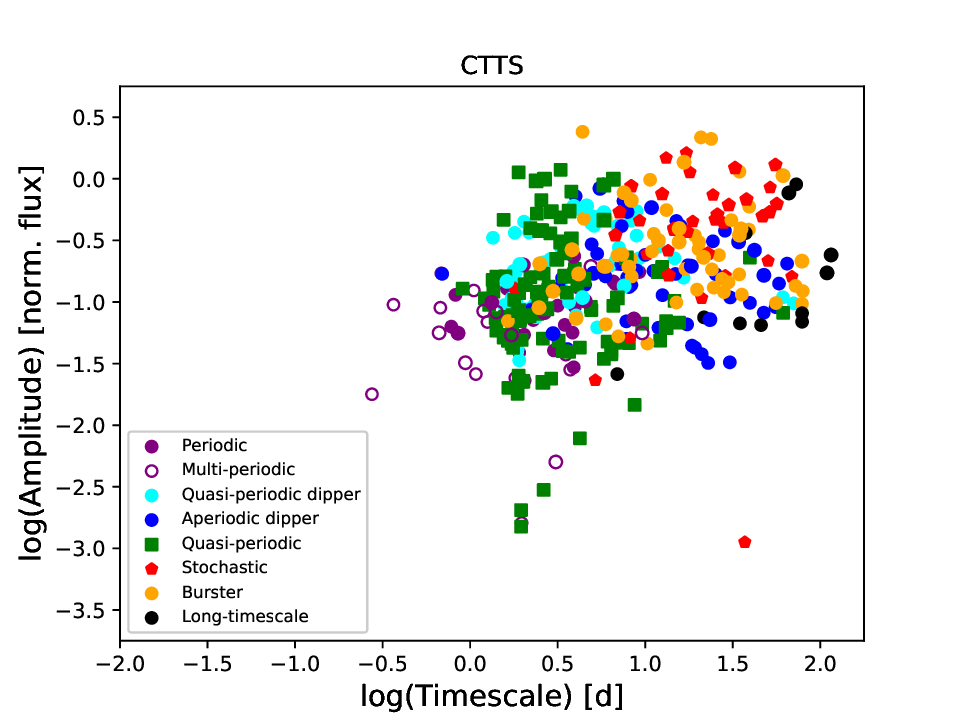}
\caption{{\em Top:} Amplitudes for the sample of Herbig Ae/Be stars presented in this work, as a function of timescale. Points are color-coded by variability type and correspond to data from individual sectors; as such, some stars appear multiple times. {\em Bottom:} We show the sample quantities, but for the low-mass T Tauri stars from \cite{2018AJ....156...71C} and \cite{2022AJ....163..212C}.}
\label{fig:AmpTimescale}
\end{figure*}

In addition, a plot of variability amplitude versus timescale is shown in the top  of Figure~\ref{fig:AmpTimescale}. Here, points correspond to individual sectors, so stars may appear multiple times in the plot.
We find that periodic and multi-periodic light curves (assumed to be associated with cool spot modulation and/or pulsation) dominate the low-amplitude space over the entire range of timescales. Quasi-periodic sources cover a narrower range, from primarily half a day to a few days duration. Longer timescales are associated with dippers, bursters, stochastic variables and, not surprisingly, the long-timescale variables. Bursters have moderate amplitudes in single sectors, from 1--10\%, while other forms of disk or accretion-dominated variables fluctuate from 1\% to over 100\%. 

At the bottom of Figure~\ref{fig:AmpTimescale} are analogous data for low-mass classical T Tauri stars in Taurus, $\rho$~Ophiuchus, and the Upper Scorpius association, presented in \cite{2018AJ....156...71C} and \cite{2022AJ....163..212C}. Several features stand out in the comparison of this plot with that of the HAeBes. First, the Herbig Ae/Be star sample extends to much lower timescale and amplitude for the periodic, multiperiodic, and some quasi-periodic stars. This is likely a combination of pulsation (particularly for the multiperiodic objects) and potentially short-timescale rotational modulation of spot patterns (see \S6.2). The pulsation instability strip crosses the H-R diagram at intermediate masses, and hence the presence of short-timescale oscillations among HAeBes but not CTTS is not surprising. 

The photometric precision of {\em TESS} light curves is generally similar to or slightly less than that of the {\em K2} light curves. There, the extension of Herbig Ae/Be stars toward short timescale should not be biased by the data quality. At longer timescales, the CTTS appear to better populate the diagram, but this is a consequence of the total time baseline of photoemtric monitoring, which was close to 90~days for the {\em K2} mission-derived light curves. In contrast, timescales for HAeBes were based on the 27-day extent of available {\em TESS} sectors, and hence cannot extend much beyond that. 

Within the range of timescales probed by both datasets, we do see a shift in the dippers toward longer timescales. Whereas for the CTTS, both quasi-periodic and aperiodic dipper light curves are present with timescales of less than $\sim$3~days, this is rare among HAeBe light curves. A logical explanation is the larger dust sublimation radii at play (see Figure~\ref{fig:Mdot_rin}) and correspondingly longer orbital timescales associated with circumstellar material that may occult the central star. 

\subsection{Multi-sector variability}

While we computed variability metrics only on single-sector light curves, additional information lies in longer timescale data for stars that were observed over multiple {\em TESS} sectors. Due to the difficulty in stitching together different sectors and calibrating the relative photometric zero point, we do not consider amplitudes on these longer timescales. However, we can assess changes in light curve morphology. 

Considering the sample of stars with well-defined variability types (i.e., not preceded by a ``C" in Table~2), 81 display the same morphological classification across multiple sectors, while 13 changed behaviors (the remaining 94 stars have either a single sector or some uncertainty in variability type). The total durations under consideration here range from 78 days (two sectors) to two years (39 sectors).  

Among the most notable changes in behavior, TIC~11199521 switches from clearly variable at the 200\% level in sector 6, to non-variable in sector 32, some 700 days later. Interestingly, the total flux that we measure in the same TESS aperture increases about 50\% over this timespan. In addition, TIC~200511734 appears to have developed a short-timescale periodicity ($\sim$31 minutes) between sector 6 and 32. However, the observing cadence changed from 30 minutes to 10 minutes in that timespan, so the short period may have simply been impossible to detect in the earlier sector.

\begin{figure*}
\centering
\includegraphics[width=0.85\textwidth]{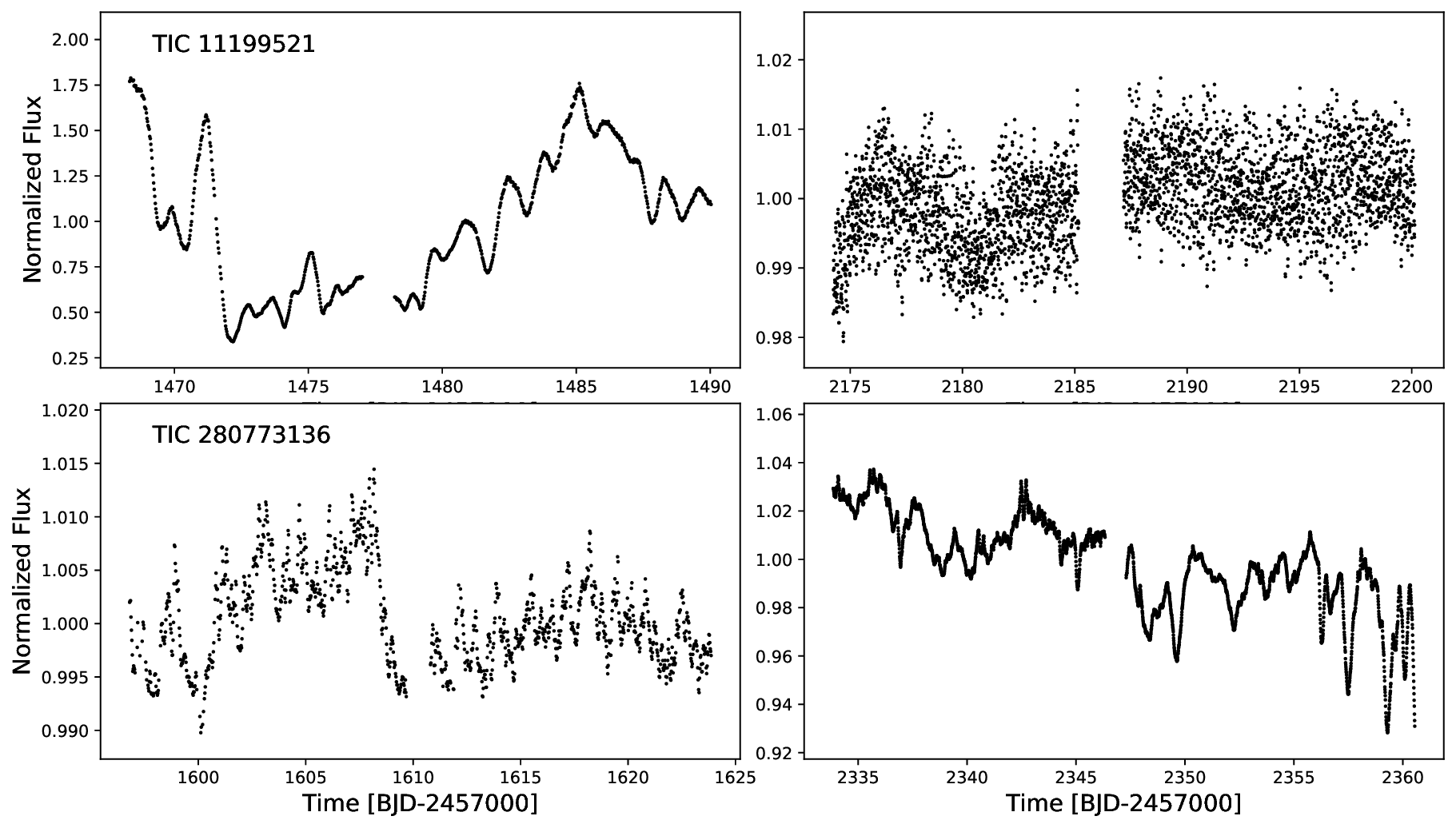}
\caption{Light curves for two stars with pronounced changes in variability morphology and amplitude over a couple years. While relatively rare, these cases highlight the possibility that HAeBe variability drivers may shift dramatically on relatively short astronomical timescales.}
\label{fig:lcs_changers}
\end{figure*}

TIC~280773136 appears as a burster in sector 11, but is classified as a dipper in sectors 37, 38 (i.e., two years later). The flux level from sector 11 to 37 decreases about 4\%, and remains stable into sector 38. While not considered as part of this dataset, a check of sectors available two years later reveals a potential reversion to bursting in sector 64, and then two prominent dips in sector 65. Finally, TIC~125641610 is stochastic in sector 7, but changes to a more steady long-term increase with higher amplitudes in sector 33 (timespan again about two years). Light curves for both TIC~11199521 and TIC~280773136 are shown as examples in Figure~\ref{fig:lcs_changers}.

\section{Variability as related to (circum-)stellar properties}
\label{sec:vartocircumstellar}

\subsection{Correlations with stellar parameters}

\begin{figure*}
\centering
\includegraphics[width=0.85\textwidth]{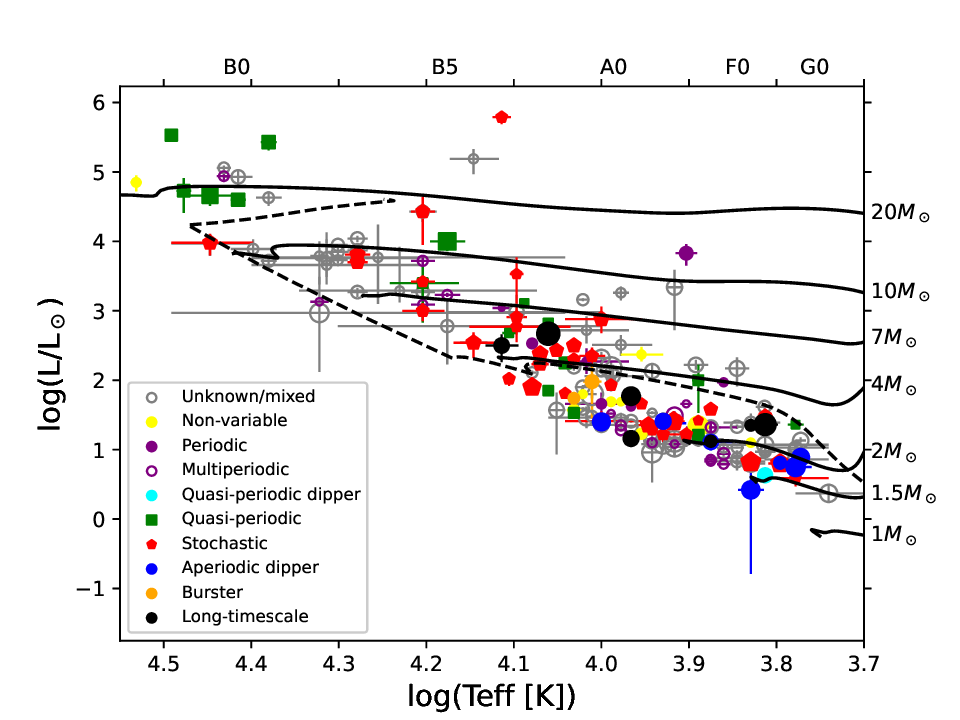}
\caption{Hertzsprung-Russell diagram for the sample, color coded by variability type. We display theoretical solar metallicity mass tracks (solid black curves) as well as a 2~Myr isochrone from \cite{2022A&A...665A.126N} and \cite{2025A&A...694A.193C}. Spectral types are shown at the top, based on effective temperatures of stars in the HarchiBe sample. Point size represents a scaling of variability amplitude to the one third power.}
\label{fig:HRdiagram}
\end{figure*}

We retrieved masses, luminosities, accretion rates, disk properties, and their uncertainties from the HArchiBe. One way to assess variability in the context of stellar properties is via the Hertzsprung-Russell diagram. We present one version of this, color coded by variability type, in Figure~\ref{fig:HRdiagram}. Mass tracks and isochrones are included, based on PARSEC V2.0 stellar models \citep{2022A&A...665A.126N,2025A&A...694A.193C}. Stars that change variability morphology or have unclear types are included in gray, while those with stable behavior across all available TESS sectors are shown in color. 

Several features stand out on this H-R diagram. First, the few dipper stars in the sample are confined to effective temperatures less than about 10,000~K. The only stable quasi-periodic dipper (TIC~287840183, otherwise known as BF~Ori) is even cooler at 8,750~K.
Thus the dipper phenomenon appears to be mass-dependent, at least on the roughly month timescales under consideration here. We suspect this is directly related to the increasing orbital periods of inner disk dust around hotter stars (see \S\ref{subsec:fewdippers} for more discussion). 

With one exception, the non-variable stars are also confined in effective temperature, to $T_{\rm eff}<$14000~K. On the other hand, multi-periodic, stochastic, and quasi-periodic sources span the entire mass/temperature range. The same goes for objects with unstable and/or unknown variability types. Mass dependencies for bursters or periodic stars are less clear due to small number statistics. 

\begin{figure*}
\centering
\includegraphics[width=0.75\textwidth]{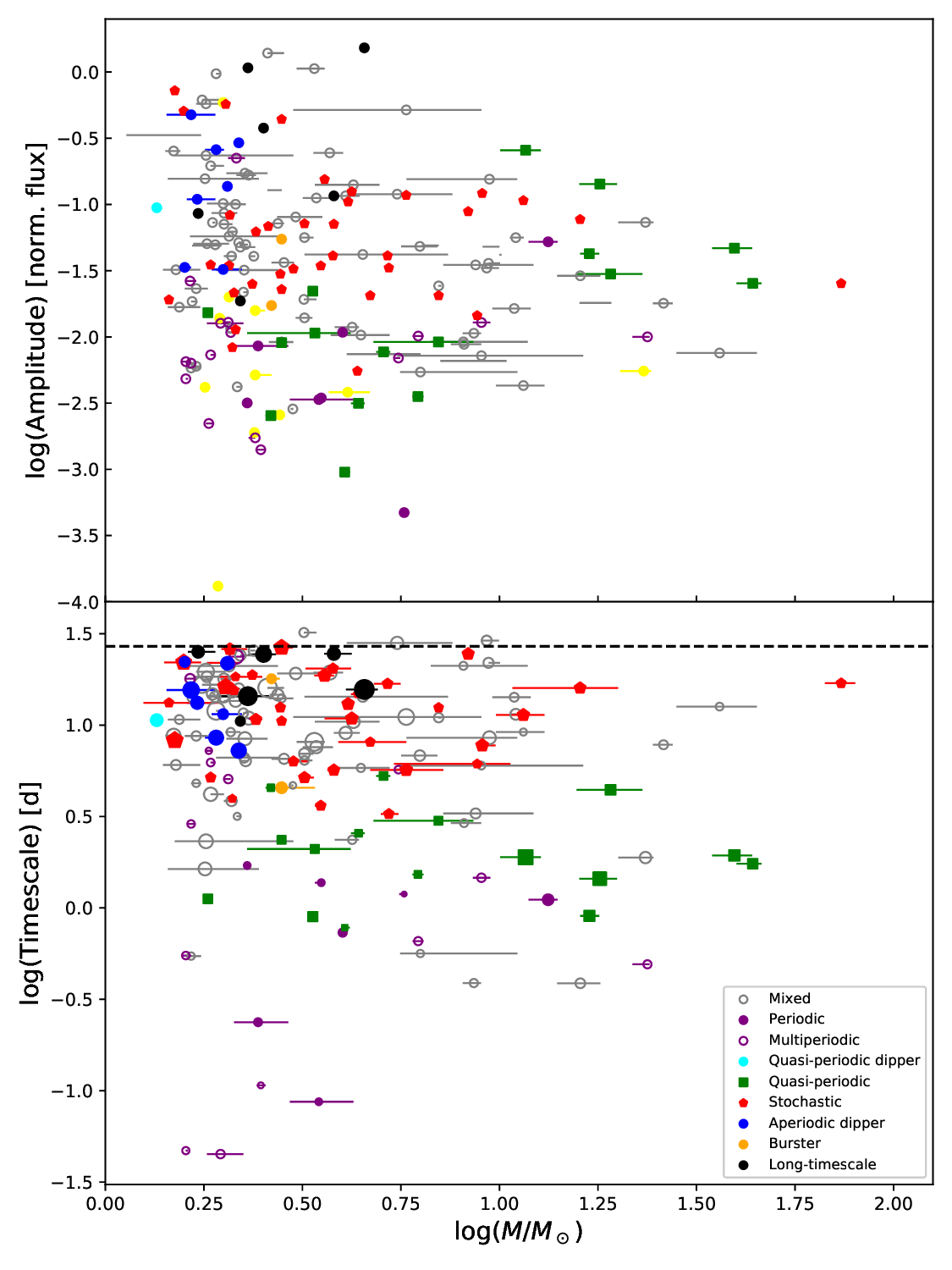}
\caption{At top, we plot amplitudes of variability versus mass, color coded by variability type. The amplitude values are taken as medians across available sectors. In the bottom panel, timescales of variability are shown versus mass, sized by amplitude to the one-third power. Amplitude and timescale values are taken as medians across available sectors. The TESS 27-day sector duration is shown as a dashed horizontal line.}
\label{fig:Amp_time_mass}
\end{figure*}

To further explore trends with stellar parameters, we also plot the median variability amplitude across available sectors (peak-to-peak normalized flux, in log space) versus mass in Figure~\ref{fig:Amp_time_mass}. Again, points are color coded by variability type; objects with uncertain or changing light curve morphology are again shown in gray. Where multiple {\em TESS} sectors were available, we show the median of all amplitudes for each object. In addition to the confinement of various variability types as mentioned above, there is at first glance a gradual decline in amplitude with mass. However, this is likely an effect of the sample size being lower for stars over 10~$M_\odot$. 

The predominant variable types at high mass are quasi-periodic (QPS) and stochastic types. A distinct jump in the typical amplitude of QPS variability appears around 10~$M_\odot$. Stochastic stars, on the other hand do not exhibit a statistically significant change in amplitude with mass.

%\begin{figure*}
%\centering
%\includegraphics[width=0.85\textwidth]
%{Timescale_M_amp_stable.eps}
%\caption{Timescales of variability versus mass, color coded by variability type and sized by amplitude to the one-third power. Amplitude and timescale values are taken as medians across available sectors. The TESS 27-day sector duration is shown as a dashed horizontal line.}
%\label{fig:Time_mass}
%\end{figure*}

\subsection{Correlations with accretion and the inner disk}
\label{correlationswithdisk}

\begin{figure*}
\centering
\includegraphics[width=0.85\textwidth]{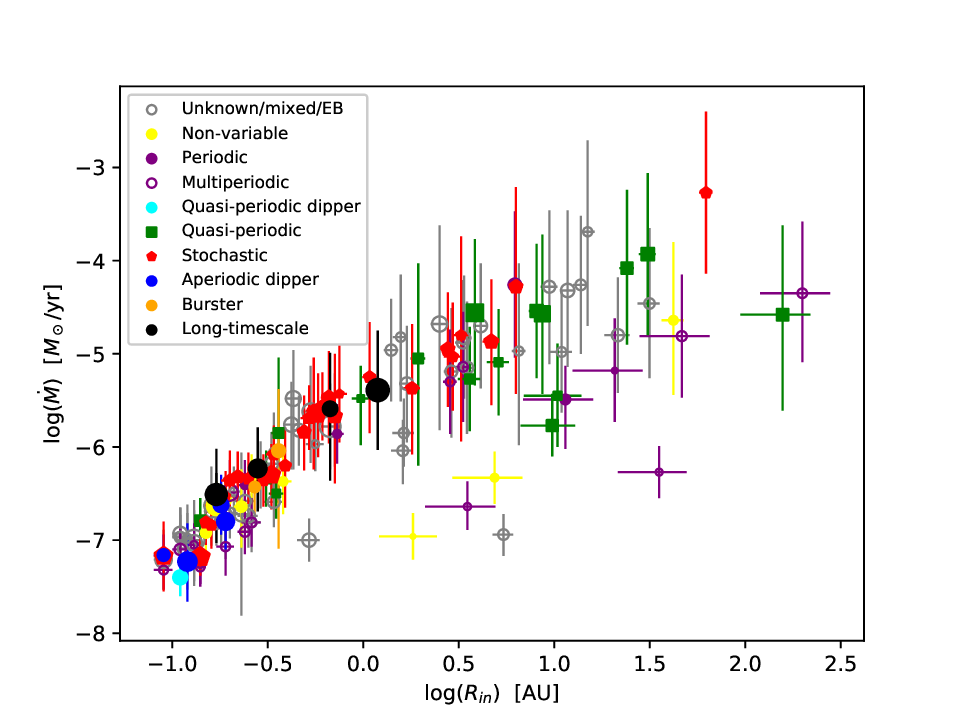}
\caption{Accretion rate versus estimated inner disk dust radius, color coded by variability type, as in Fig. \ref{fig:Amp_time_mass}. Point size scales with the amplitude of variability to the one-third power.}
\label{fig:Mdot_rin}
\end{figure*}

To test associations of variability with circumstellar properties, we consider the accretion rates and inner disk dust radii estimated by \citet{2021AA...650A.182G} for stars in the sample. The dust radius, $R_{\rm in}$, corresponds to the location in the spectral energy distribution where an infrared excess begins, while accretion rate was derived based on empirical relations between accretion luminosity and stellar luminosity \citep{2020MNRAS.493..234W}. These quantities are plotted in Fig. \ref{fig:Mdot_rin} according to variability type and amplitude. There is a correlation in accretion rate and inner disk dust radius, which is due to the connection of both quantities to the stellar mass. However, there are a number of outliers from this general trend. These outliers all have lower estimated accretion rates than expected for their relatively large radii ($R_{\rm in} > 2$~AU). Furthermore, all of these outliers have some level of periodicity, suggestive of either starspot modulation at the stellar rotation period or pulsation at shorter periods. Our interpretation of this phenomenon is that in cases for which the accretion rate is low relative to the dust destruction radius (and correspondingly, stellar mass/luminosity), accretion-related stochastic variability is no longer present in time series photometry, at least on the roughly month sector timescales available from {\em TESS} monitoring. This may happen if, e.g., the inner disk has been cleared out well beyond the point of dust sublimation.

\begin{figure*}
\centering
\includegraphics[width=0.85\textwidth]
{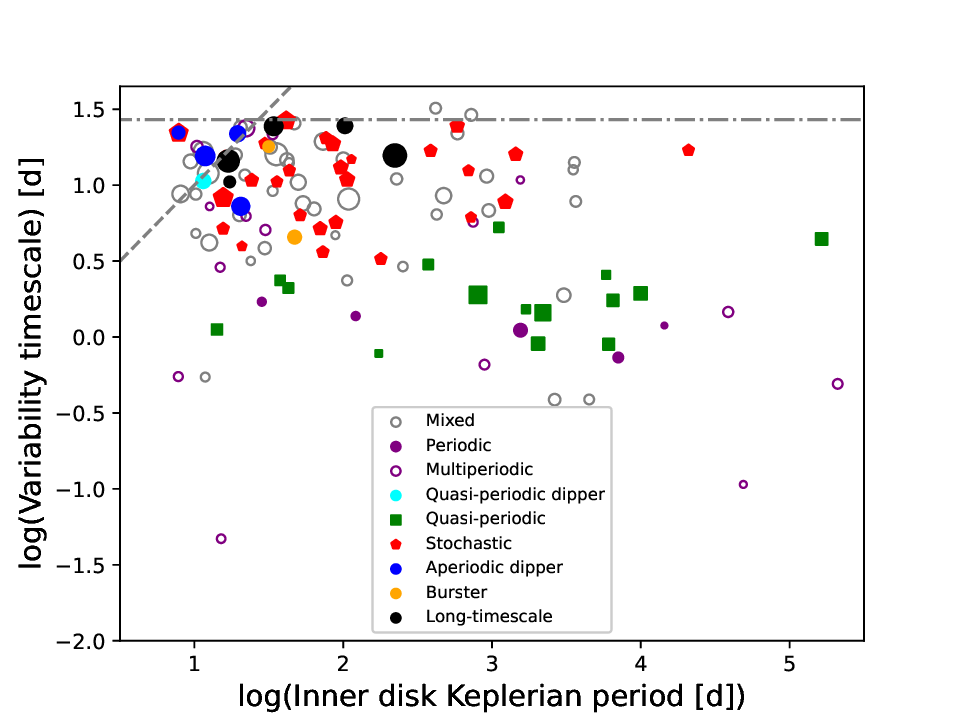}
\caption{Keplerian orbital period of the inner dust disk versus variability timescale, scaled and color coded by variability type, as in Fig. \ref{fig:Mdot_rin}. Non-variable objects and unknown types are excluded since they do not have a well-definied variability timescale. Eclipsing binaries are also omitted. The gray dot-dashed line indicates the duration of a single {\em TESS} sector, while the dashed line indicates equality of the two timescales.  We find that the vast majority of Herbig Ae/Be star variability observed here is dominated by timescales shorter than the inner disk orbital period.}
\label{fig:Pkep_timescale}
\end{figure*}

To probe further connections between variability and inner disk size, we calculate the Keplerian orbital period associated with the estimated inner disk dust radius ($R_{\rm in}$; see \S\ref{correlationswithdisk}), from the SED analysis of \cite{2021AA...650A.182G}, and for stars where they provide mass estimates. 
%This is given by 
%$$P_{\rm Kep} = 2\pi\left(\frac{{GM_*}}{R_{\rm in}%^{3}}\right)^{-1/2},$$
%where $M_*$ is the stellar mass, $R_{\rm in}$ is the inner %disk dust radius, and $G$ is the gravitational constant.
In some cases, it is consistent with sublimation, whereas for others it lies farther out.  We plot the associated period against the variability timescales determined in \S\ref{subsec:timescales} in Figure~\ref{fig:Pkep_timescale}. It is clear that the vast majority of the variability timescales are much shorter than the expected orbital period of the inner disk. The exceptions are mainly dippers, for which the two timescales are similar. This is not surprising, as the leading theory for explaining dipper variability is occultation of the central star by material at the inner disk edge. Remaining variability types involve fluctuations on shorter timescales, suggesting that they originate either at the stellar surface (e.g., starspots) or in the circumstellar gas accreting through an inner disk hole. 

\begin{figure*}
\centering
\includegraphics[width=0.85\textwidth]
{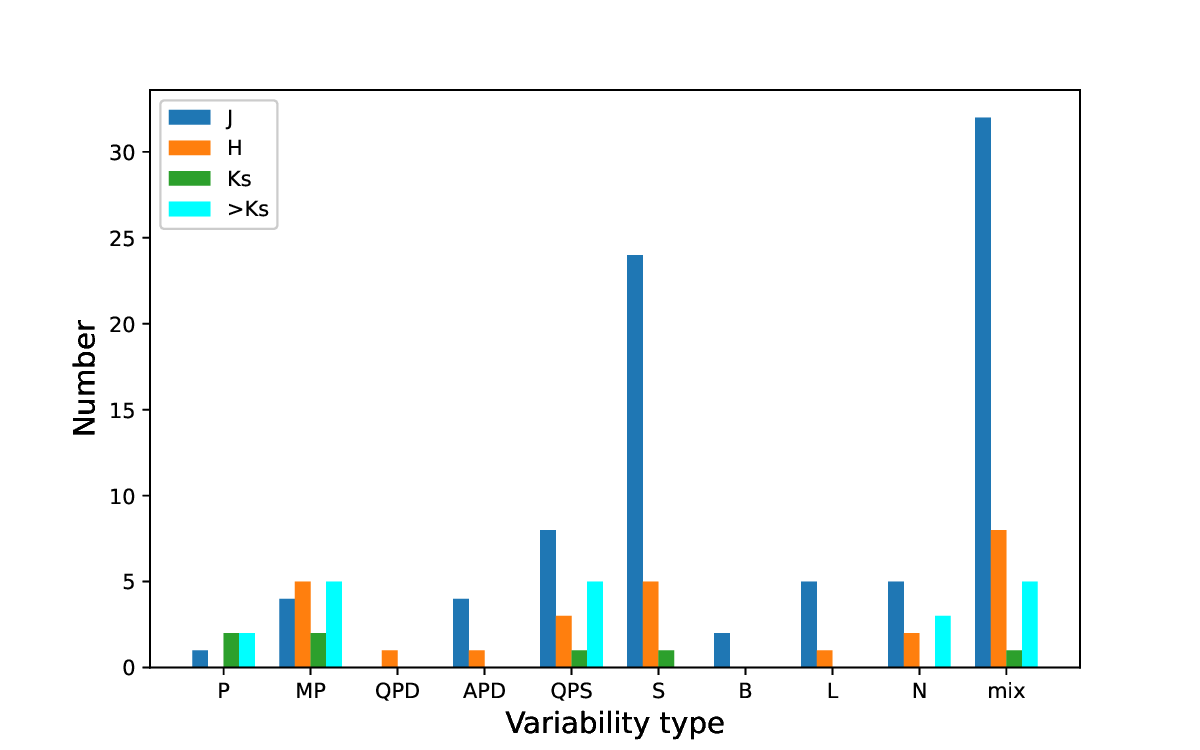}
\caption{Histogram of disk SED types, according to \cite{2021AA...650A.182G}. In brief, the J/H/Ks label is an indication of the shortest wavelength at which an infrared excess appears. These types are broken down by variability class, as shown in the labels. 
%Show errorbars?
}
\label{fig:JHKhist}
\end{figure*}

For many stars in our sample, the HArchiBe also provides classifications of inner disks via spectral energy distribution (SED) analysis, in terms of the wavelength where the infrared excess begins. As discussed in \cite{2021AA...650A.182G}, SEDs are tagged with a $J$, $H$, $Ks$, or $>Ks$, after the band associated with that wavelength. Excesses starting at $Ks$ band or redward are considered to reflect ``transitional" disk status, with those systems having inner dust cavities. To check whether variability behavior correlates with the status of inner disk dust, we show in Fig.\ \ref{fig:JHKhist} the break-down of light curve morphology type with the so-called $JHK$ category. We find that the majority of variability types have infrared excesses beginning at or close to the J band, indicating relatively small inner disk holes. This is somewhat expected, as the majority (89/188) of the stars in our {\em TESS} dataset have excesses starting here. 
The periodic and multi-periodic variables stand out as exceptions, with roughly equal distributions of $JHK$ types. This suggests that starspot modulation and pulsation are easier to observe when the disk rim lies farther out from the central star. However, the small total number of such sources makes this result a tentative one. We find that none of the variability types are dominated by infrared excesses that are exclusively long-wavelength (``$>$Ks"). In particular, several of the classes (bursters, dippers, long-timescale variables) are completely devoid of such transitional disks. 

\begin{figure*}
\centering
\includegraphics[width=0.85\textwidth]
{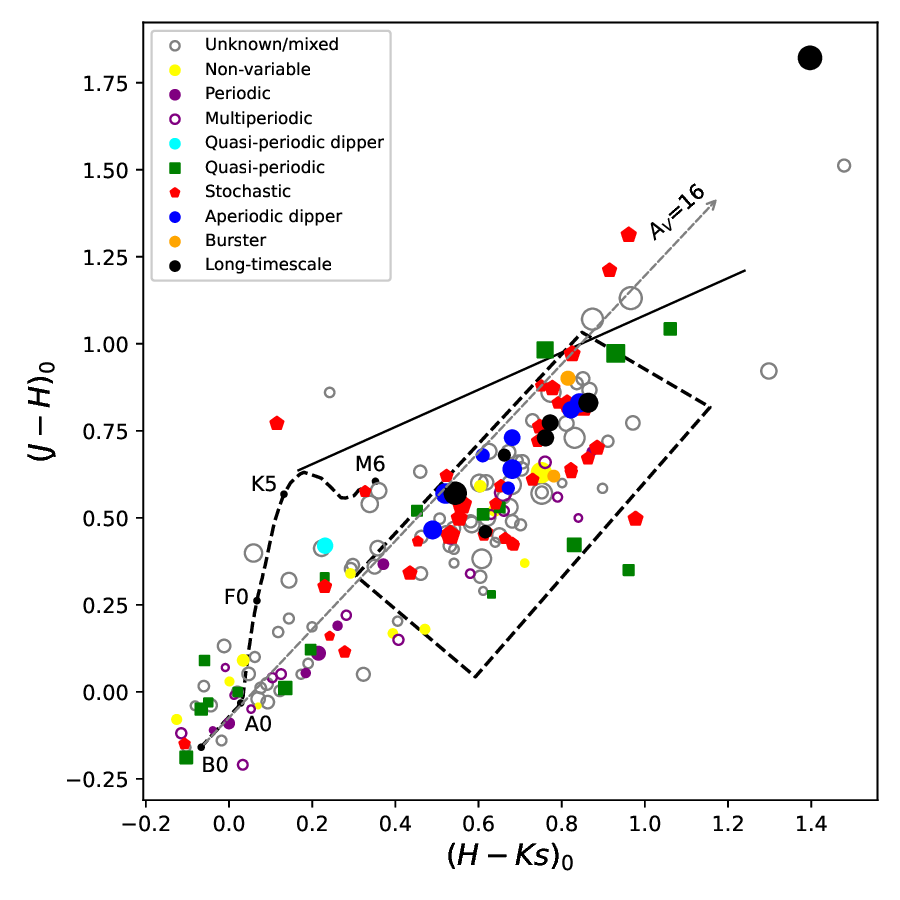}
\caption{Dereddened $J-H$ versus $H-Ks$ for our sources. The dashed quadrilateral region indicates the locus of HAeBes defined by \cite{2005AJ....129..856H}. The corresponding locus for CTTS \citep{1997AJ....114..288M} is shown as a solid line, while the sequence of dwarf stars from \cite{2013ApJS..208....9P} is included as a dashed curve. A gray dashed vector marks the colors of an extincted B0 dwarf with various values of $A_V$ up to 16. All colors have been transformed to the 2MASS system.}
\label{fig:JHHK}
\end{figure*}

In addition to disk types via spectral energy distributions, we can search for connections between HAeBe variability and circumstellar properties via a 2MASS color-color diagram. In Fig. \ref{fig:JHHK}, we display the sequence of $(J-H)_0$ versus $(H-K_S)_0$ after correcting for extinction using the $A_V$ values reported in the HarchiBe and the wavelength dependence of \cite{1989ApJ...345..245C} for $R_V=3.1$. % (although 5 may be more appropriate; see Hernandez et al 2005).
The locus of HAeBes defined by \cite{2005AJ....129..856H} is shown as a dashed quadrilateral. We find that the majority of the highly variable stars (in terms of amplitude) lie within the HAeBe locus, predominantly with redder colors than that of an extincted A or B dwarf. Several light curve types, including the aperiodic dippers and bursters, appear exclusively within this region. In contrast, a number of stars with periodic or quasi-periodic patterns appear on the bluer area of the diagram, which may be more consistent with contamination by main sequence classical Be stars \citep{2005AJ....129..856H}.

\section{Discussion}
\label{sec:discussion}

\subsection{A paucity of dipper variables}
\label{subsec:fewdippers}

As mentioned in \S\ref{subsec:timescales}, there is a lack of dipper variables in our sample, relative to previously observed populations of CTTS. This finding applies to both the quasi-periodic and aperiodic types. Among the former, TIC~367420555 (BO Cep) is perhaps the most prominent. With a spectral type of F3, it is on the border between the HAeBes and lower mass CTTS. \cite{2015JAVSO..43...35P} found it to be a UX Orionis star (``UXor"), and earlier, \cite{1999A&AS..140..293G} found a period of 10.7~d. This is consistent with the timescales that we have derived, in the 10.0-11.4~d range, depending on sector. The only other quasi-periodic variable is TIC~427395473 (NV~Ori) in sector 6. This star likewise has an F spectral type (F0-F2). TIC~287840183 (BF~Ori) remains a candidate quasi-periodic dipper in the early A spectral type range. 

There are a handful more aperiodic dippers and candidates, as listed under the ``APD" category in Table~2. These include TIC~261539498 (spectral type A8), TIC~280773136 (A0-A1), TIC~285920558 (F1-F4), TIC~312493387 (A3), TIC~335623526 (G0), TIC~339989392 (F7-G5), TIC~358901633 (F), TIC~372070676 (F1-F4), TIC~385894679 (A3-A4), TIC~4206238 (F5-F8), TIC~427395473 (F0-F2), and TIC~428476066 (B9). What stands out about this subset is that 
more than half have spectral types of G or F. This is in stark contrast to the overall distribution shown in Fig.\ \ref{fig:SpTdist}, for which only 15\% of our sample falls at these types. 

A simple explanation for these findings is that, as proposed by others \citep[][and references therein]{2003A&A...409..169B,2017MNRAS.470..202B}, dipper variability is caused by inner disk dust warps or other structures occulting the central star along the line of sight. Higher mass stars have larger dust sublimation radii, and hence the minimum inner disk Keplerian orbital period (assumed equivalent to timescale of quasi-periodic dippers) eventually becomes longer than the {\em TESS} single sector observing duration for stars hotter than a late F spectral type \citep[see][]{2021AJ....162..101V}. Where inner disk radius estimates are available, this conclusion is borne out by the data in Fig.\ \ref{fig:Pkep_timescale}. Here, the dashed line shows that most of the dipper variables have variability timescales similar to the inner disk Keplerian period.  

Aperiodic dippers are slightly more challenging to explain, but may be associated with dusty material at or beyond the inner disk edge. For higher mass stars, single-sector {\em TESS} observations could encompass a single occultation event, but would be increasingly unlikely to do so for the larger sublimation radii, if just one or two inner disk warps were responsible for such occultations. Overall, the picture of both quasi-periodic and aperiodic dippers is consistent with variability arising from stuctures at or just beyond the dust sublimation radius. It explains why the UXor variables, many of which have timescales of weeks to months \citep{1994ASPC...62...23T,1999AJ....118.1043H}, would only be identified in photometric observing campaigns with durations longer than the {\em TESS} sectors analyzed here. 

\subsection{The meaning of periodic modulations}
\label{subsec:starspots}

Sinusoidal variability in young stars is typically attributed to starspots (either chemically peculiar or lower in temperature relative to the surrounding photosphere), which is in turn thought to originate in magnetic activity. It is estimated that 7\% of HAeBes are magnetic \citep[][and reference therein]{2019ASPC..518...18H}, but the origin of such fields remain controversial due to the lack of a convective envelope. They may be fossil relics of an earlier convective phase \citep{2009ARA&A..47..333D}, or the result of mergers \citep{2009MNRAS.400L..71F} or representative of a dynamo generated by sub-surface turbulence \citep{1995AA...301..155B}. In the pre-main sequence cluster NGC~2264, \cite{2014A&A...562A.143F} identified a connection between sinusoidal photometric modulations and large-scale magnetic fields ($\sim$400~G) in two early B-type stars. However, they could not pinpoint the origin of the observed fields. Given that we have also detected a handful of periodic stars among our Herbig sample, it is natural to ask whether their variability may also be the result of magnetic effects. 

The two most prominent periodic HAeBe stars with {\em TESS} light curves are TIC~469534433 (AS~470) and TIC~139715411 (HD~59319). The light curve of TIC~469534433 displays clear sinusoidal modulations with a period of 1.1 days and normalized flux semi-amplitude of $\sim$0.025 in {\em TESS} sectors 16 and 17. It is categorized as an A4--A7 star in \citet{2021AA...650A.182G}, but its listed mass of 13.3~$M_\odot$ is at odds with that. Based on the spectral type range, we suspect that the mass is closer to 2--4~$M_\odot$ (i.e., in line with the data on other stars in the HArchiBe). Of additional concern, the light curve of TIC~469534433 (TESS magnitude 11.5) is blended with neighboring star TIC~469534429 (magnitude 13.1). No literature information is available on this neighbor, but if it were at the same distance, it would be $\sim$23\% of the brightness of our Herbig target. If brightness variations originate here, then the semi-amplitude would instead be 13\%. Thus without higher resolution observations, we cannot rule out that these modulations are caused by spots on a lower mass neighbor. 

Our second periodic star, TIC~139715411, is of spectral type B7--B8 and TESS magnitude 8.4. All other stars in the photometric aperture are at minimum five magnitudes fainter. We conclude that the $\sim$0.5\% sinusoidal photometric fluctuations in the light curve of TIC~139715411 are from this star. The period is 0.73 days, and no additional signals other than harmonics appear in the periodogram. The variability is thus consistent with stellar spot modulation. With an estimated radius of 4.26~$R_\odot$, the associated equatorial rotational velocity would be 295~km~s$^{-1}$. 
In an analysis of $v\sin i$ distributions, \cite{2013MNRAS.429.1027A} found that magnetic Herbig Ae/Be stars tend to rotate slower ($<$100~km~s$^{-1}$) on average than their non-magnetic counterparts (up to 350~km~s$^{-1}$). % (the reason could be related to a magnetic braking phenomenon.) 
If the variability of TIC~139715411 is magnetic in origin, then it is in tension with their finding. Follow-up studies of this object would be beneficial, as we find no prior information on magnetic field status.

Additional stars in the periodic category are TIC~199757113 (sectors 6 and 32), TIC~336518731 (sectors 11, 12, and 38), TIC~302824830 (sectors 43 and 44), and TIC~341984965 (sectors 16 and 17). However, the first was previously classified as a delta Scuti pulsator \citep{2000AA...355L..35M}. The latter three have relatively low amplitudes and are listed as blends in Table~1, so we cannot be confident that the signals are coming from these Herbig Ae/Be stars themselves.

%TIC~302824830 is a B6--B7 stars with periodic modulations in {\em TESS} sectors 43 and 44, albeit at a much lower amplitude of $10^{-4}$. The variability seen here is low enough that the object could feasibly be blended with a spotted lower mass star. 

One may also ask the question of whether magnetic field hosting stars necessarily show spot-dominated variability. There is a correlation for main sequence A- and B-type stars (i.e., an association of magnetism with chemical peculiarities), but the same pattern has not been borne out on the pre-main sequence \citep{2012MNRAS.422.2072F}. 

We can test this idea for the few stars in our sample reported to host significant magnetic fields: TIC~388935225 (V380~Ori; Alecian et al. 2007), TIC~427395300 (LP~Ori; \cite{2013MNRAS.429.1027A}), and TIC~254387071 (HD 139614; \cite{2004A&A...428L...1H}). An additional potential magnetic star, TIC~387813185 (HD 200775), is discounted since it has been found to be a spectroscopic binary, with the magnetic field {\em not} originating in the disk-bearing Herbig B3 component \citep{2013A&A...555A.113B}. The A7-A9 star TIC~254387071 has both a 450~G longitudinal magnetic field strength \citep{2004A&A...428L...1H} and multiperiodic variability in two sectors of {\em TESS} data. However, the very short dominant period of $\sim$1.2~h in our light curves is more suggestive of a pulsation mechanism than starspots. For the remaining two objects (each with one sector available), the variability that we observe is definitively aperiodic; TIC~388935225 shows bursting behavior, while the pattern for TIC~427395300 is tentatively stochastic. 
Furthermore, the star TIC 320547523 (HD~101412) was previously detected as having chemical spots (presumably of magnetic origin), modulating the light at a stellar rotation period of 13.86~d \citep{2010AN....331..361H}. We do not detect variability in any of the {\em TESS} light curves available for this HAeBe.

Thus while a small portion of our sample is known to host magnetic fields, and overall, approximately 7\% (i.e., $\sim$13 magnetic stars of the 188) is expected, the fraction that appears spotted in our available {\em TESS} light curves is much smaller. It is possible that a number of our targets do host spots, but the associated variability is drowned out by accretion effects. Alternatively, magnetic fields may not necessarily generate spots at these early stages of stellar evolution. At the very least, we expect that spectroscopic and spectropolarimetric studies of the two relatively unusual periodic stars TIC~469534433 and TIC~139715411 could illuminate whether their variability may be attributed to large-scale magnetic fields.

\subsection{Timescales of variability}

In this study, we are somewhat limited to the 27-day timescales of each {\em TESS} sector, although we have assessed variability morphology changes over longer periods where multiple sectors are available. 
In contrast, previous photometric monitoring studies have covered over a decade, albeit with much sparser time sampling. Previous work \cite[e.g.,][]{1999AJ....118.1043H} has suggested that HAeBe variability timescales are longer than those observed for CTTS. We do not recover this finding, although it is potentially due to the aforementioned {\em TESS} sector baseline. In Figure~\ref{fig:Amp_time_mass}, we see divisions between different light curve morphology types, with the periodic, multiperiodic, and quasi-periodic variables at the short end (sub-hour to a few days). This is likely a reflection of stellar rotation and pulsation periods. The stochastic, burster, and dipper variables span a larger range of timescales from several days to nearly a month. Not surprisingly, the long-timescale variables are at the upper end of this range. 

To understand the distribution of timescales, we have plotted the inner dust disk Keplerian orbital period in Fig.\ \ref{fig:Pkep_timescale}. Apart from the few dippers, most of the variability that we observe in individual {\em TESS} sectors is significantly faster than this orbital timescale (but again limited by our observing duration). We suspected that the physical drivers here are not only stellar rotation and pulsation, but also short-timescale fluctuations in the accretion flow well within the disk sublimation radius. In Fig. \ref{fig:AmpTimescale}, we see similar behavior among the bursters and stochastic sources when comparing the HAeBes monitored here against CTTS previously observed with the {\em K2} mission. Thus, on day to month timescales, there is no appreciable evidence for a change in variability properties in going from CTTS (e.g., GKM spectral types) to HAeBes. As discussed below, however, there may be a transition at higher masses.

\subsection{Mass dependence of variability}

There is previous evidence for a transition in variability behavior, and associated accretion dynamics, in the B spectral type range \citep[e.g.,][]{2015MNRAS.453..976F}. In a study undertaken with Gaia DR2 data, \citet{2018A&A...620A.128V} found $\sim$25\% of HAeBes to be variable at at least the 0.5-magnitude level on a 22-month timescale. They %suggested that much of this variability is due to occultation by dust structures in the inner regions of nearly edge-on disks, and 
noted lower amplitudes for stars of mass greater than 7~$M_\odot$. 

To look for a similar transition in our own dataset, we have plotted normalized variability amplitude versus stellar mass compiled from \citet{2021AA...650A.182G} in Figure~\ref{fig:Amp_time_mass}. Among the stars with definitive and stable variability morphologies, we see that dipper stars are confined to a mass range less than 2.2~$M_\odot$, bursters have $M<2.8$~$M_\odot$, and the long-timescale variables do not exceed masses of 4.5~$M_\odot$. Likewise, periodic behavior only persists up to 5.7~$M_\odot$, with the exception of TIC~469534433, which may have an erroneously high mass as discussed above in \S6.2. We therefore split the sample into two mass regimes, with a boundary at $\sim$6--7~$M_\odot$. 

Of the stars reportedly more massive than 6~$M_\odot$, there are a number that either change variability types across the sectors or could not be confidently classified (``C" types in Table~2). These objects are represented by gray circles in Figure~\ref{fig:Amp_time_mass}). In taking a closer look at this set, we find that none of the variability classes in any sector are dippers (``QPD" or ``APD'' in Table~2), bursters, or candidates thereof (``CAPD" or ``CQPD"). In addition, only one object (TIC~125641610) in this putative higher mass set shows long-timescale variability, but its listed mass of 9.43~$M_\odot$ may be too high given the A4--A6 spectral type. 

%CS, CMP, CMP, QPS/S, TIC 410951921: CP, TIC 125641610: S/L, CS, CEB, CQPS, TIC 272142683: L/CL/S but M=5.8, CQPS, CS, CQPS, CMP, CMP, CQPS, CMP

In terms of amplitudes, there is a distinct jump at 7~$M_\odot$ for the quasi-periodic symmetric variables (i.e., green in Figure~\ref{fig:Amp_time_mass}), with values exclusively less than 1\% for the lower-mass stars and values exclusively greater than 2\% for higher mass stars. The stochastic variables, on the other hand, span the 7~$M_\odot$ boundary, and no clear amplitude trends are visible.

Our conclusion from this analysis is that dipper, burster, periodic, or long-timescale variability in HAeBes appears to be limited to relatively lower masses, at least when considered on timescales shorter than one month. Based on the disappearance of these variability types at higher masses and the jump in amplitude of quasi-periodic sources around 7~$M_\odot$, we concur that there is evidence for a change in accretion and/or inner disk structure at this mass boundary, which corresponds to early or mid-B spectral types. 

\section{Conclusions}
\label{sec:conclusions}

With the high photometric precision and continuous mnonitoring of the {\em TESS} mission, we found that a large fraction (roughly 95\% according to Table~2) of Herbig Ae/Be stars are variable on timescales of hours to months. Overall, this is an increase over the proportion of HAeBes identified as ``strongly" variable ($\gtrsim 50\%$) in longer term studies \citep[$\sim 25\%$][]{2018A&A...620A.128V}. Part of this difference can be attributed to {\em TESS's} sensitivity to low-amplitude variability down to 0.1\%.

To assign variability categories to the Herbig star sample, we employed the same classification scheme as developed by \cite{2014AJ....147...82C} for lower mass young disk-bearing stars. Many of the stars were readily classifiable by eye, although roughly 40\% presented challenges due to longer timescale or more unusual behavior (i.e., those marked with ``C'' in front of the variability type Table~2). For light curves that were confidently classified, the $Q$-$M$ diagram (Fig. \ref{fig:QMdiagram}) offered insights into differences in variability subtype prevalence among HAeBes versus CTTS. The most prominent change was a paucity of dipper stars at higher masses. We explained this by invoking the larger radius to which dust sublimination extends around hotter stars. With our extensive 188-star sample we also identified tentative mass dependences for the burster, periodic, and long-timescale subsets of HAeBe variables, uncovering further evidence for a change in the accretion structure of young stars beyond 7~$M_\odot$. 

{\em TESS} observed many stars in our HAeBe sample over multiple sectors, offering a window into the longer term photometric behavior of these stars. Among the stars with clear variability subtype, 86\% displayed the same light curve morphology in all available sectors. While the timescales in question vary from star to star, this result points to a tendency toward consistency over months to years.

In general, connecting variability to circumstellar properties is a challenge, with no clear correlations appearing in light curve morphology versus near-infrared colors (as shown in Fig.\ \ref{fig:JHHK}). The majority of disks in the HAeBe sample appear with excesses starting in the $J$ band, particularly for bursters, dippers, and long-timescale variables. Periodic and multi-periodic variables, on the other hand, are associated with relatively higher numbers of $H$ and $K$-band excesses. This is reminiscent of the division between classical and weak-lined T Tauri stars, the latter of which have low or no detectable accretion and tend to display periodic behavior associated with the stellar surface.

Finally, we speculated on the origin of periodicities observed in several Herbig Ae/Be stars observed with {\em TESS}. In particular, we identified two targets with single photometric periods that are most consistent with spot modulation. Such behavior is unusual among the A and B spectral type, and warrants follow-up to determine whether spots are indeed present, and if so, whether they are generated by large-scale magnetic fields.

\begin{acknowledgments}
We thank the referee for helpful comments. This work was supported by the National Aeronautics and Space Administration (NASA) under grants No. 80NSSC18K0141 and 80NSSC19K0670 issued through the TESS Guest Observer Program, as well as NASA SMD grant No. 80NSSC21K2077. This research made use of Lightkurve, a Python package for Kepler and TESS data analysis \citep{2018ascl.soft12013L}.
\end{acknowledgments}


\vspace{5mm}
\facilities{TESS}


\software{Matplotlib \citep{matplotlib}, NumPy \citep{numpy}, astrocut \citep{2019ASPC..523..397B}, lightkurve \citep{2018ascl.soft12013L}}



\appendix

We plot here the {\em TESS} light curves created for each target from sectors 1--45. 

\begin{figure*}
\figurenum{12}
\epsscale{0.9}
\plotone{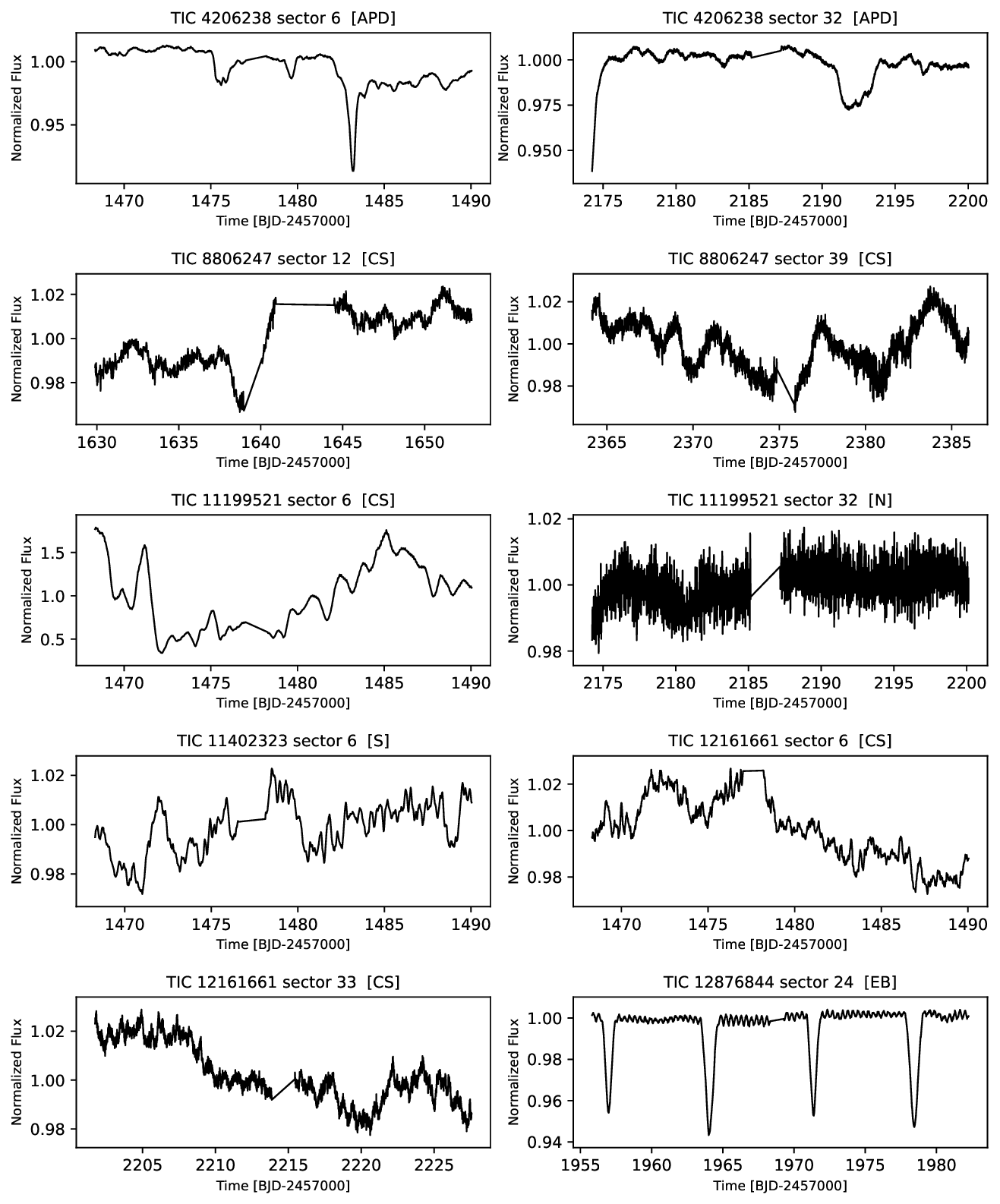}
\caption{Median-normalized light curves of intermediate-mass stars observed with {\em TESS}, in order of TIC identifier. Figure labels include in brackets the variability type from Table~2, namely "P" = strictly periodic behavior, "MP" = multiple distinct periods, "QPD" = quasi-periodic dippers, "QPS" = quasi-periodic symmetric, "APD" = aperiodic dippers, "B" = bursters, "S" = stochastic stars, "L" = long-timescale behavior that doesn't fall into the other categories. The complete figure set is available in the online journal.}
\label{alllcs}
\end{figure*}

\clearpage
%\addtocounter{figure}{-1}
\begin{figure*}
\epsscale{0.90}
\plotone{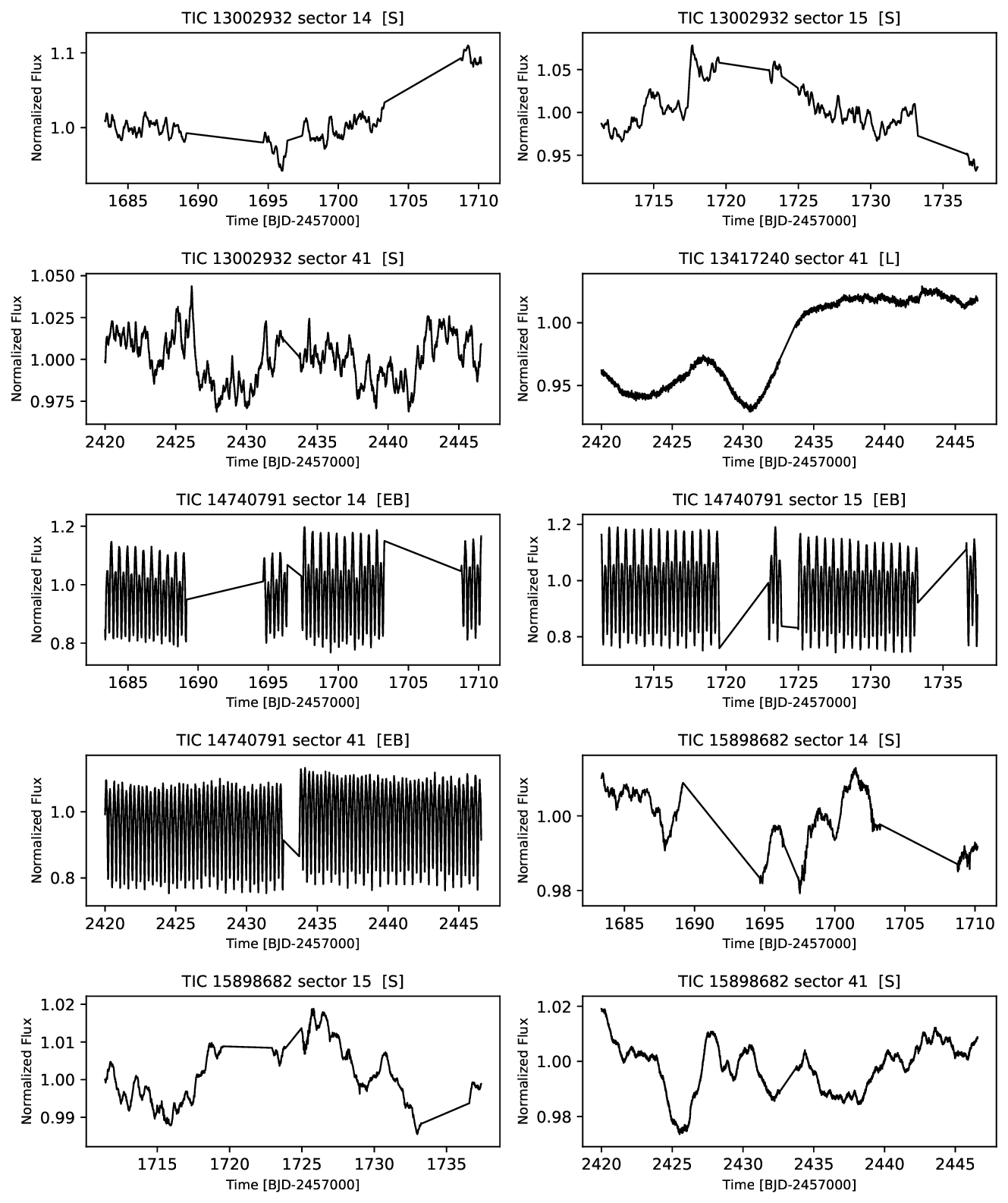}
\caption{Cont.}
\end{figure*}

\clearpage
\addtocounter{figure}{-1}
\begin{figure*}
\epsscale{0.90}
\plotone{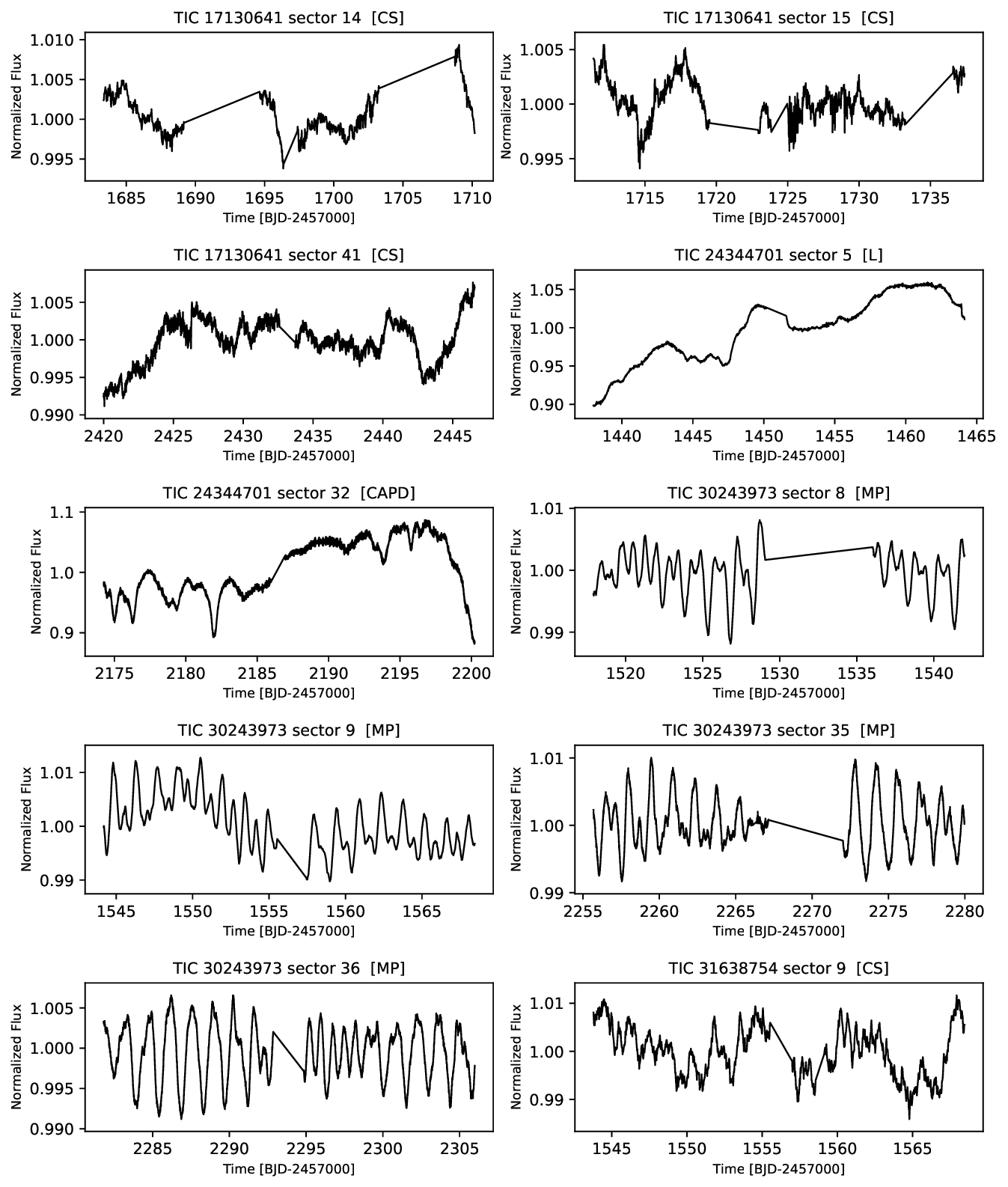}
\caption{Cont.}
\end{figure*}

\clearpage
\addtocounter{figure}{-1}
\begin{figure*}
\epsscale{0.90}
\plotone{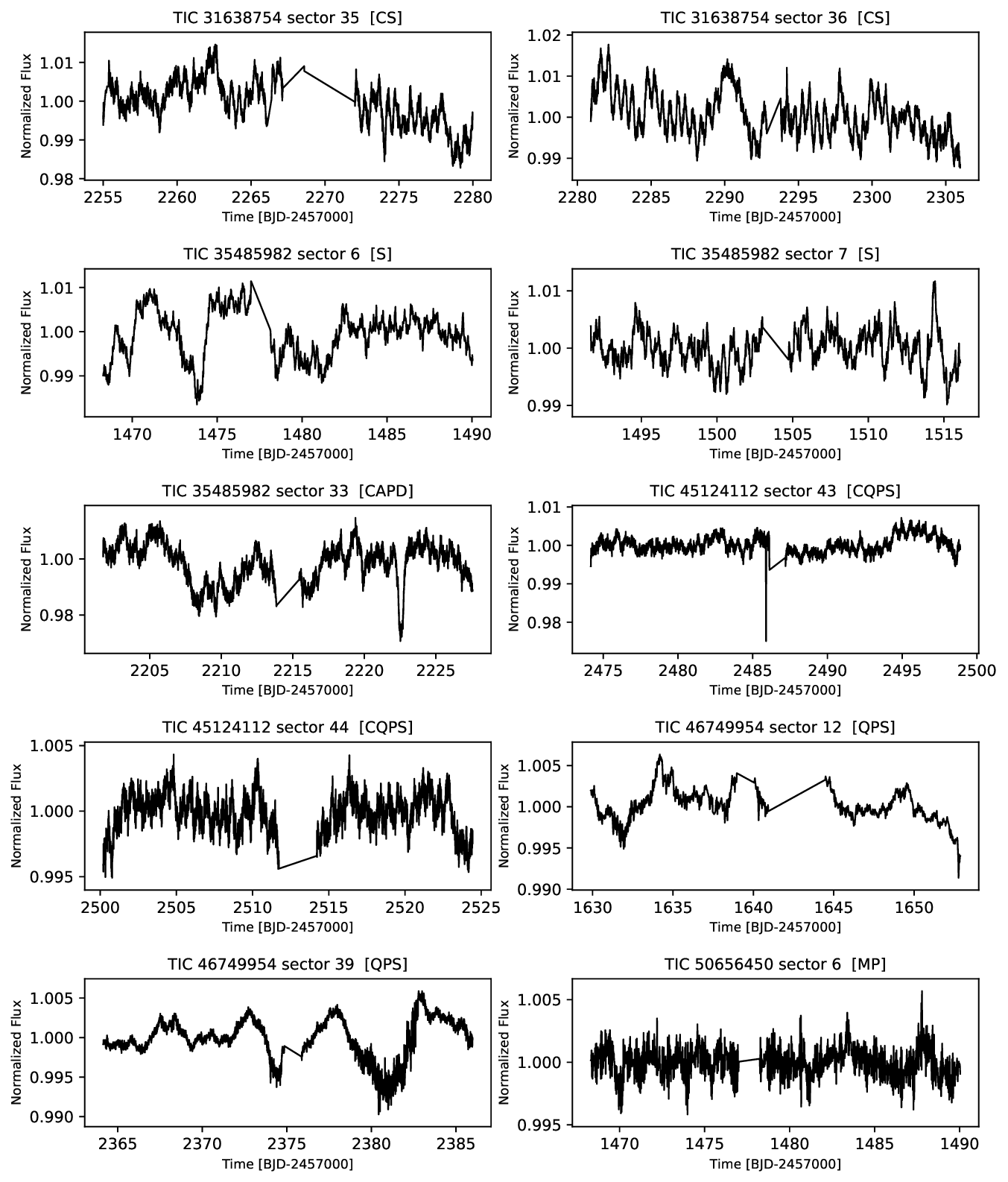}
\caption{Cont.}
\end{figure*}

\clearpage
\addtocounter{figure}{-1}
\begin{figure*}
\epsscale{0.90}
\plotone{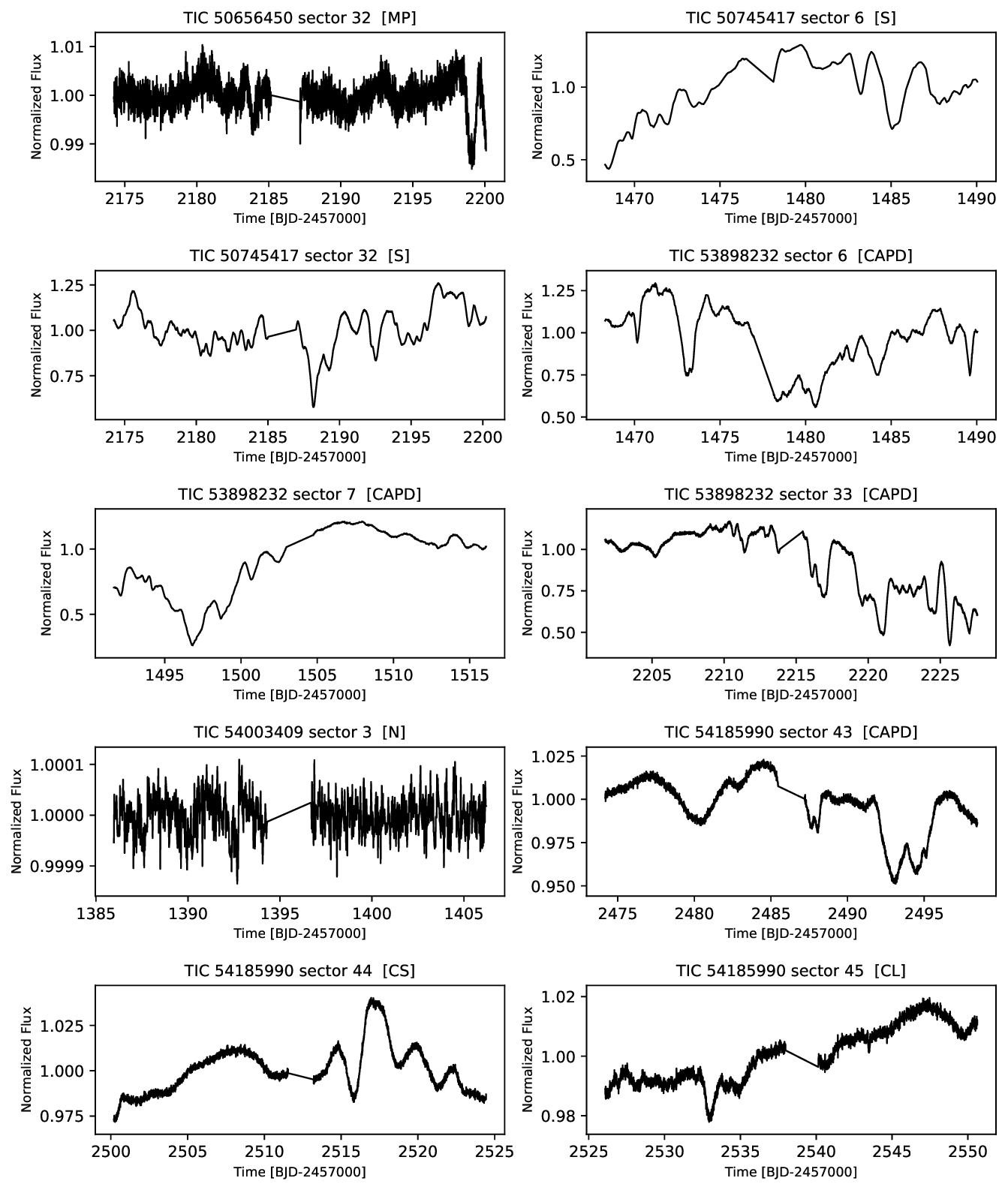}
\caption{Cont.}
\end{figure*}

\clearpage
\addtocounter{figure}{-1}
\begin{figure*}
\epsscale{0.90}
\plotone{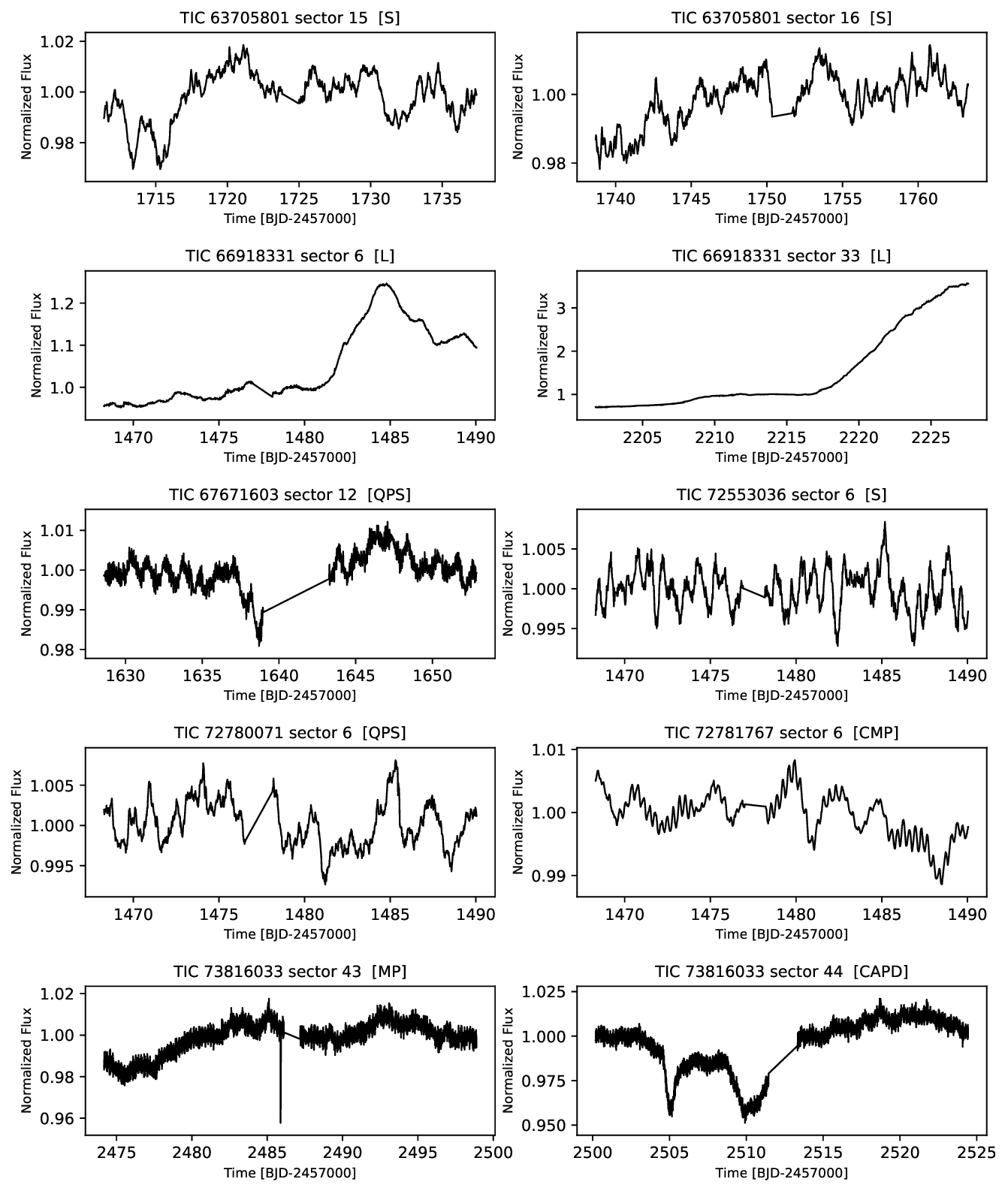}
\caption{Cont.}
\end{figure*}

\clearpage
\addtocounter{figure}{-1}
\begin{figure*}
\epsscale{0.90}
\plotone{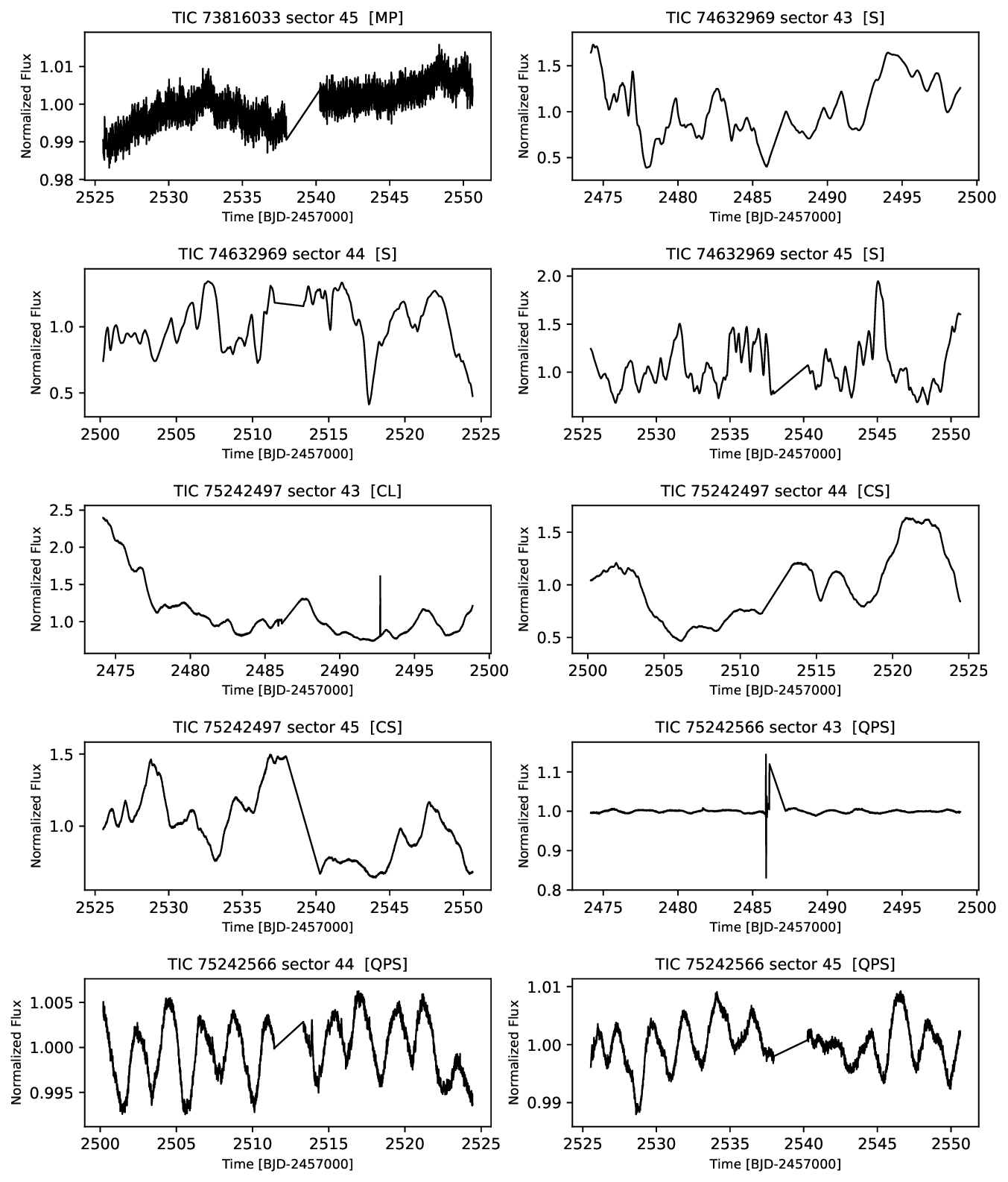}
\caption{Cont.}
\end{figure*}

\clearpage
\addtocounter{figure}{-1}
\begin{figure*}
\epsscale{0.90}
\plotone{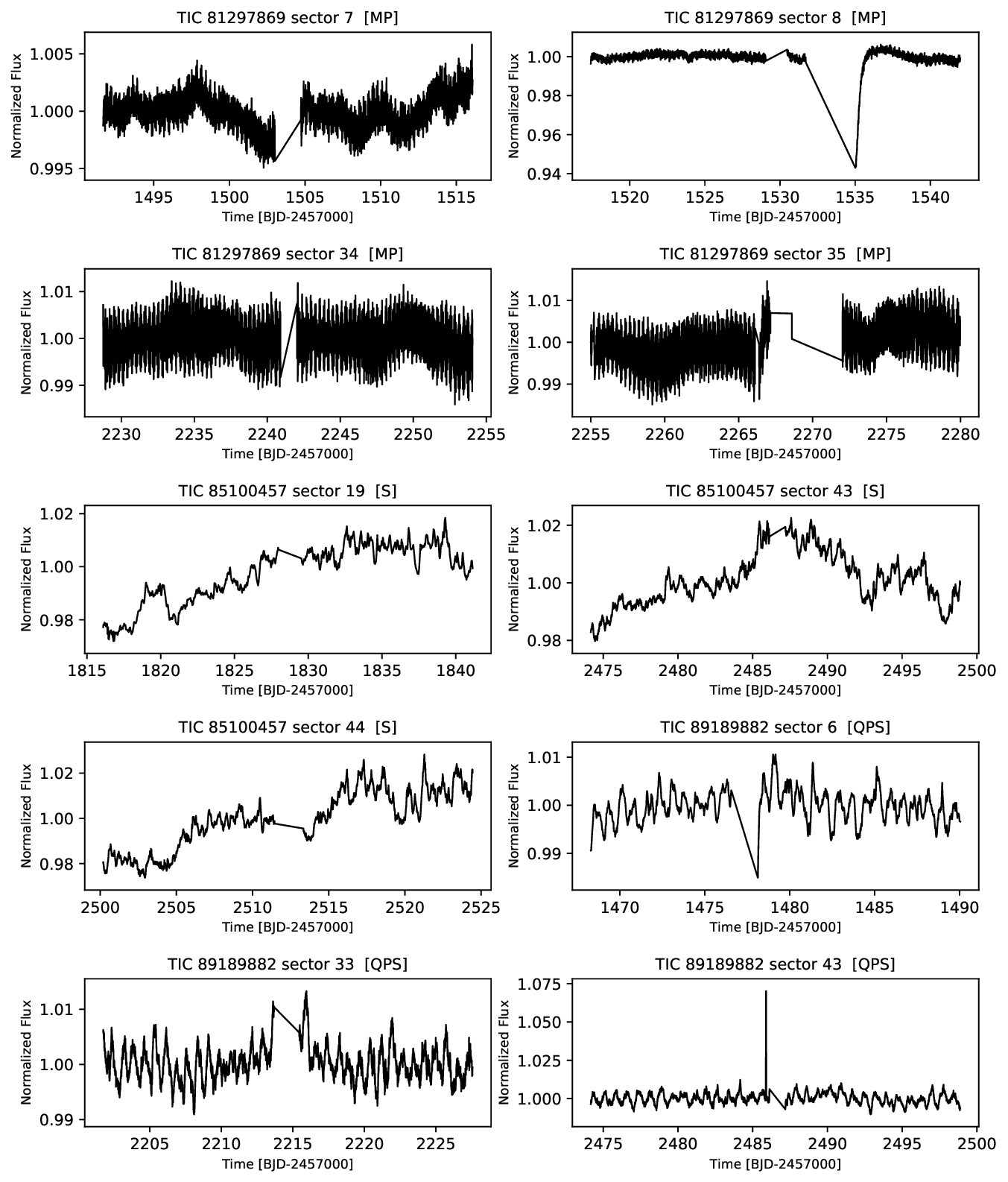}
\caption{Cont.}
\end{figure*}

\clearpage
\addtocounter{figure}{-1}
\begin{figure*}
\epsscale{0.90}
\plotone{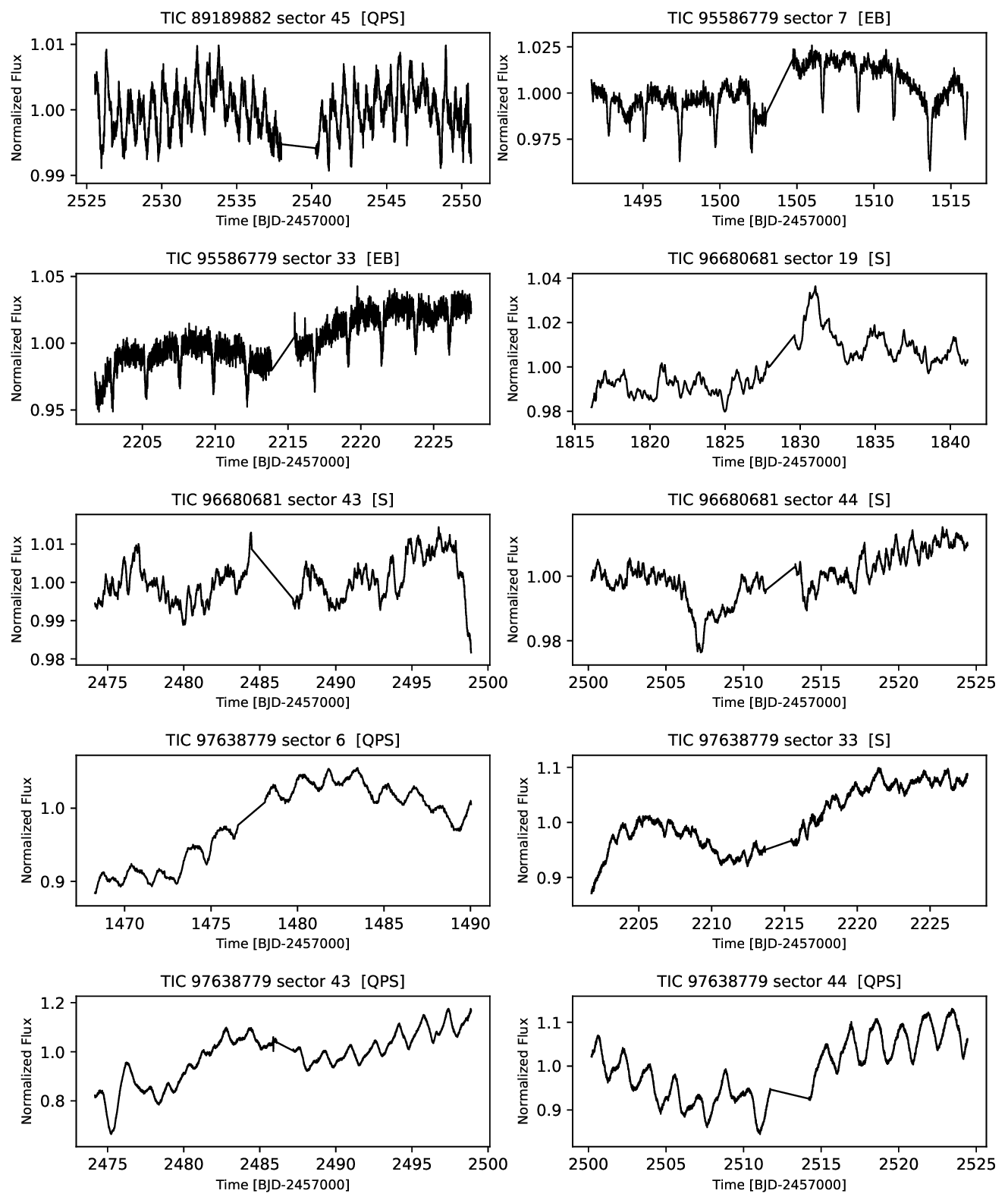}
\caption{Cont.}
\end{figure*}

\clearpage
\addtocounter{figure}{-1}
\begin{figure*}
\epsscale{0.90}
\plotone{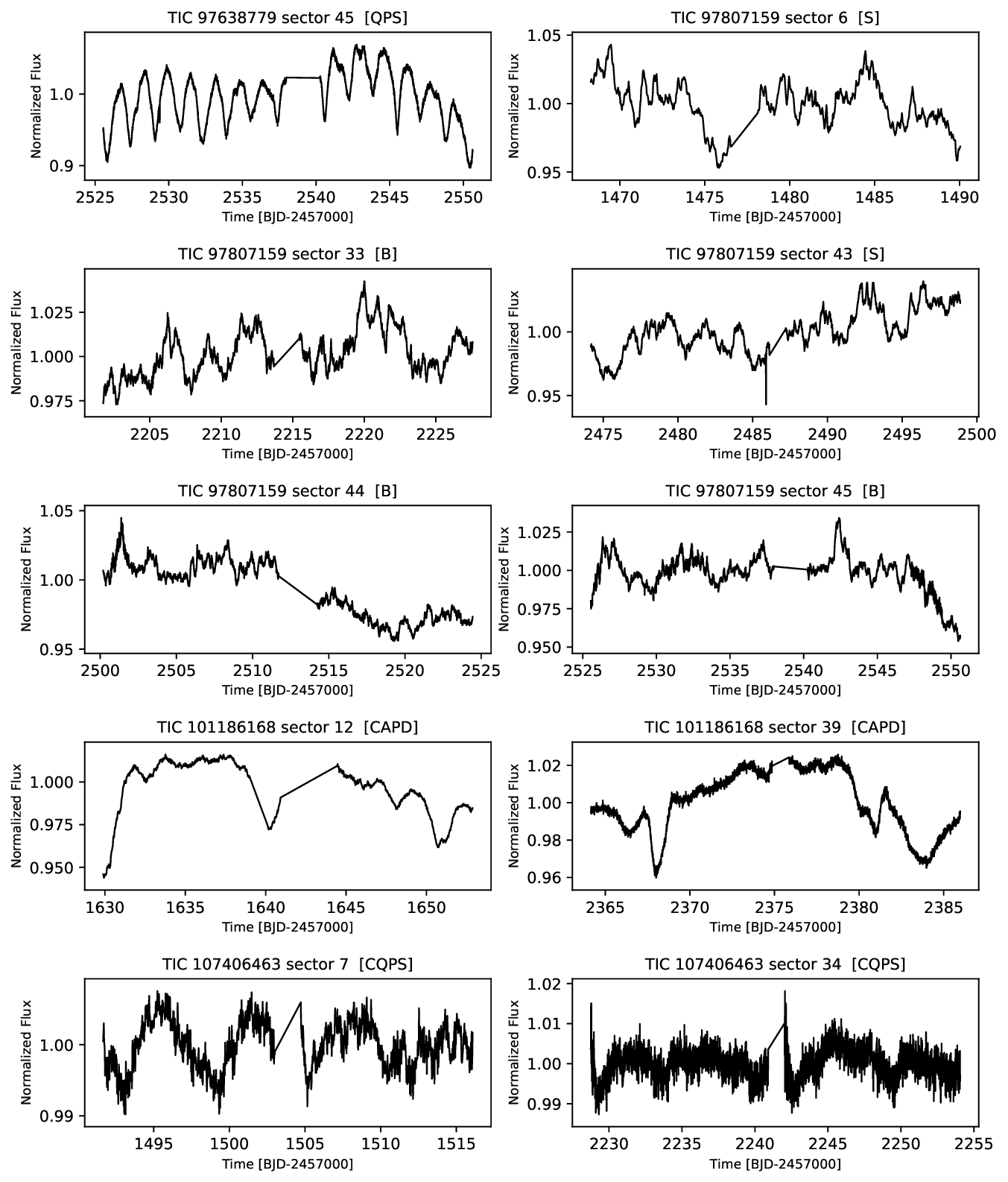}
\caption{Cont.}
\end{figure*}

\clearpage
\addtocounter{figure}{-1}
\begin{figure*}
\epsscale{0.90}
\plotone{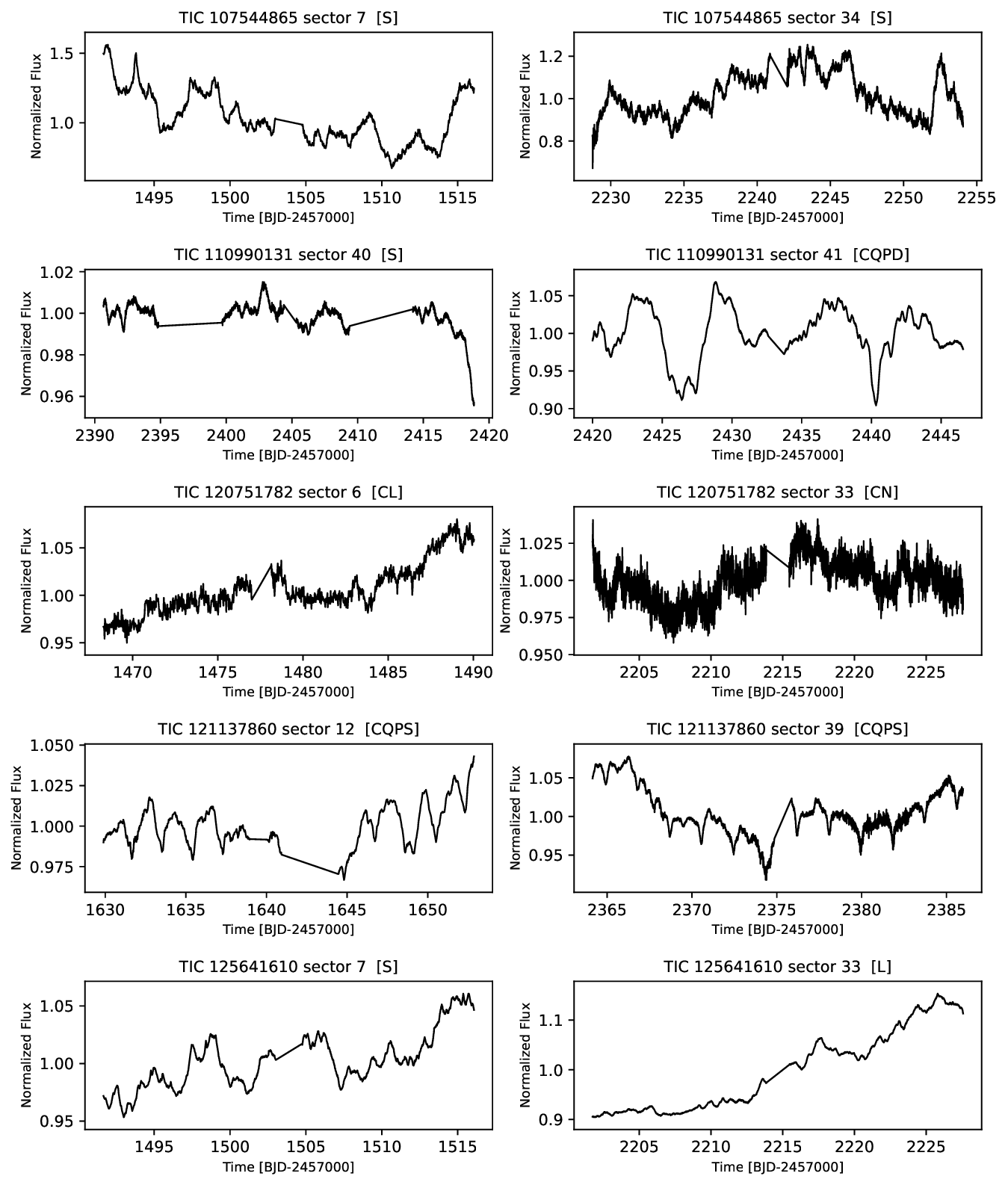}
\caption{Cont.}
\end{figure*}

\clearpage
\addtocounter{figure}{-1}
\begin{figure*}
\epsscale{0.90}
\plotone{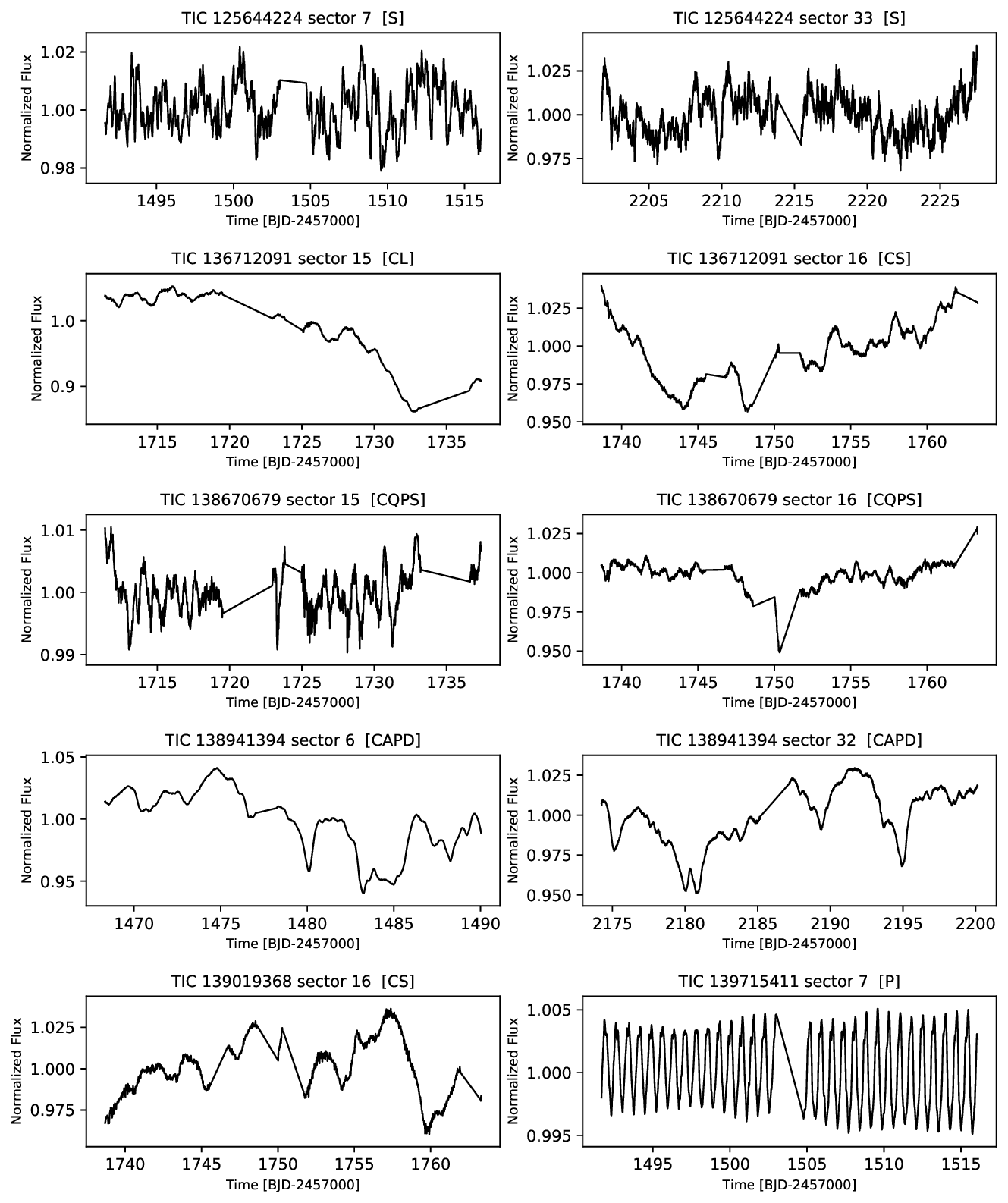}
\caption{Cont.}
\end{figure*}

\clearpage
\addtocounter{figure}{-1}
\begin{figure*}
\epsscale{0.90}
\plotone{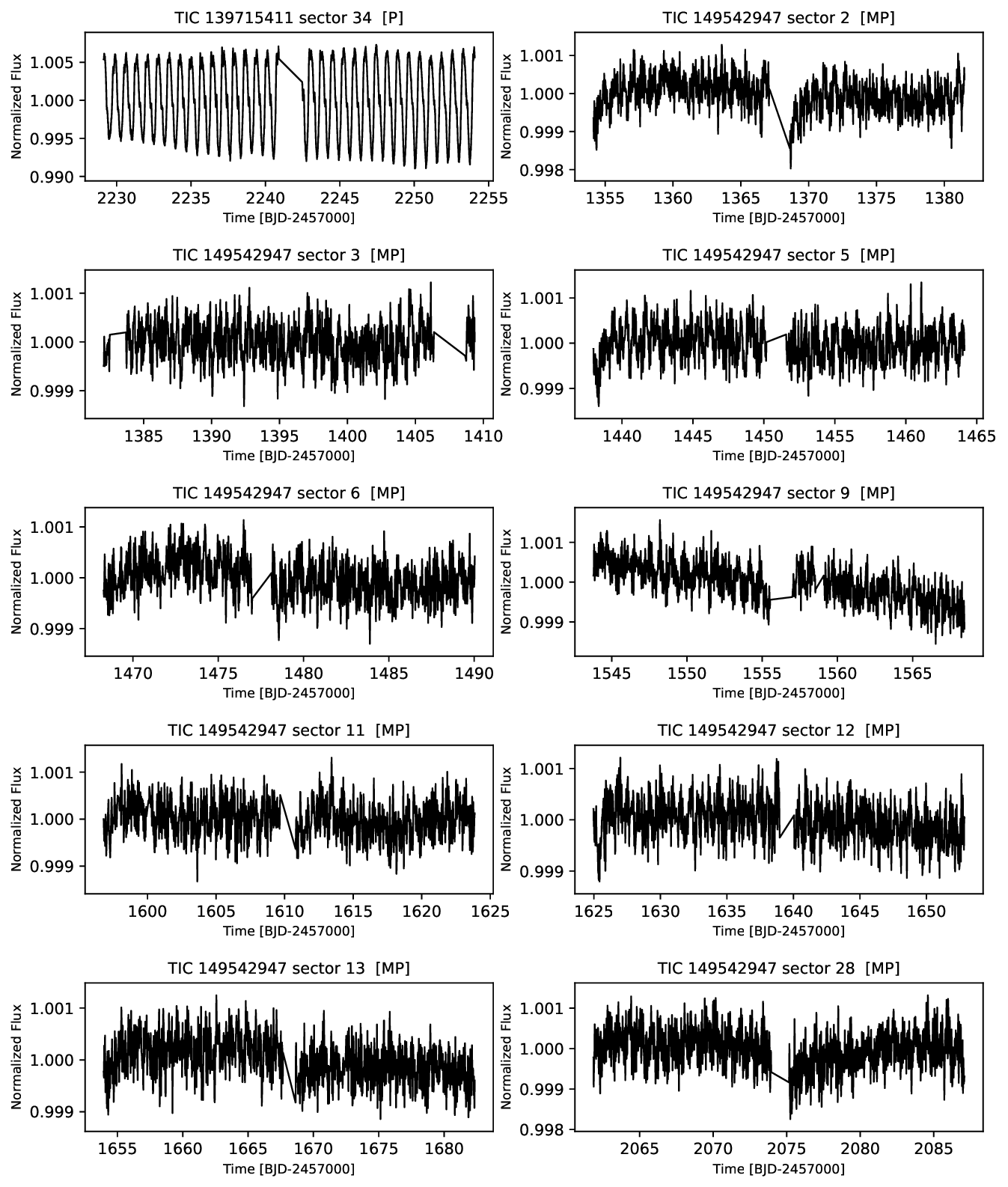}
\caption{Cont.}
\end{figure*}

\clearpage
\addtocounter{figure}{-1}
\begin{figure*}
\epsscale{0.90}
\plotone{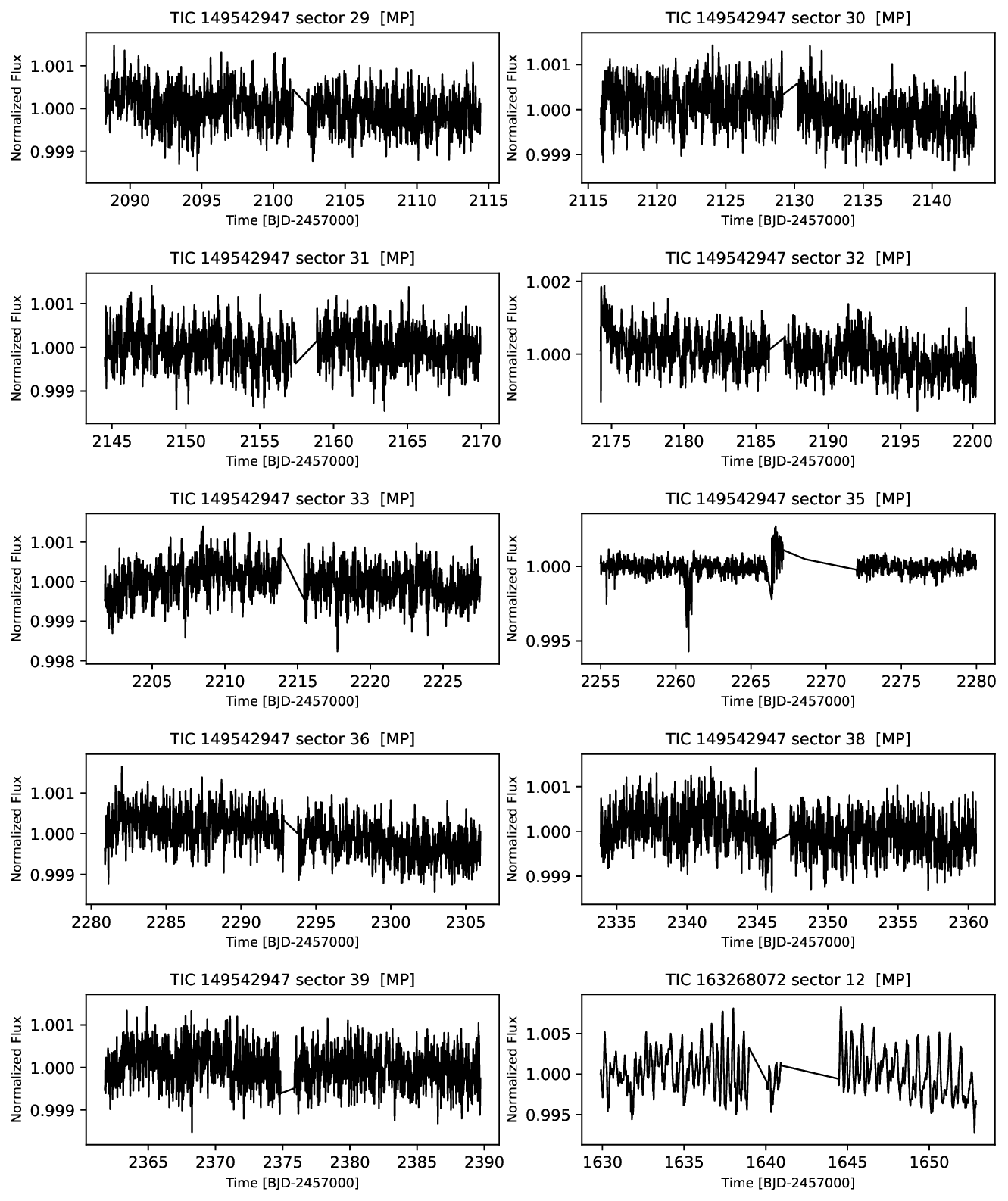}
\caption{Cont.}
\end{figure*}

\clearpage
\addtocounter{figure}{-1}
\begin{figure*}
\epsscale{0.90}
\plotone{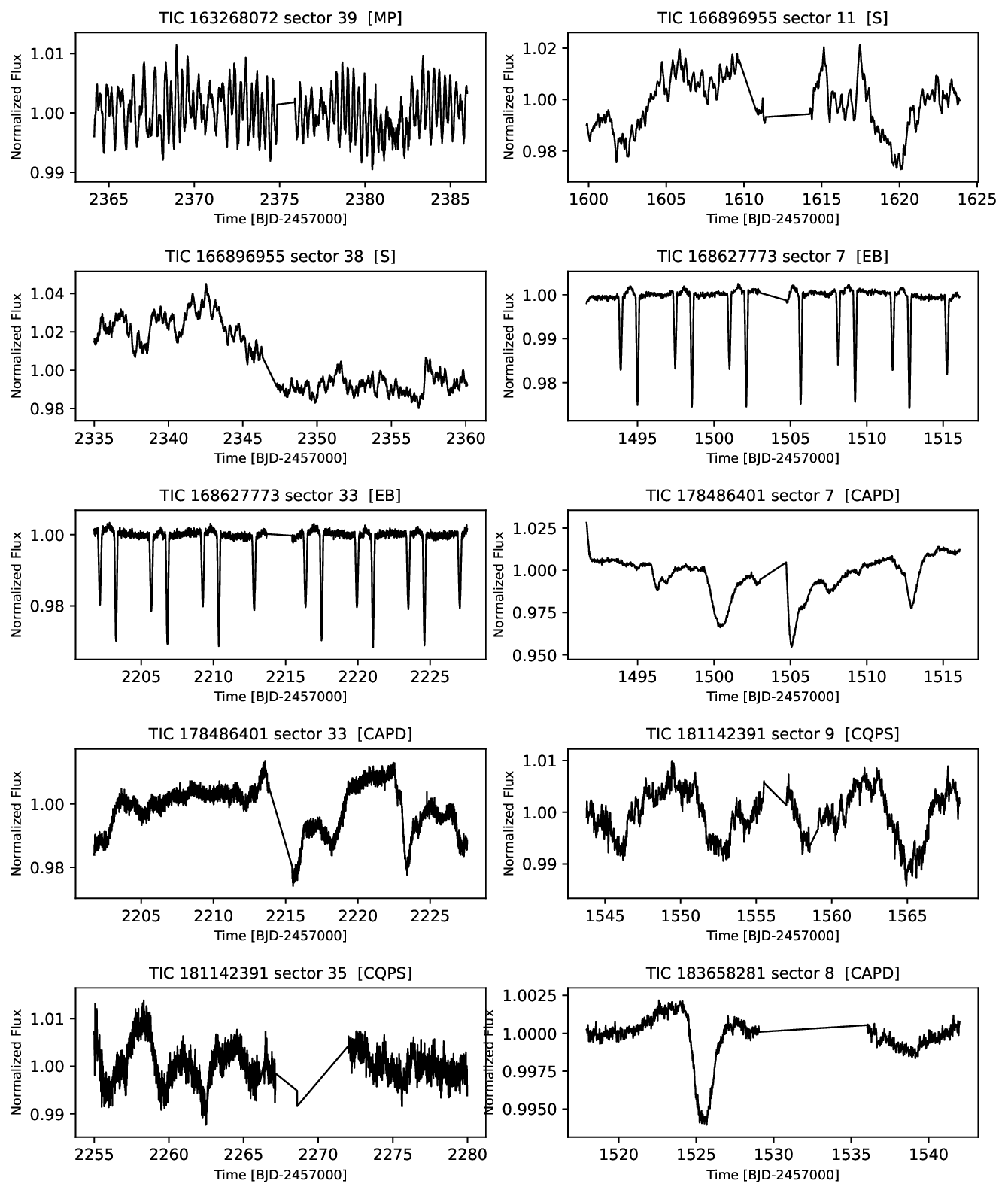}
\caption{Cont.}
\end{figure*}

\clearpage
\addtocounter{figure}{-1}
\begin{figure*}
\epsscale{0.90}
\plotone{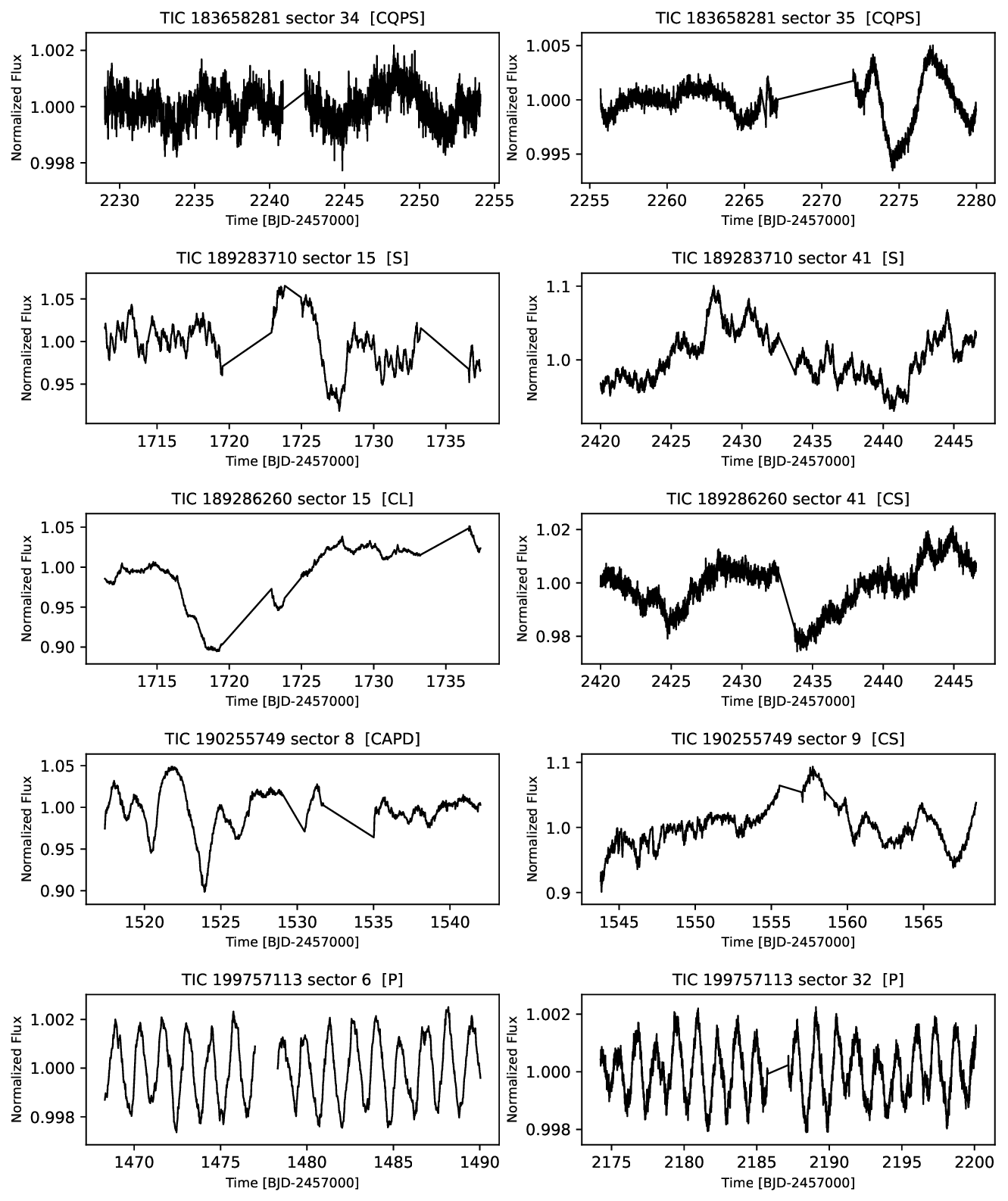}
\caption{Cont.}
\end{figure*}

\clearpage
\addtocounter{figure}{-1}
\begin{figure*}
\epsscale{0.90}
\plotone{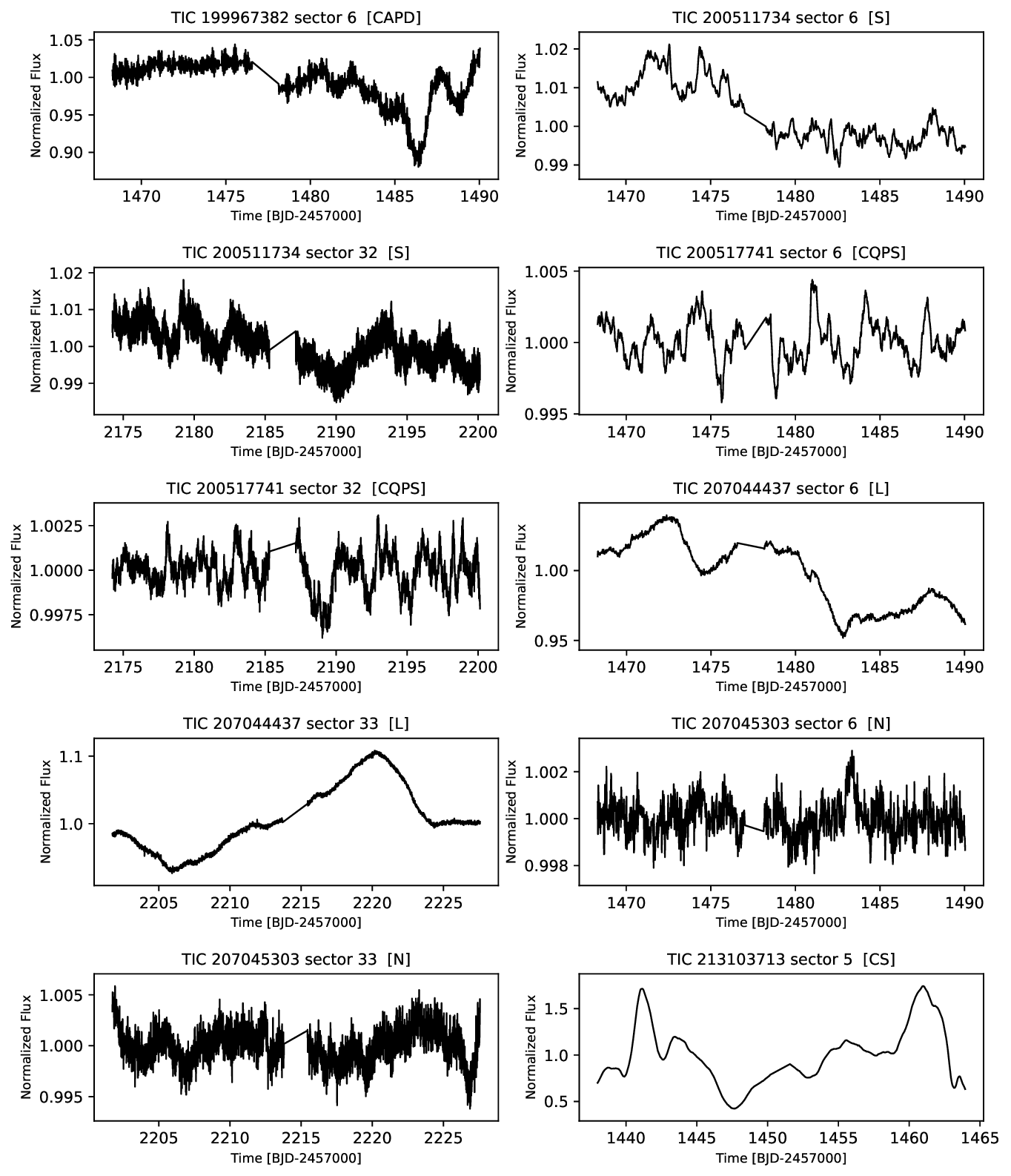}
\caption{Cont.}
\end{figure*}

\clearpage
\addtocounter{figure}{-1}
\begin{figure*}
\epsscale{0.90}
\plotone{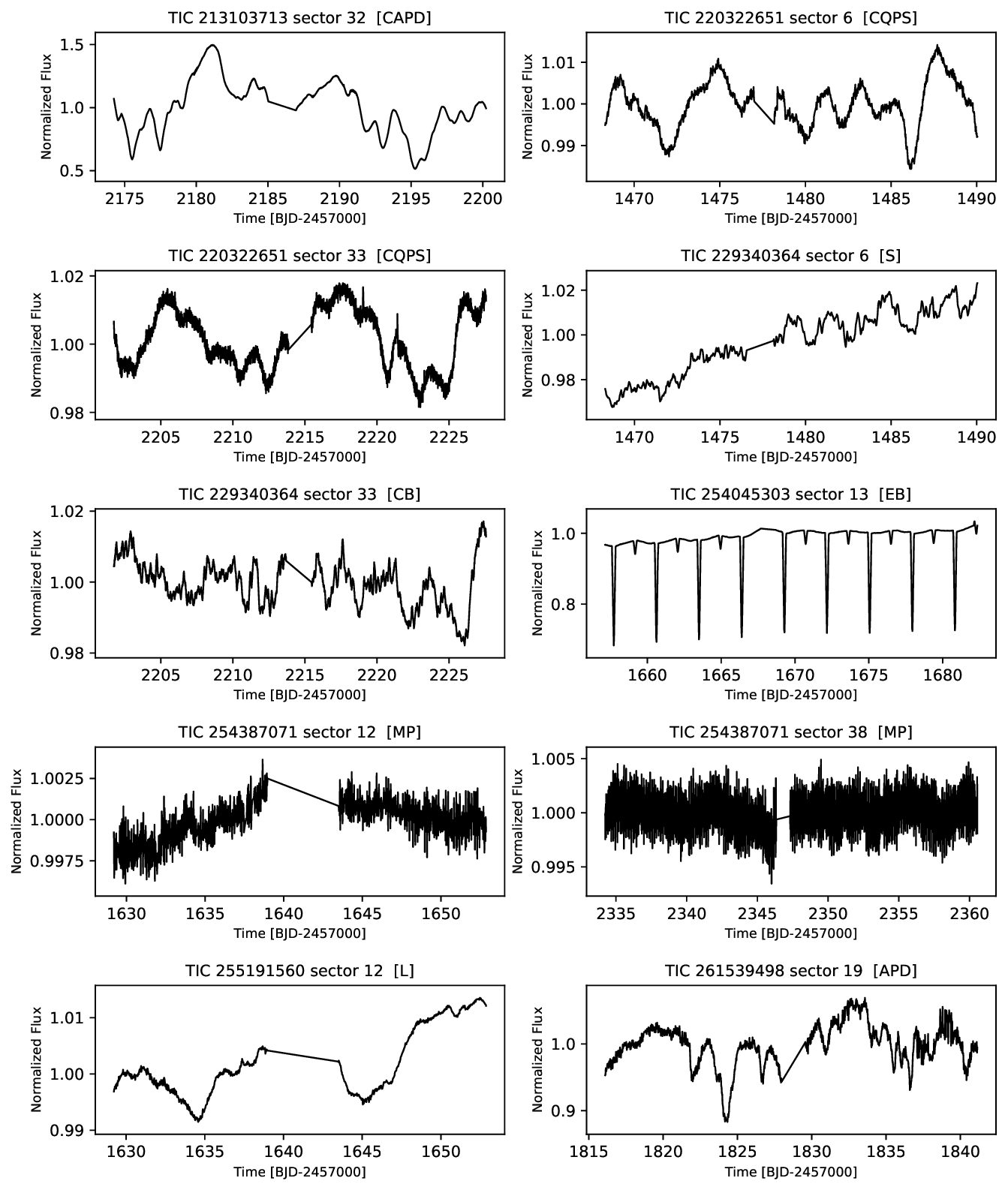}
\caption{Cont.}
\end{figure*}

\clearpage
\addtocounter{figure}{-1}
\begin{figure*}
\epsscale{0.90}
\plotone{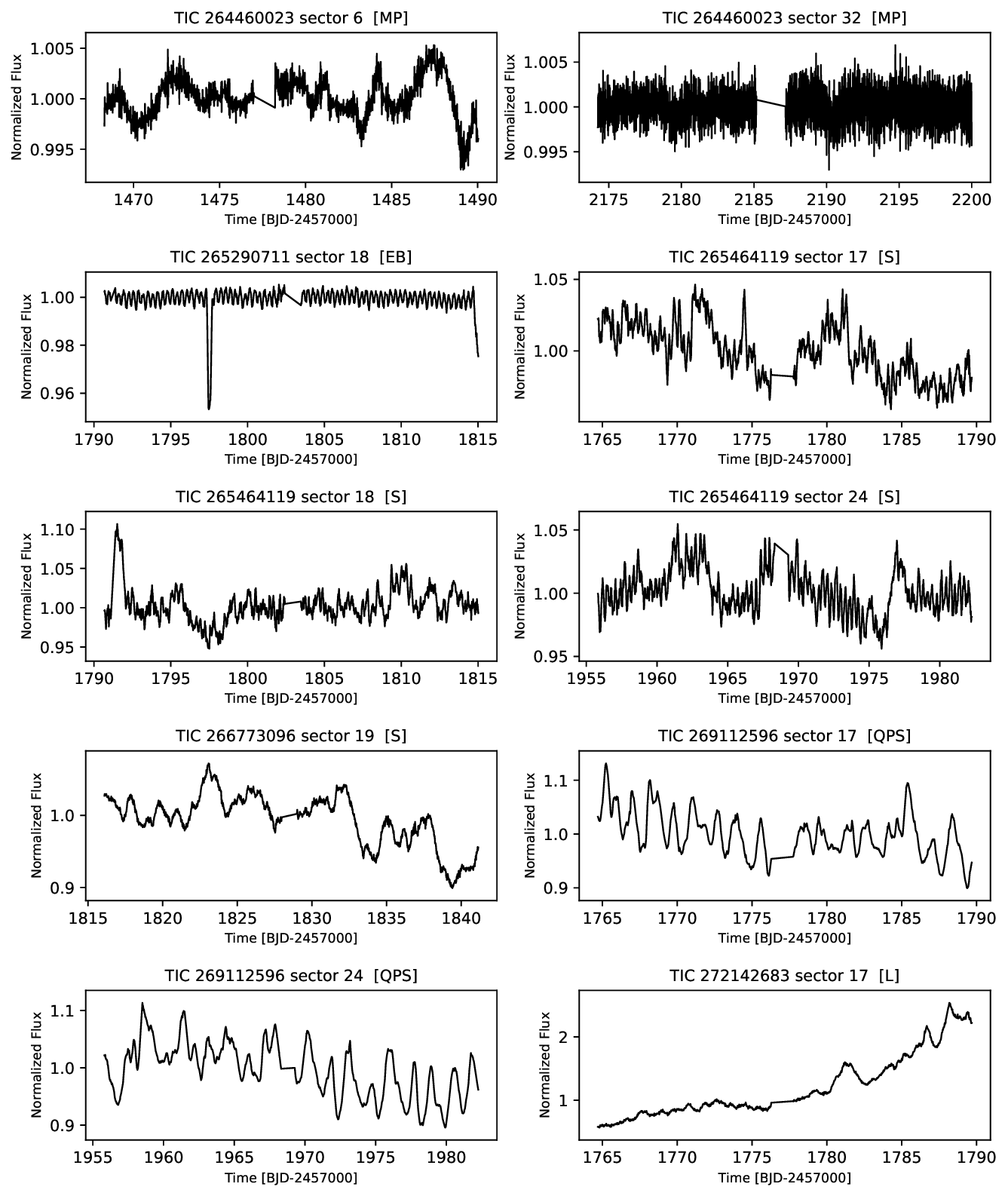}
\caption{Cont.}
\end{figure*}

\clearpage
\addtocounter{figure}{-1}
\begin{figure*}
\epsscale{0.90}
\plotone{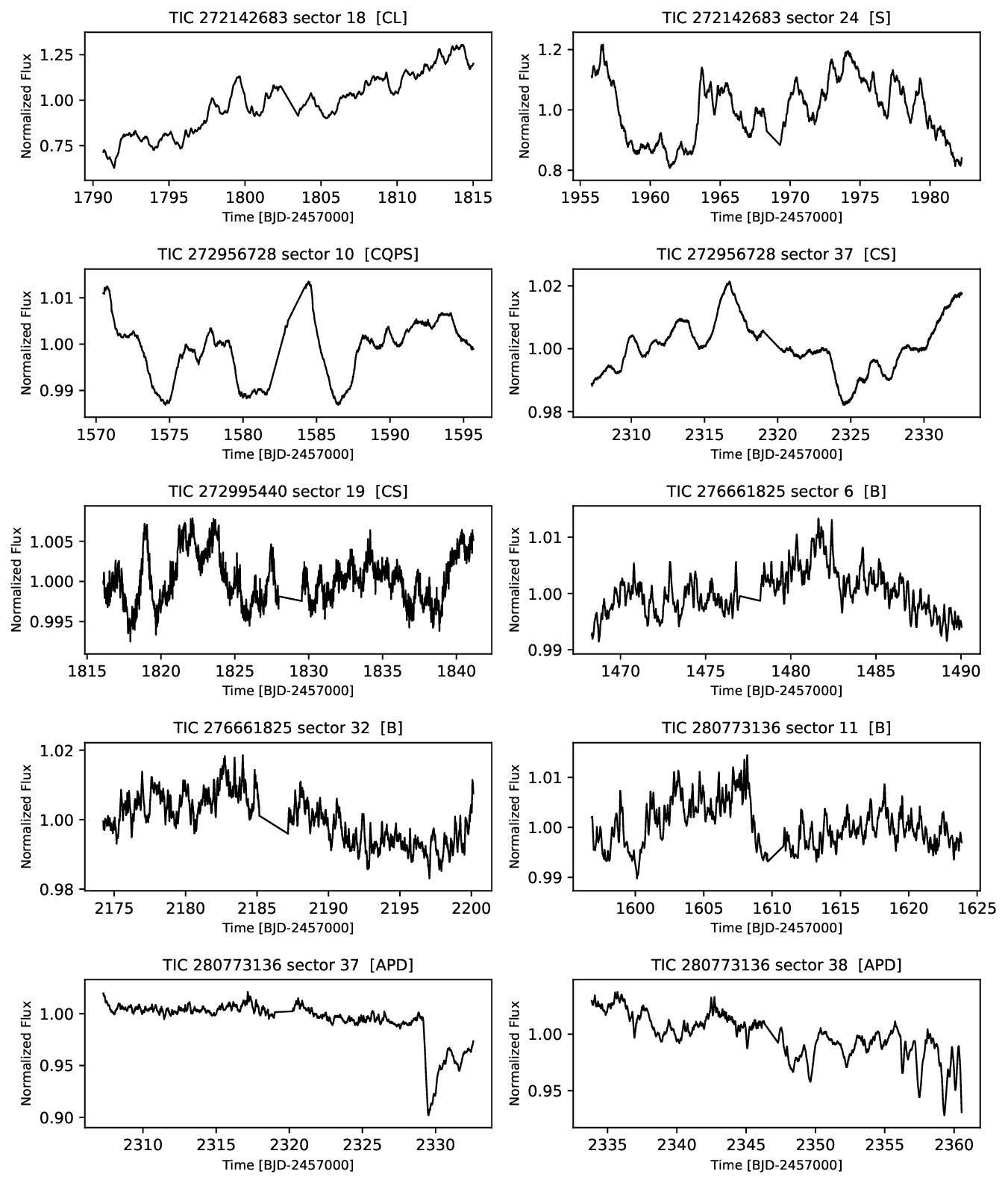}
\caption{Cont.}
\end{figure*}

\clearpage
\addtocounter{figure}{-1}
\begin{figure*}
\epsscale{0.90}
\plotone{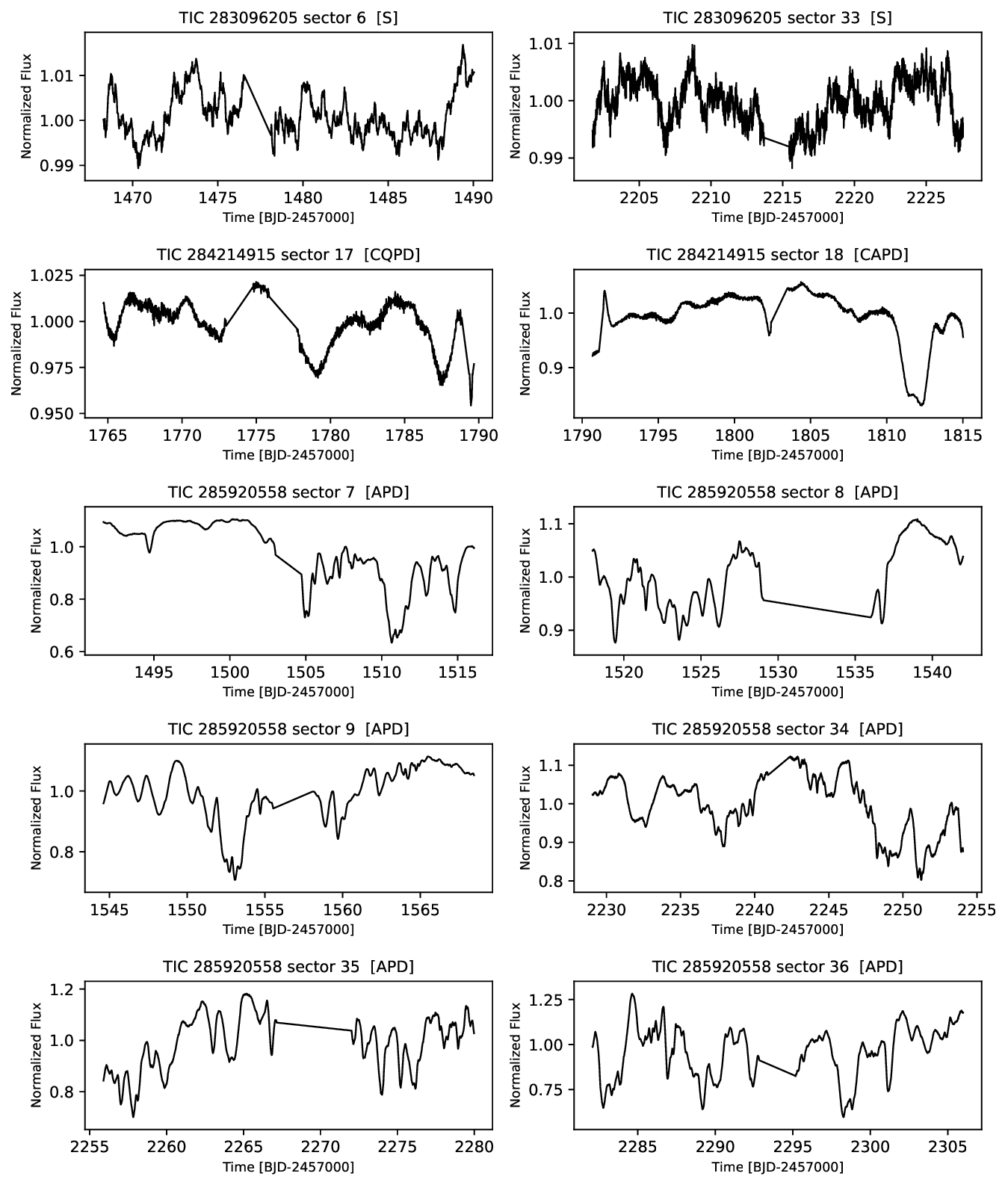}
\caption{Cont.}
\end{figure*}

\clearpage
\addtocounter{figure}{-1}
\begin{figure*}
\epsscale{0.90}
\plotone{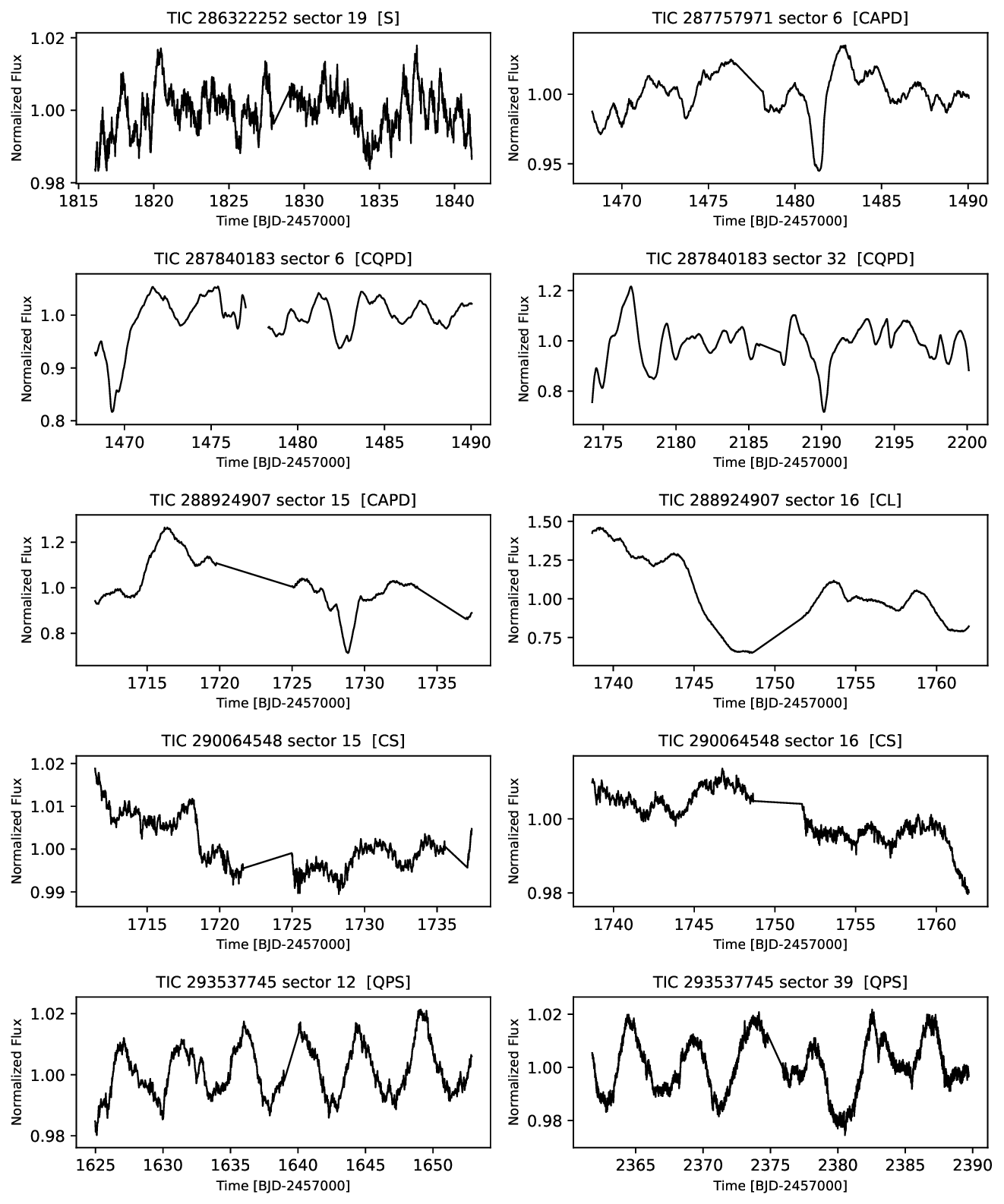}
\caption{Cont.}
\end{figure*}

\clearpage
\addtocounter{figure}{-1}
\begin{figure*}
\epsscale{0.90}
\plotone{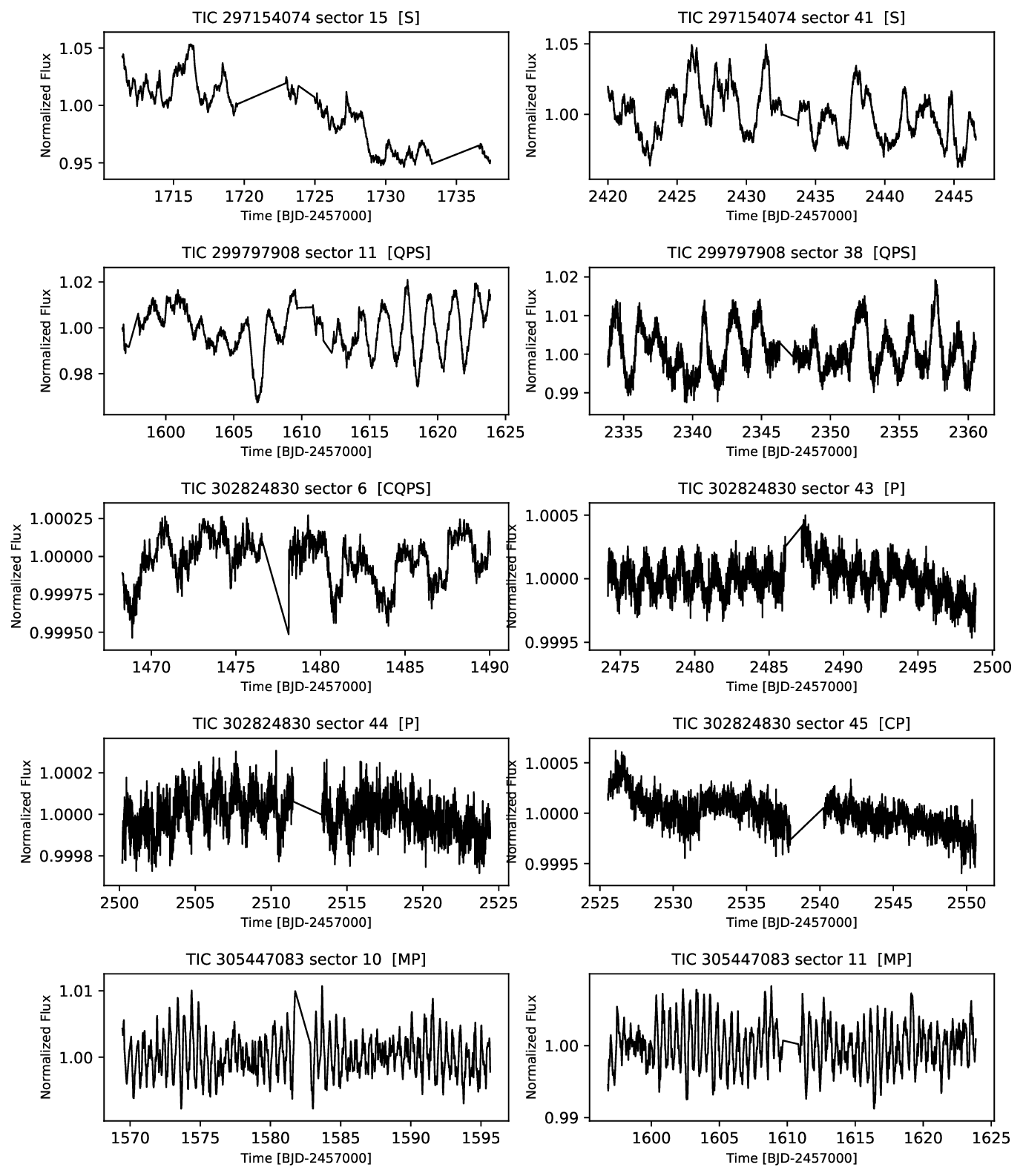}
\caption{Cont.}
\end{figure*}

\clearpage
\addtocounter{figure}{-1}
\begin{figure*}
\epsscale{0.90}
\plotone{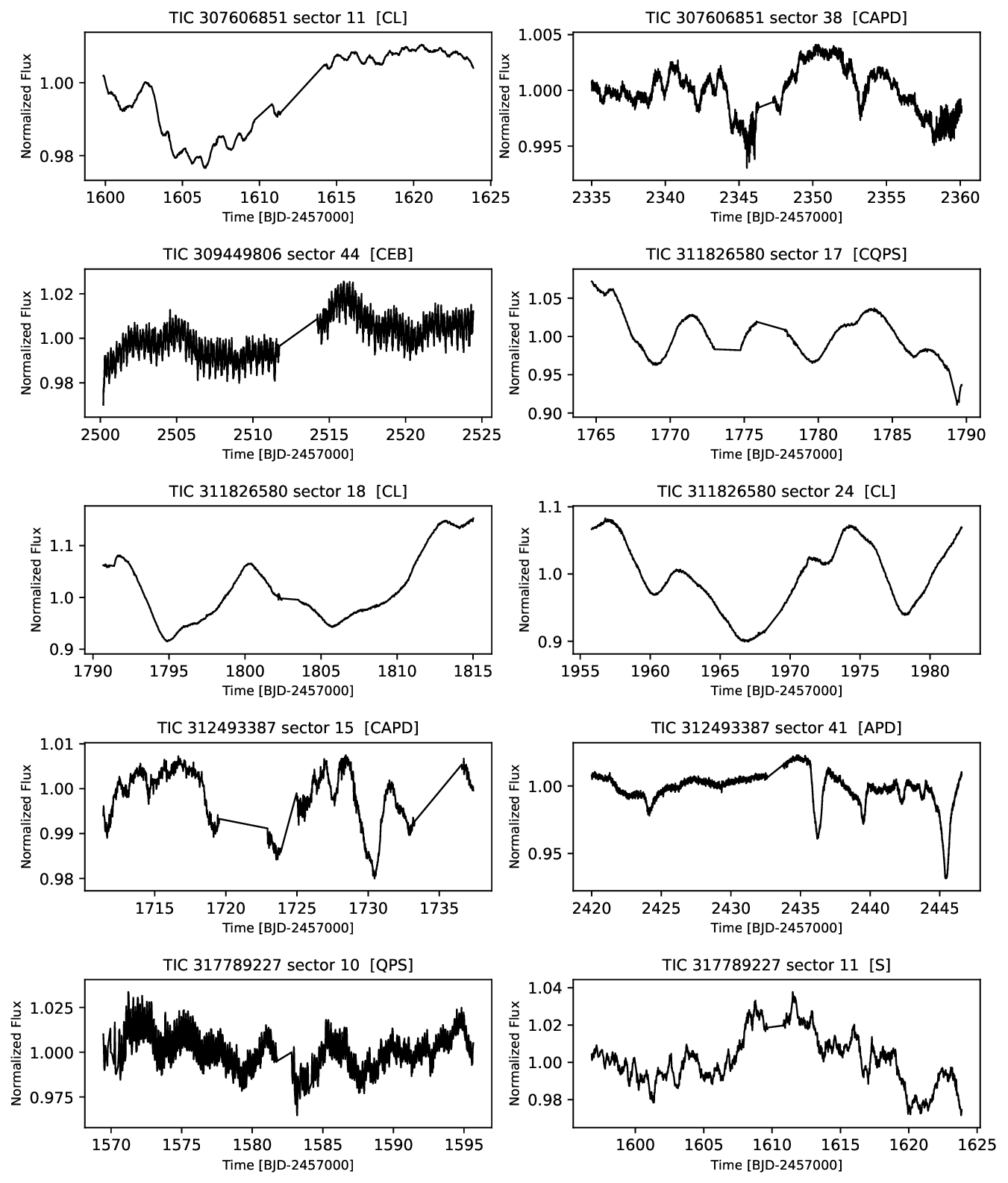}
\caption{Cont.}
\end{figure*}

\clearpage
\addtocounter{figure}{-1}
\begin{figure*}
\epsscale{0.90}
\plotone{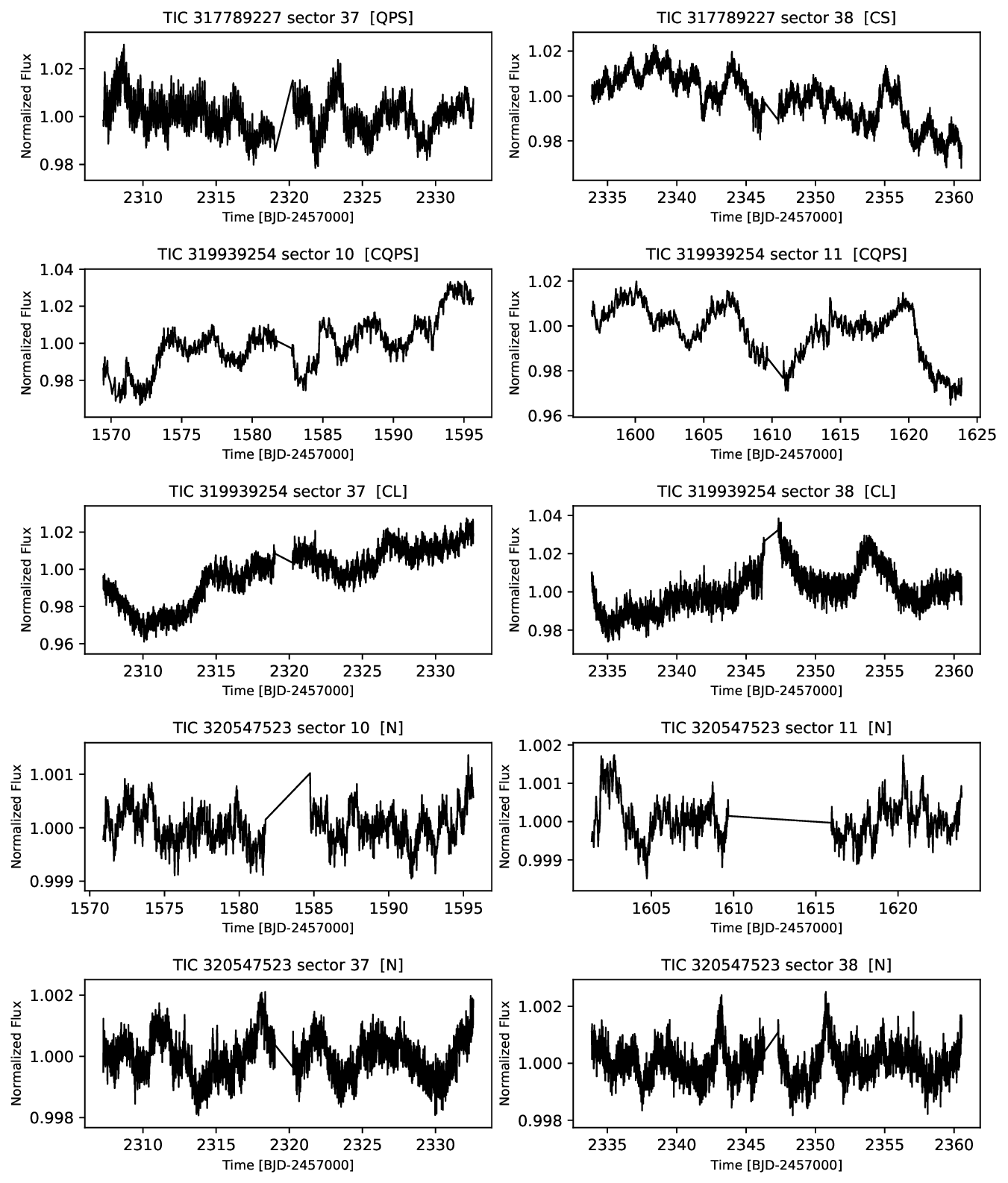}
\caption{Cont.}
\end{figure*}

\clearpage
\addtocounter{figure}{-1}
\begin{figure*}
\epsscale{0.90}
\plotone{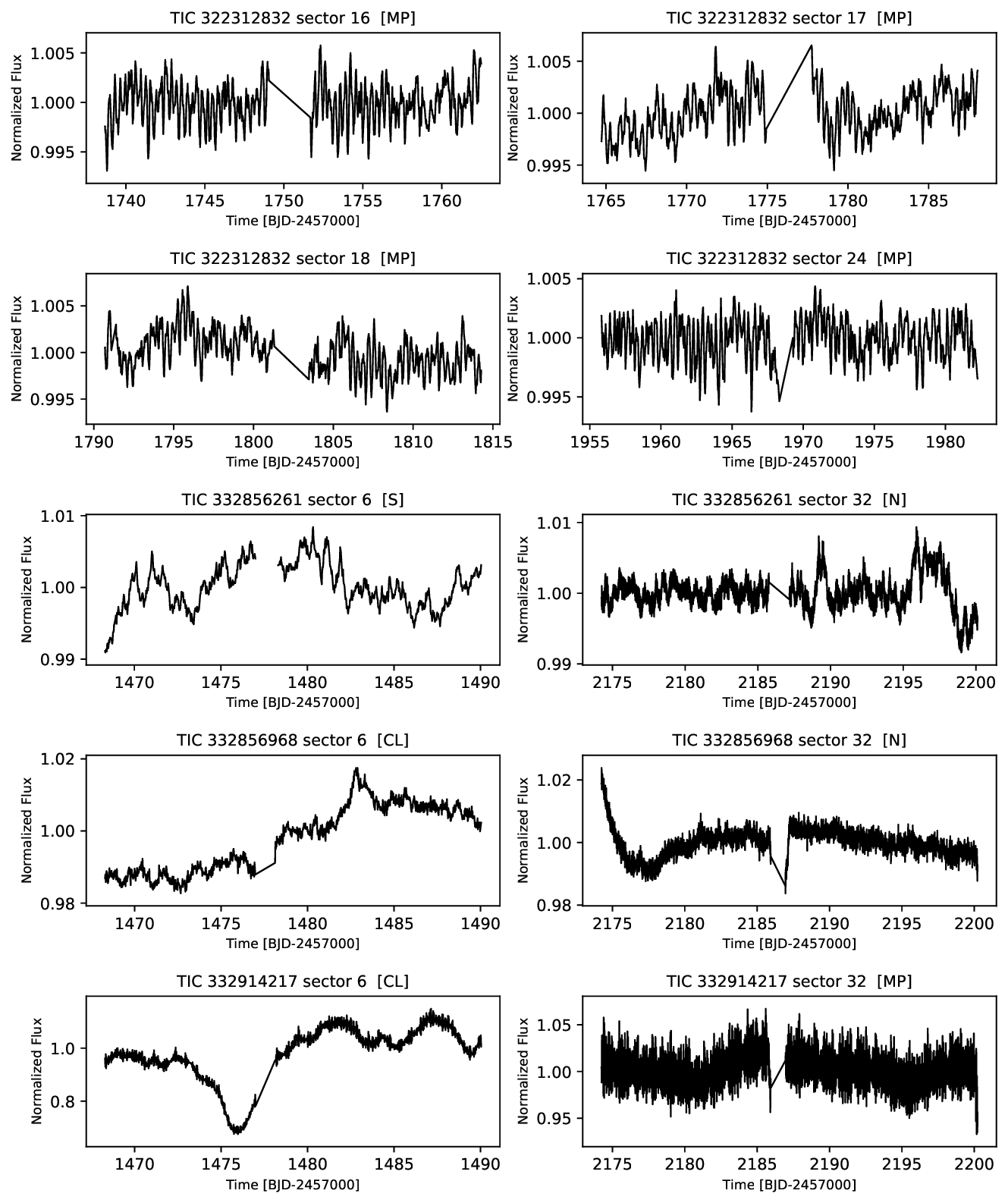}
\caption{Cont.}
\end{figure*}

\clearpage
\addtocounter{figure}{-1}
\begin{figure*}
\epsscale{0.90}
\plotone{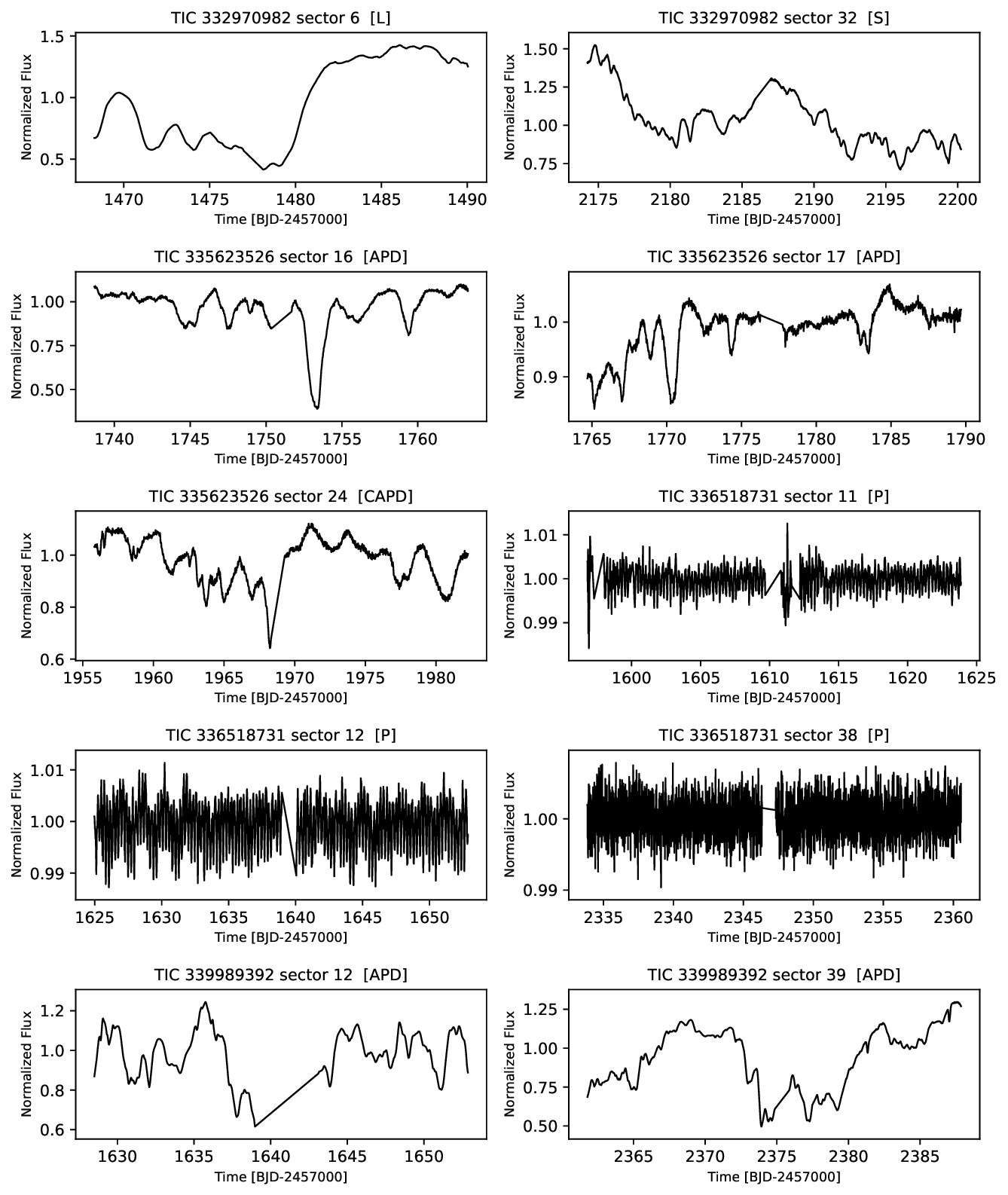}
\caption{Cont.}
\end{figure*}

\clearpage
\addtocounter{figure}{-1}
\begin{figure*}
\epsscale{0.90}
\plotone{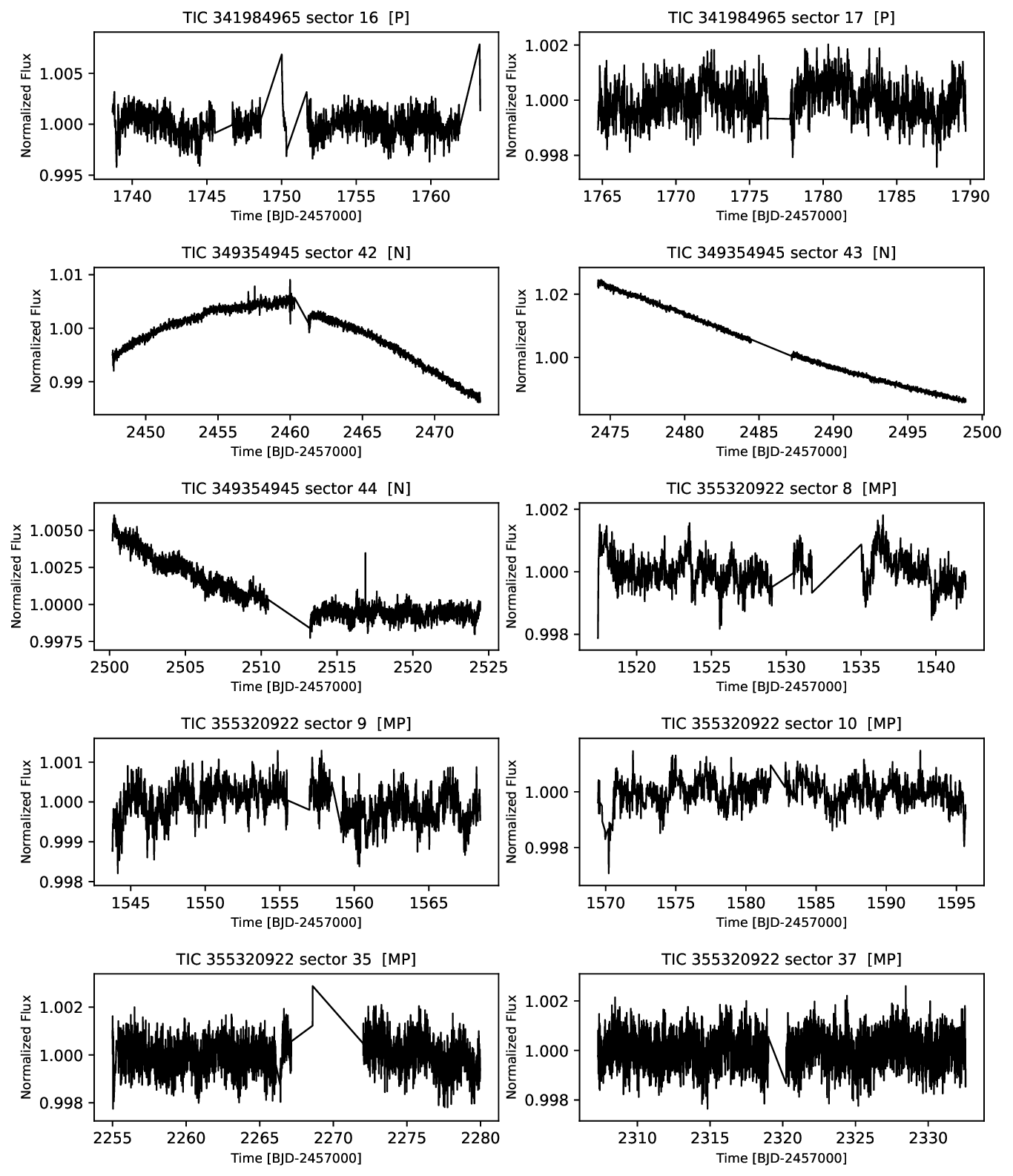}
\caption{Cont.}
\end{figure*}

\clearpage
\addtocounter{figure}{-1}
\begin{figure*}
\epsscale{0.90}
\plotone{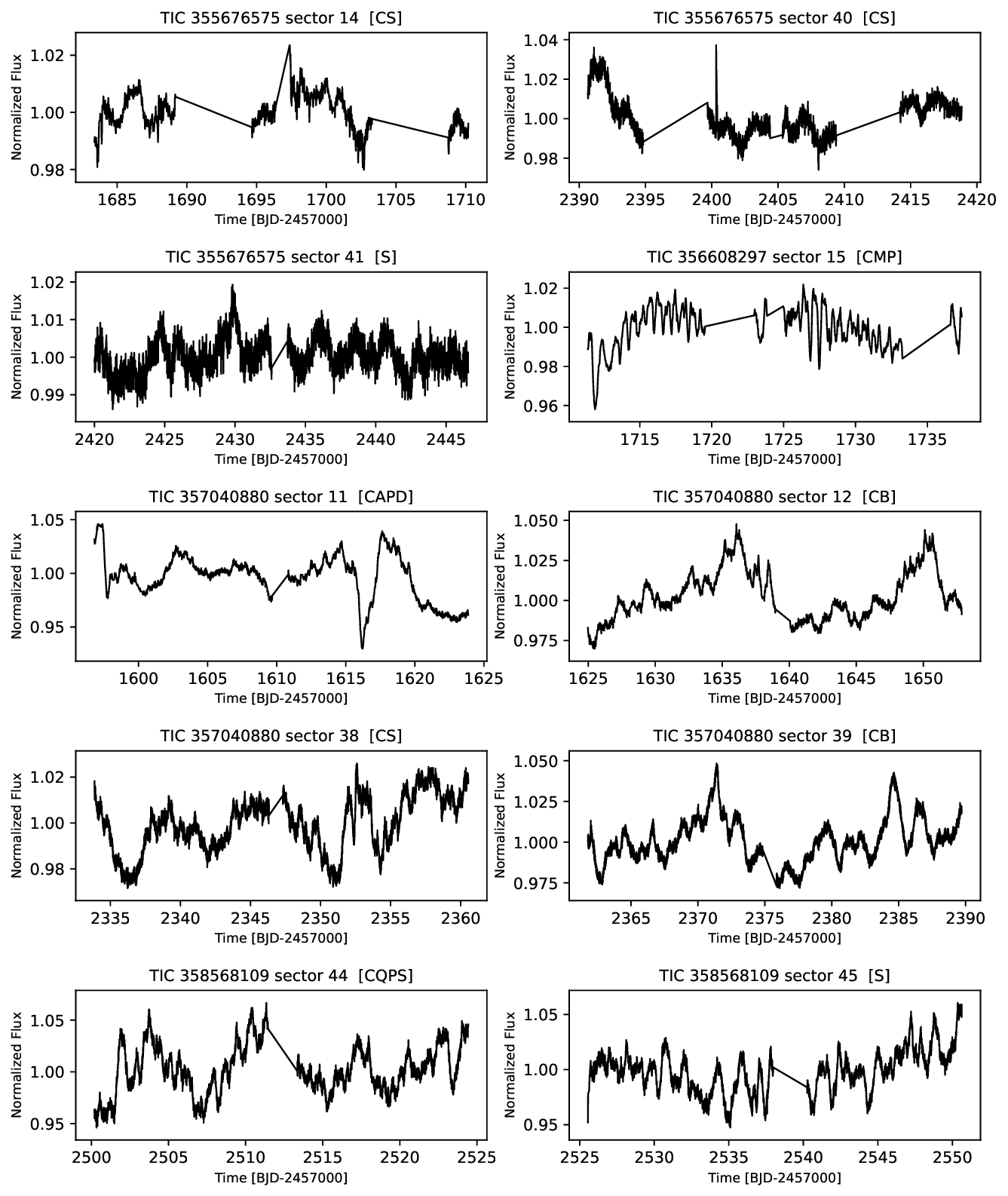}
\caption{Cont.}
\end{figure*}

\clearpage
\addtocounter{figure}{-1}
\begin{figure*}
\epsscale{0.90}
\plotone{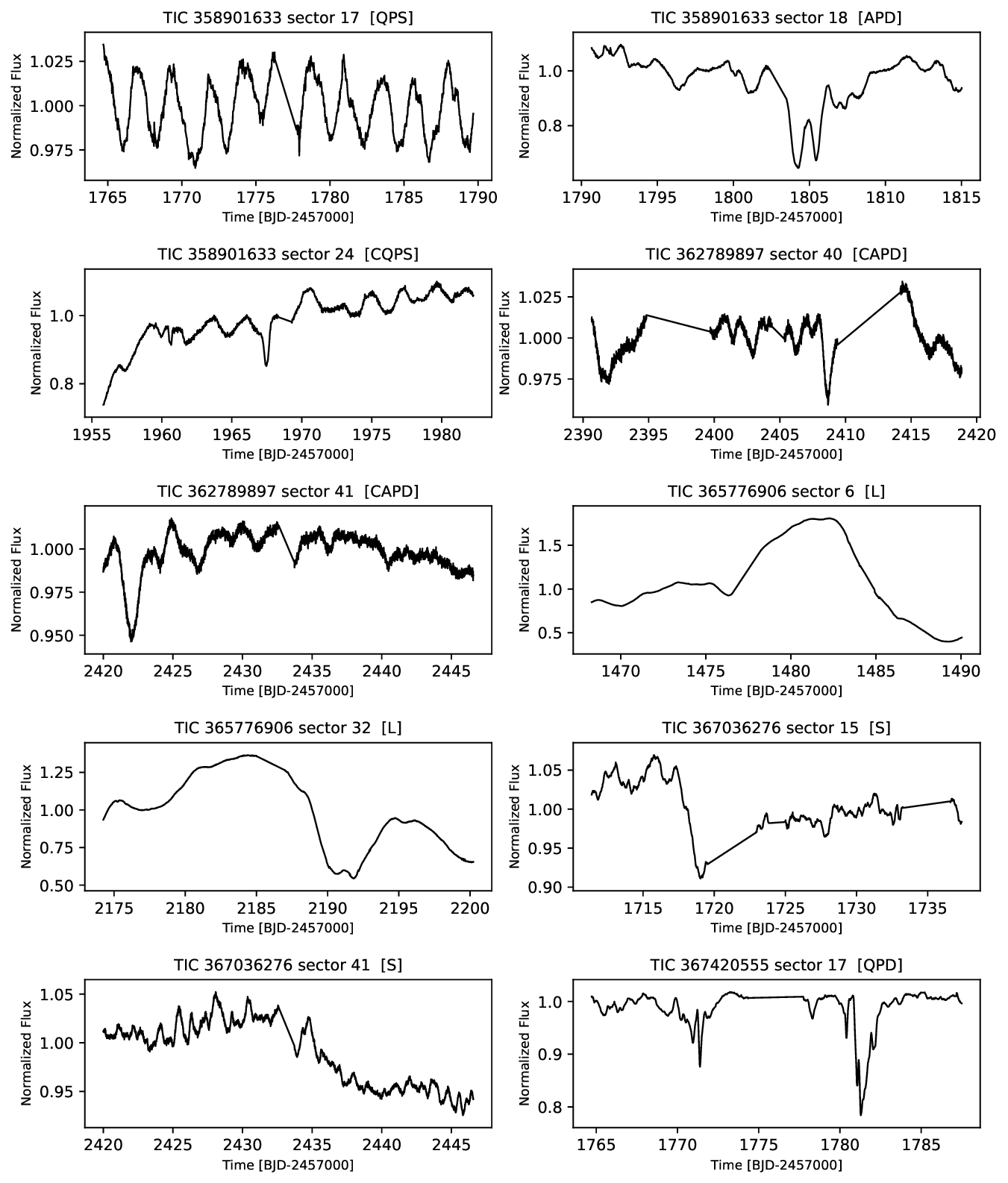}
\caption{Cont.}
\end{figure*}

\clearpage
\addtocounter{figure}{-1}
\begin{figure*}
\epsscale{0.90}
\plotone{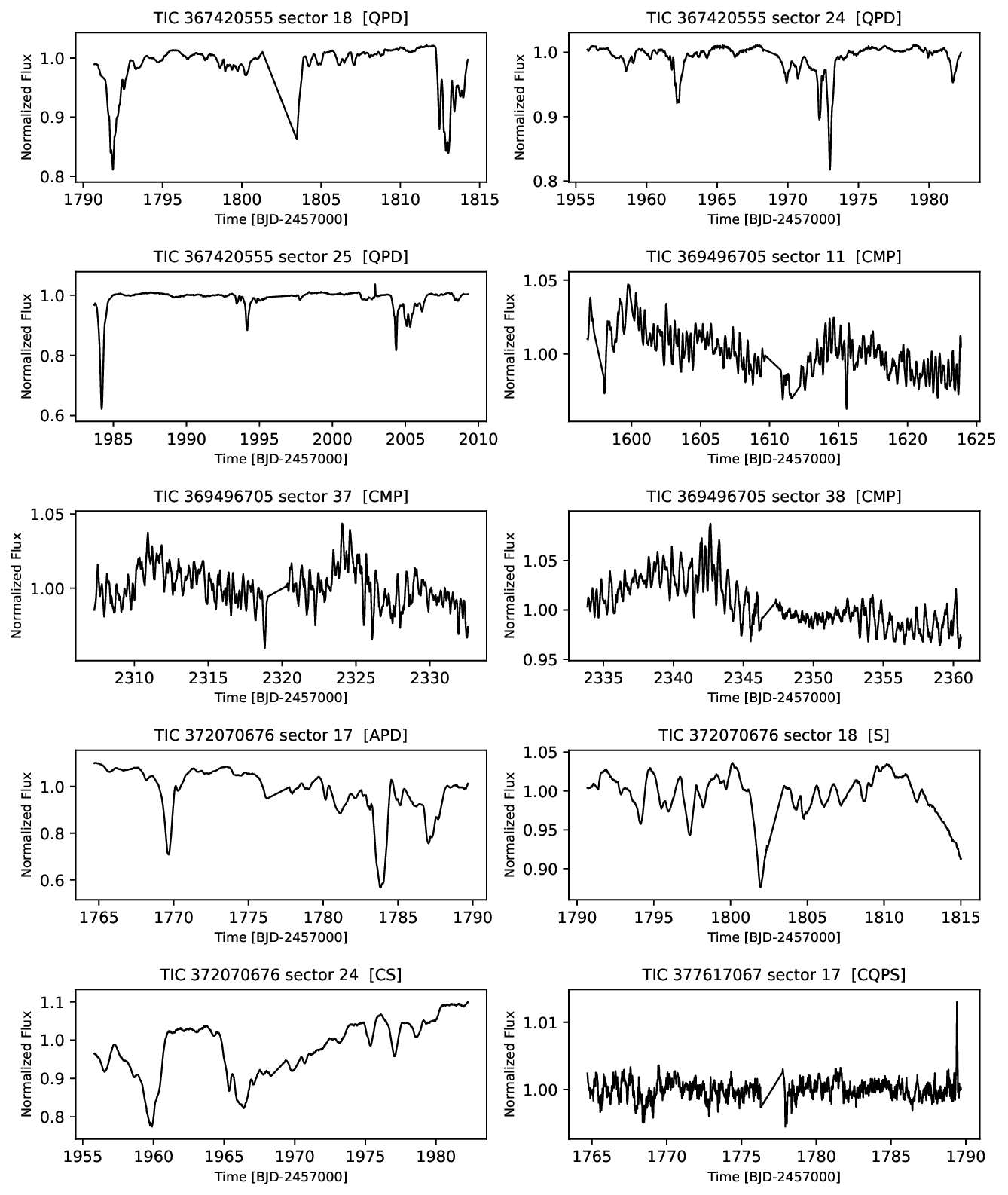}
\caption{Cont.}
\end{figure*}

\clearpage
\addtocounter{figure}{-1}
\begin{figure*}
\epsscale{0.90}
\plotone{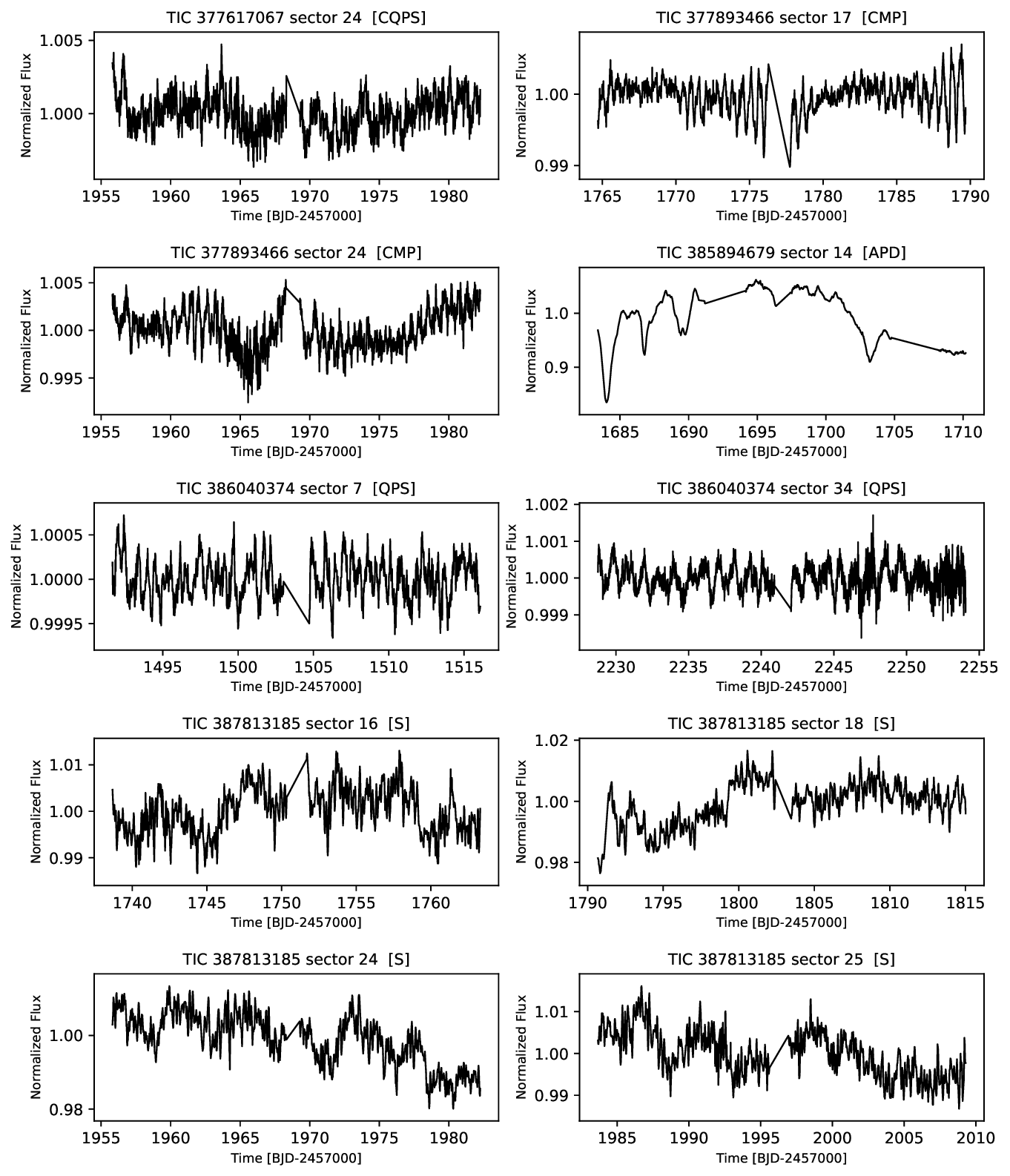}
\caption{Cont.}
\end{figure*}

\clearpage
\addtocounter{figure}{-1}
\begin{figure*}
\epsscale{0.90}
\plotone{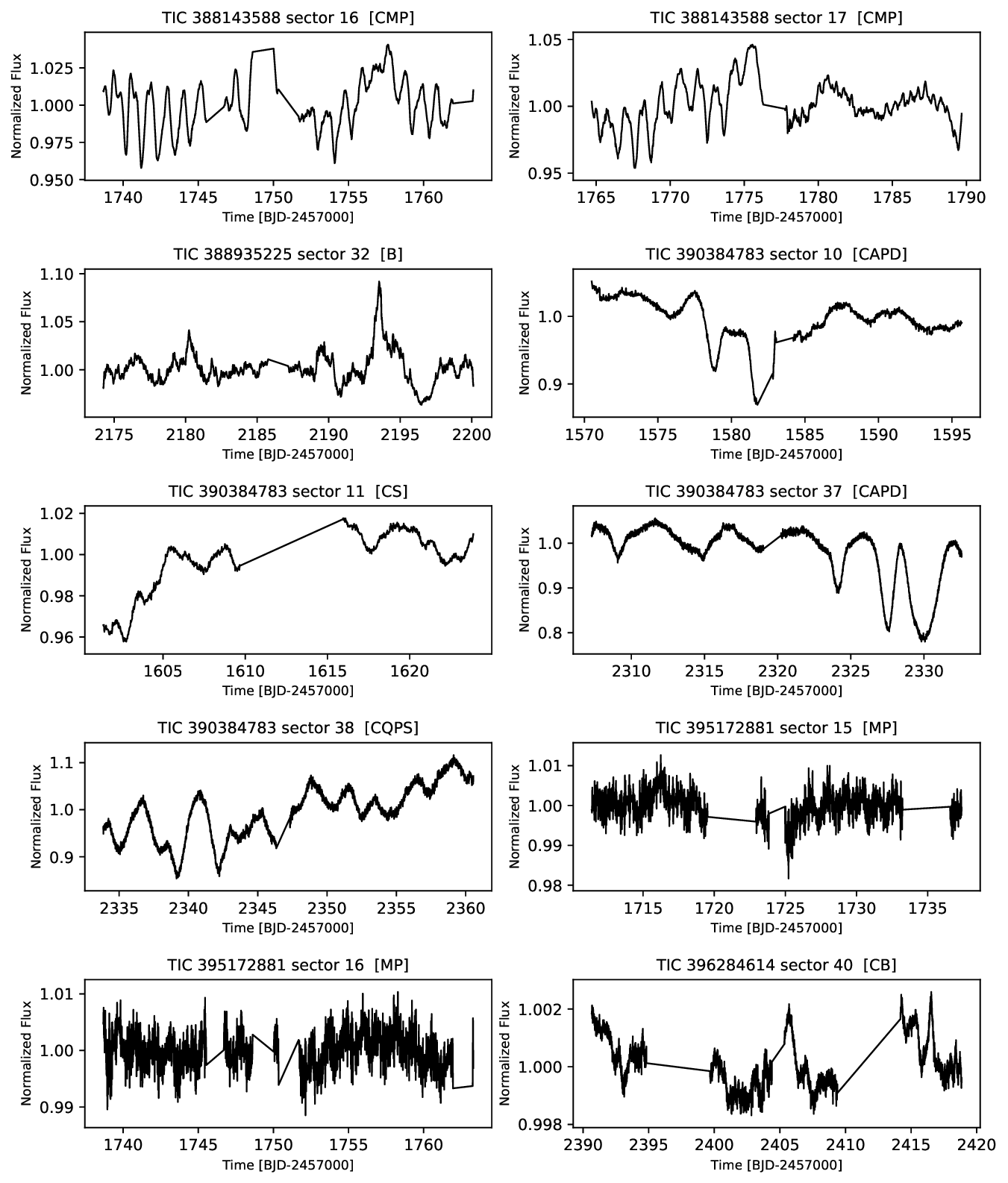}
\caption{Cont.}
\end{figure*}

\clearpage
\addtocounter{figure}{-1}
\begin{figure*}
\epsscale{0.90}
\plotone{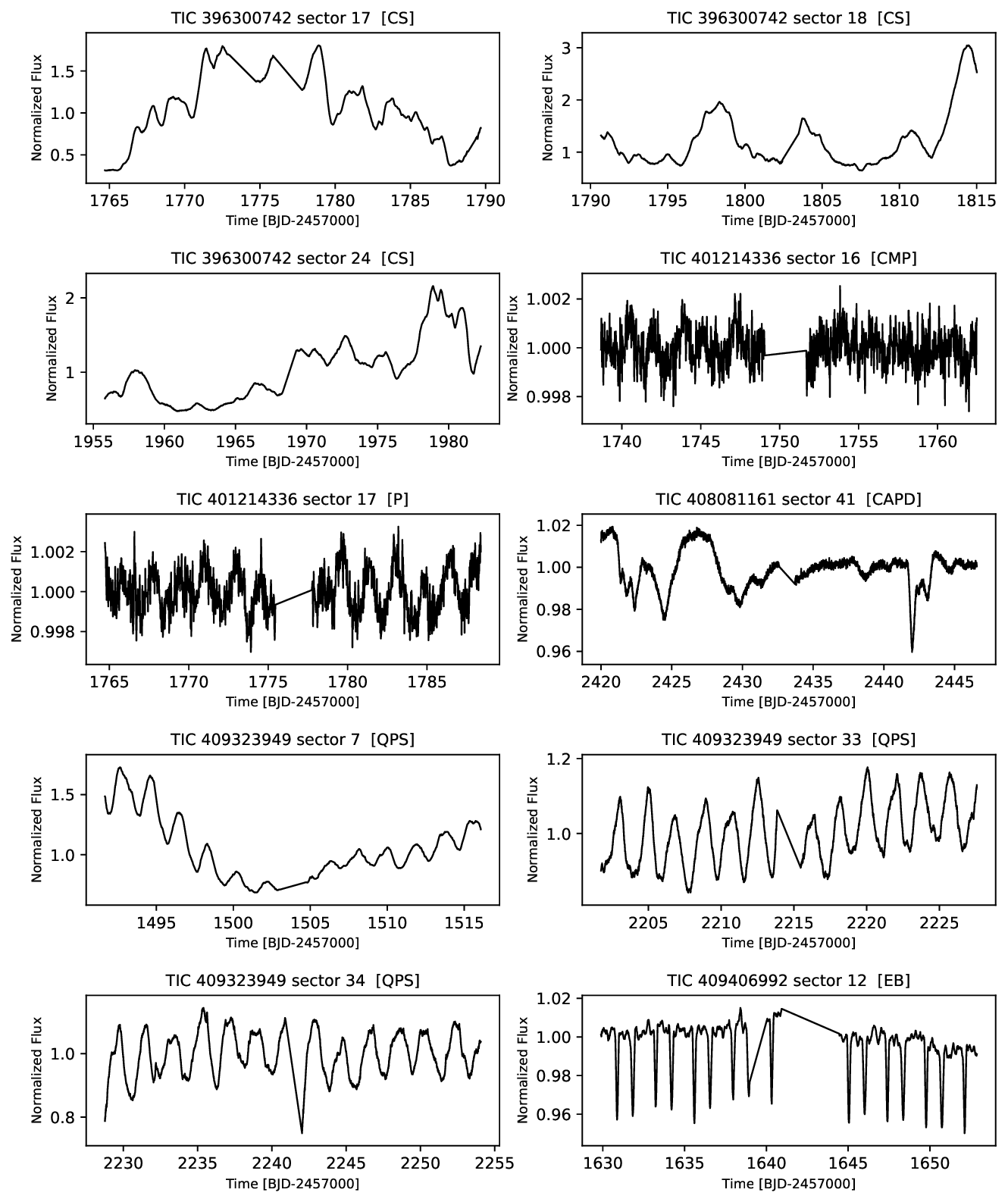}
\caption{Cont.}
\end{figure*}

\clearpage
\addtocounter{figure}{-1}
\begin{figure*}
\epsscale{0.90}
\plotone{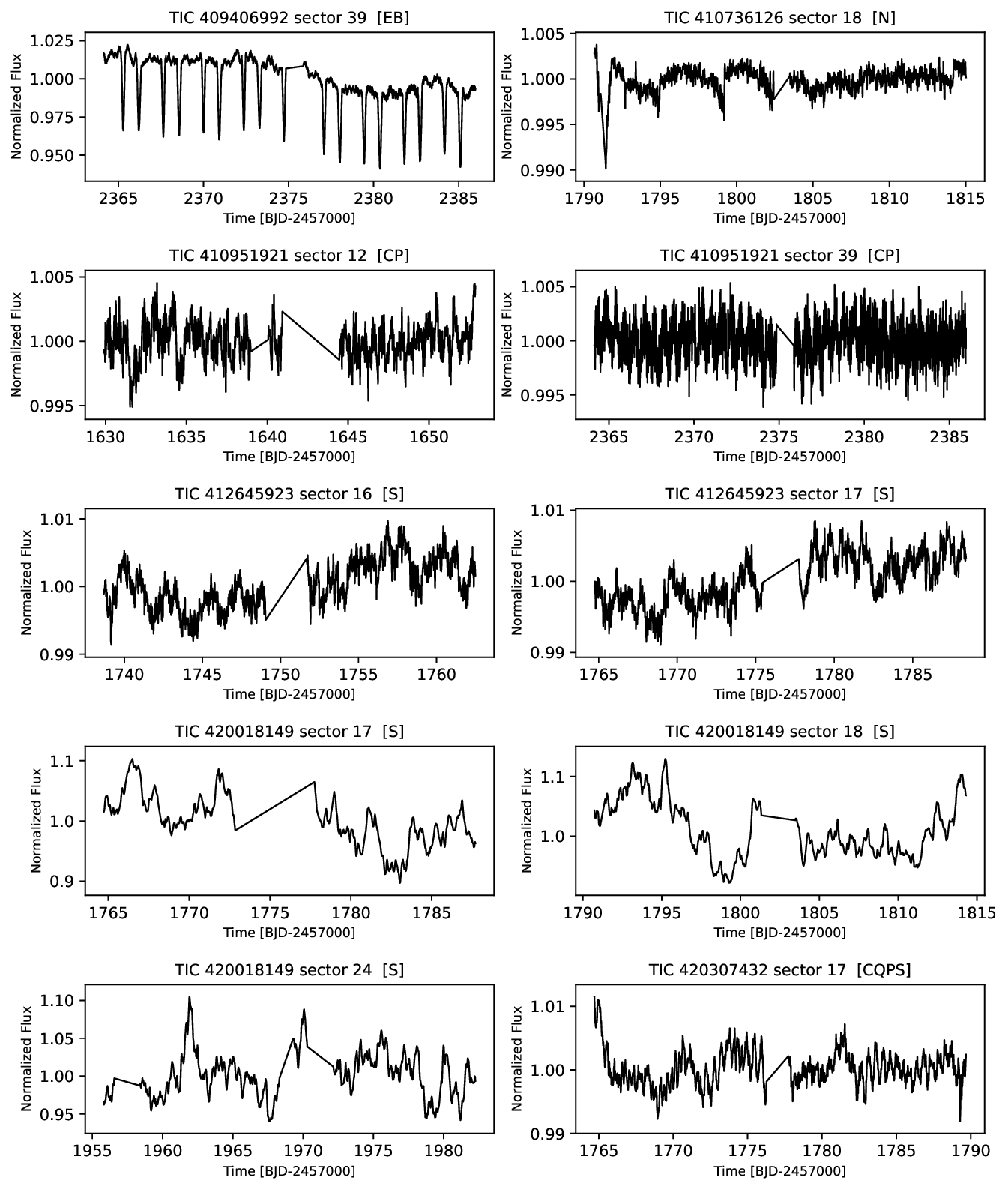}
\caption{Cont.}
\end{figure*}

\clearpage
\addtocounter{figure}{-1}
\begin{figure*}
\epsscale{0.90}
\plotone{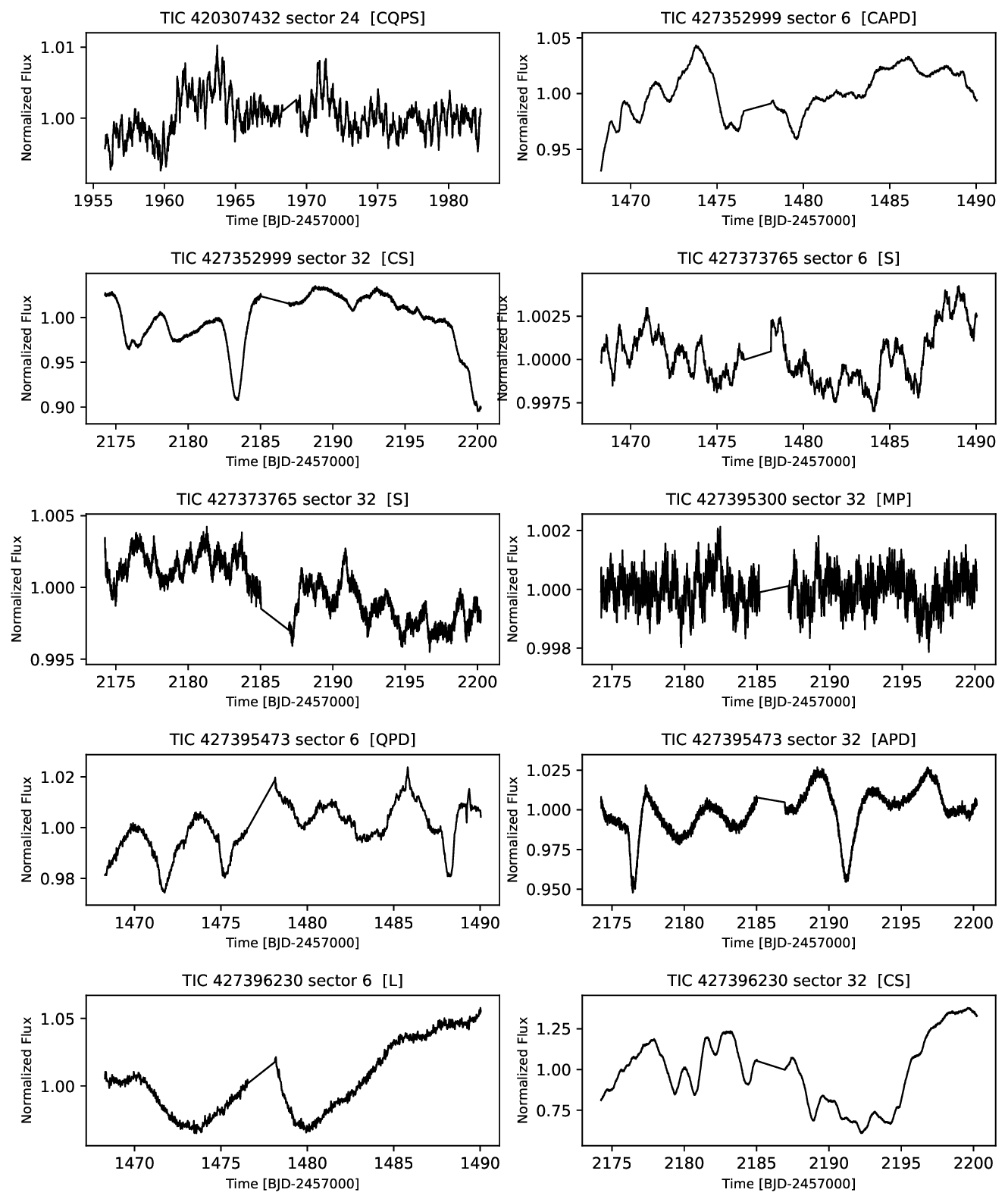}
\caption{Cont.}
\end{figure*}

\clearpage
\addtocounter{figure}{-1}
\begin{figure*}
\epsscale{0.90}
\plotone{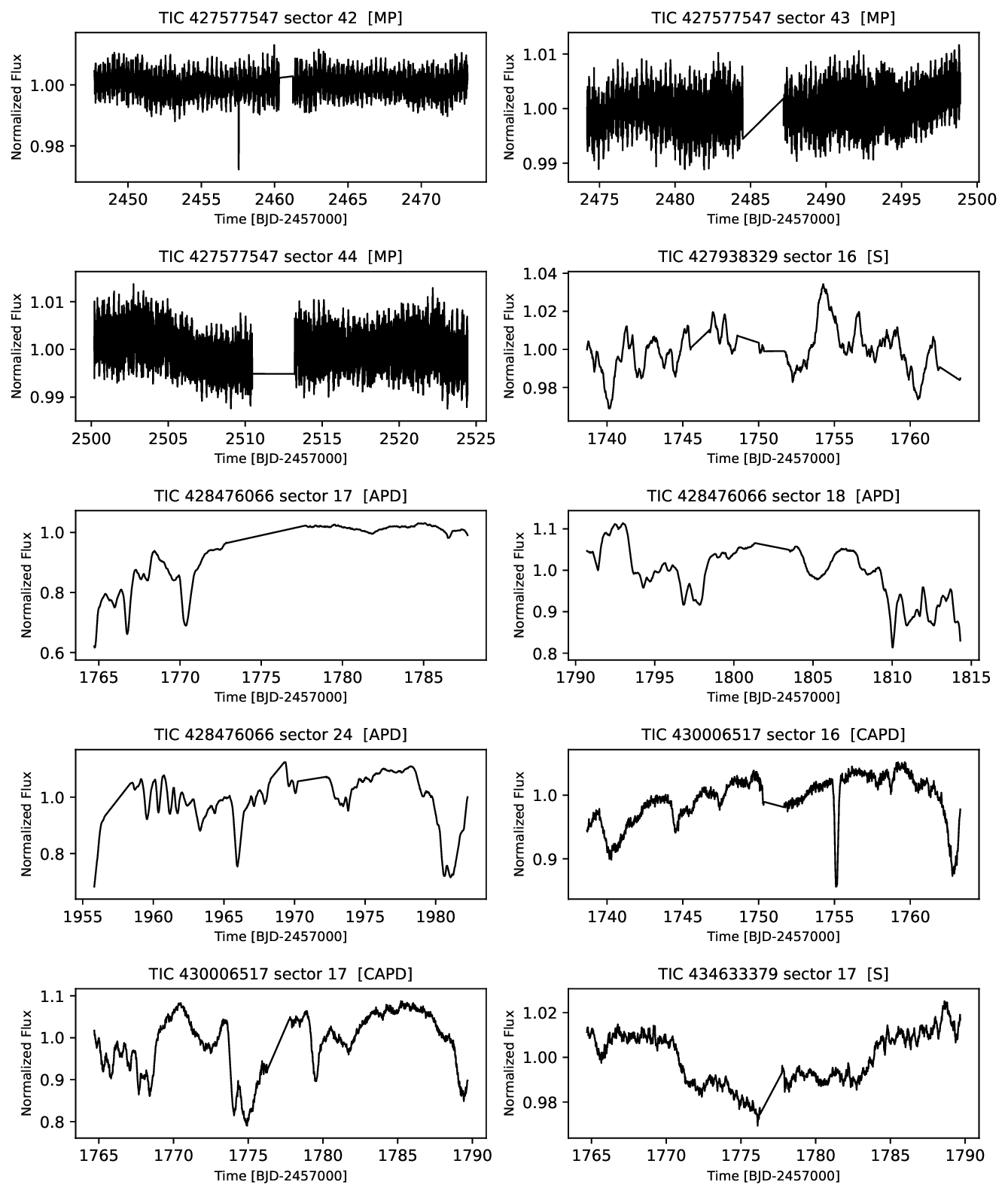}
\caption{Cont.}
\end{figure*}

\clearpage
\addtocounter{figure}{-1}
\begin{figure*}
\epsscale{0.90}
\plotone{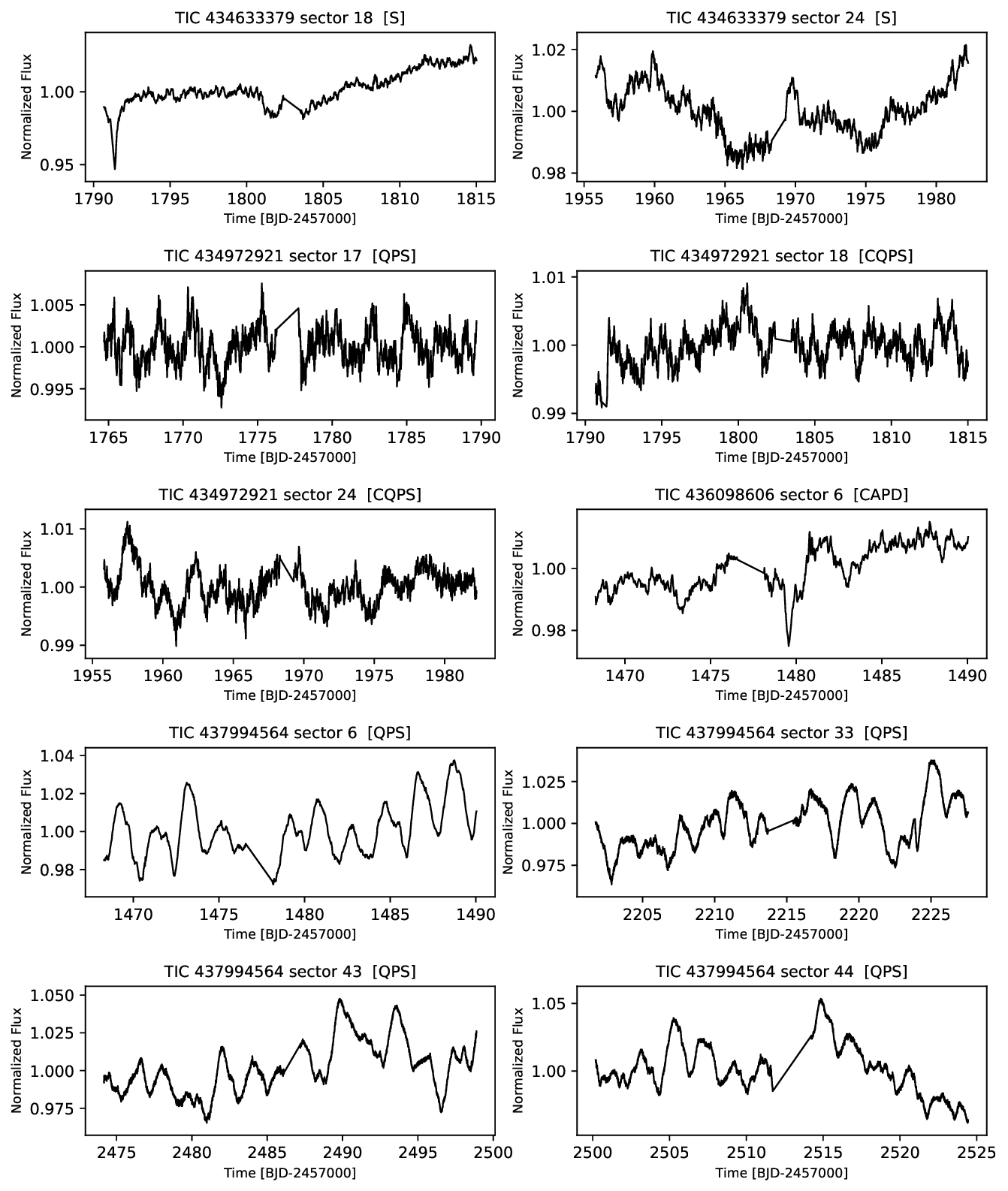}
\caption{Cont.}
\end{figure*}

\clearpage
\addtocounter{figure}{-1}
\begin{figure*}
\epsscale{0.90}
\plotone{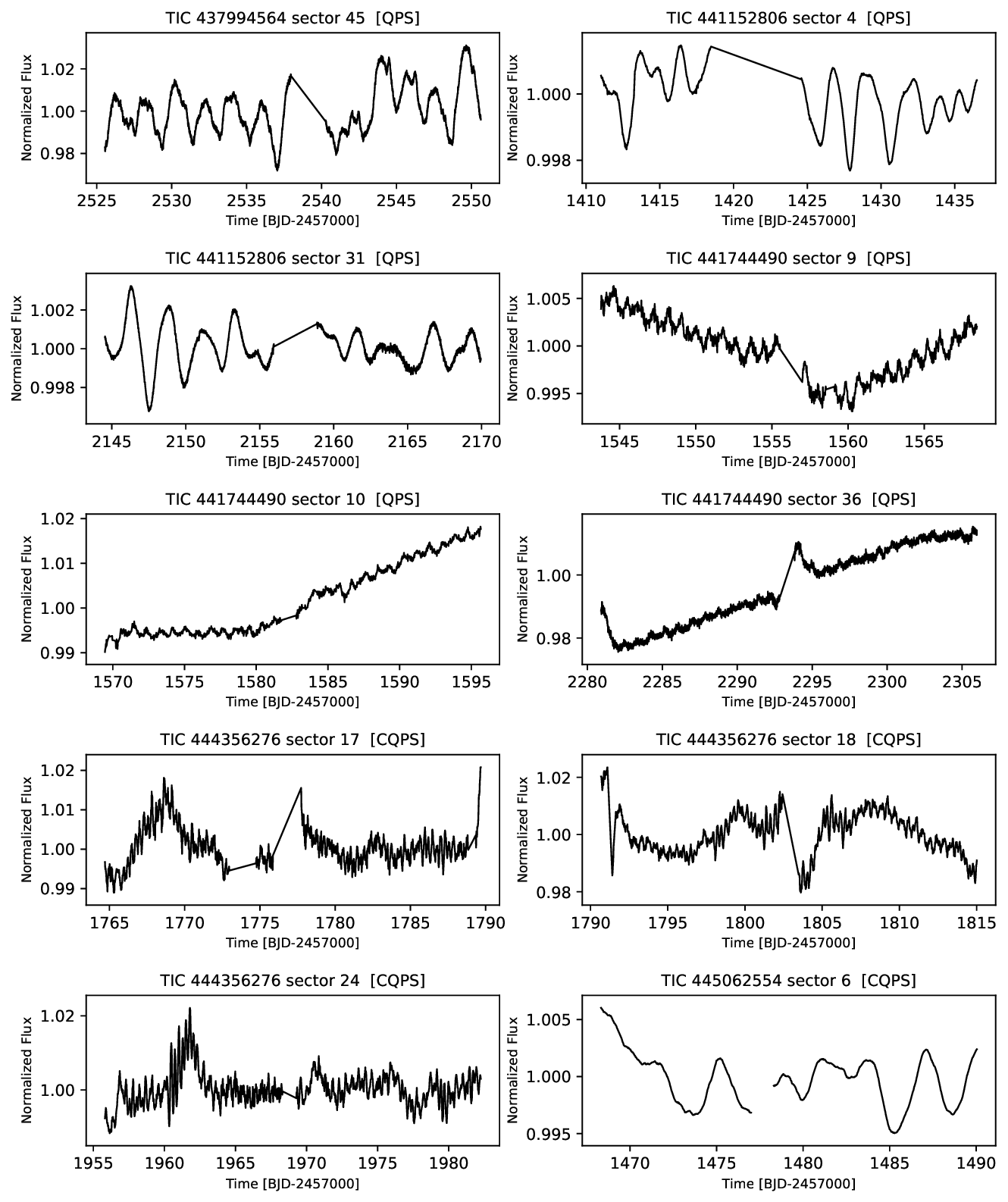}
\caption{Cont.}
\end{figure*}

\clearpage
\addtocounter{figure}{-1}
\begin{figure*}
\epsscale{0.90}
\plotone{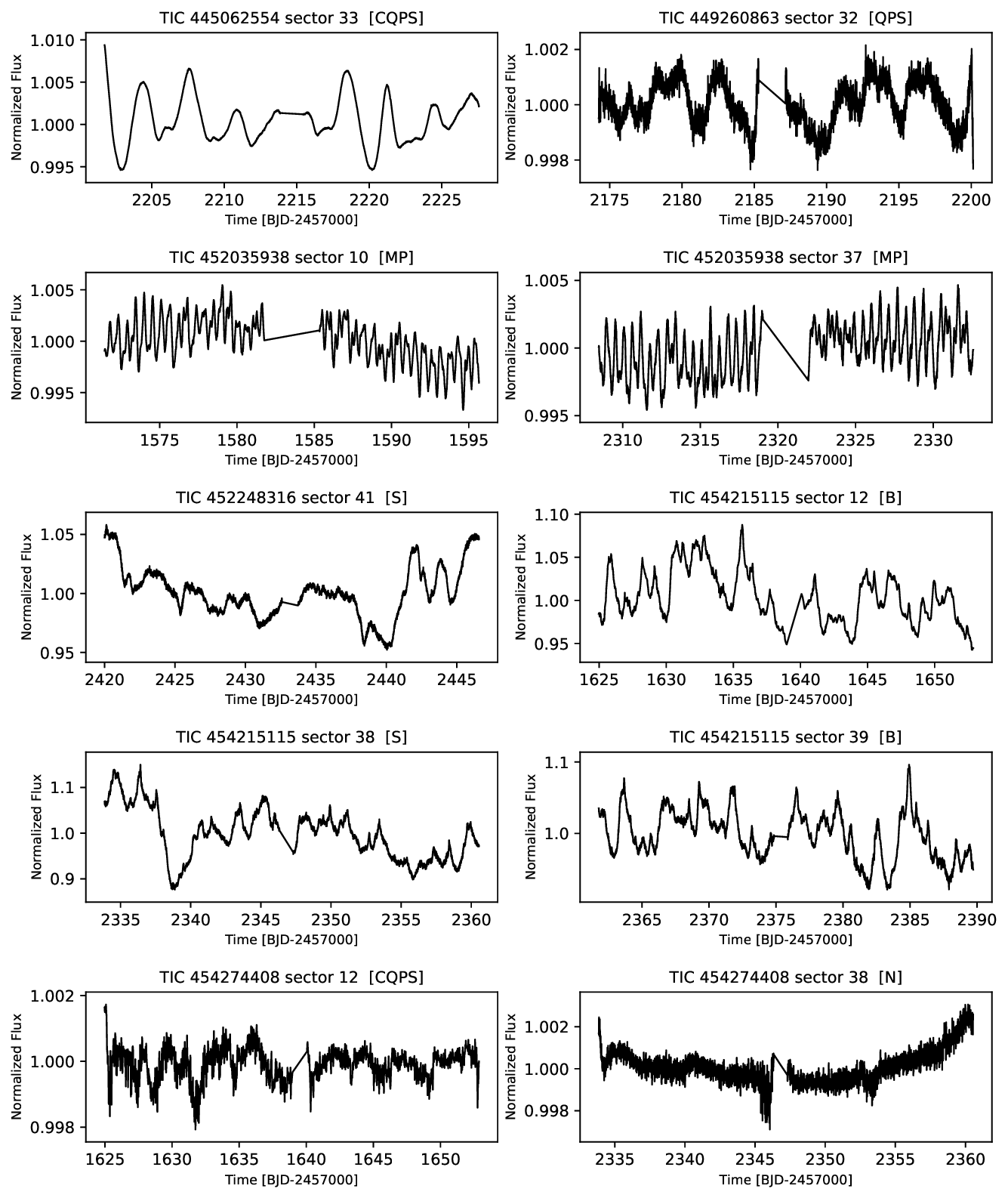}
\caption{Cont.}
\end{figure*}

\clearpage
\addtocounter{figure}{-1}
\begin{figure*}
\epsscale{0.90}
\plotone{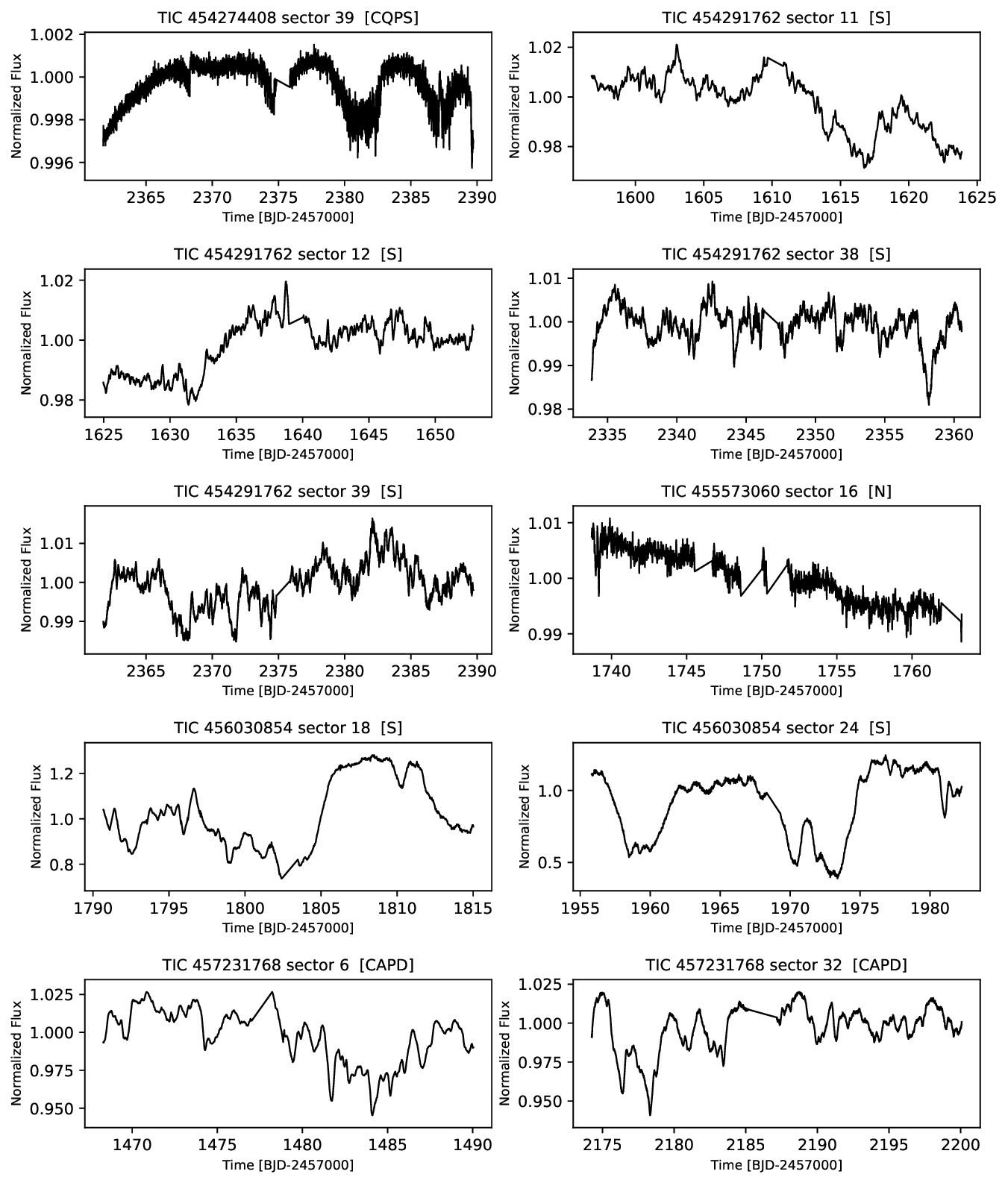}
\caption{Cont.}
\end{figure*}

\clearpage
\addtocounter{figure}{-1}
\begin{figure*}
\epsscale{0.90}
\plotone{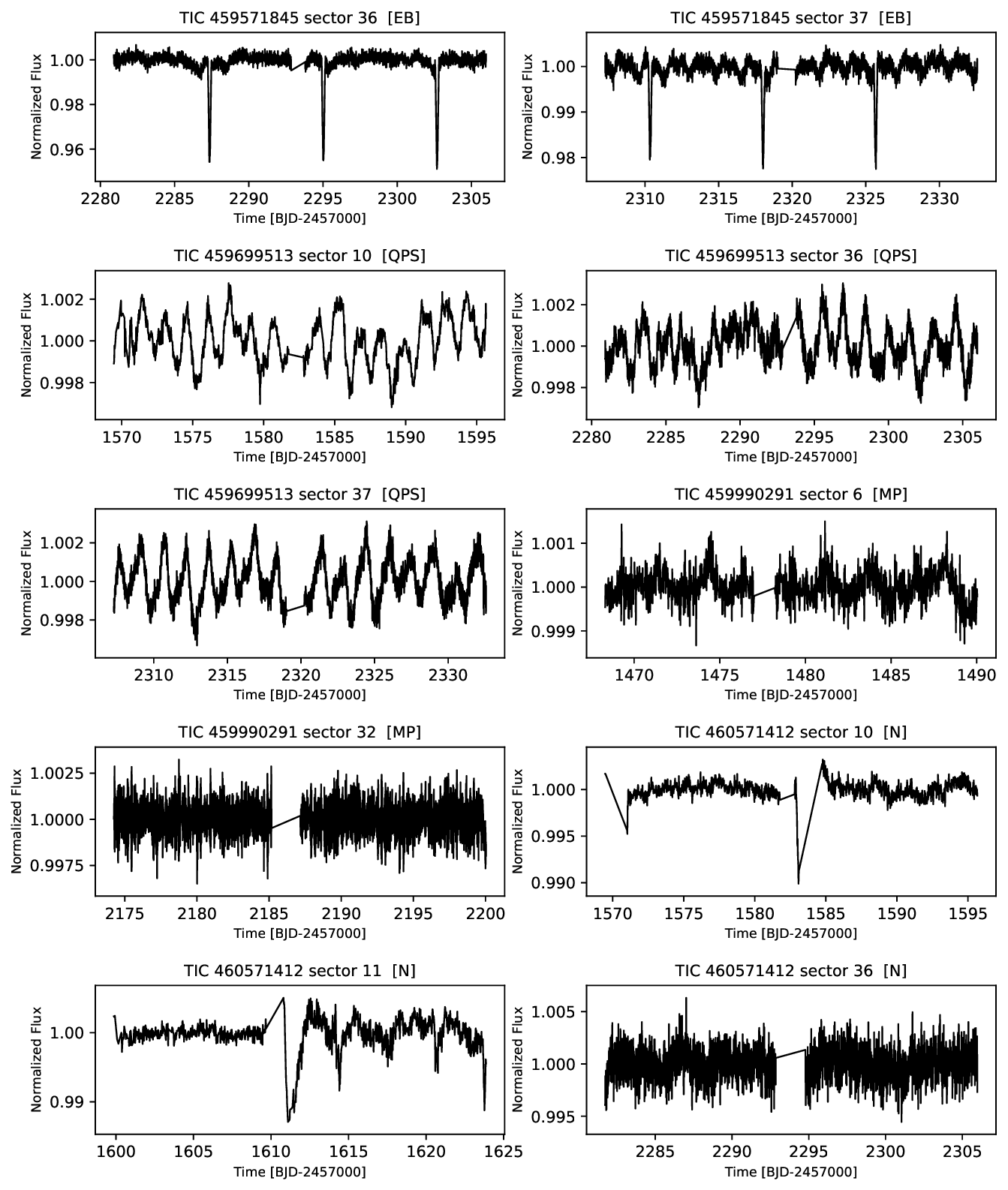}
\caption{Cont.}
\end{figure*}

\clearpage
\addtocounter{figure}{-1}
\begin{figure*}
\epsscale{0.90}
\plotone{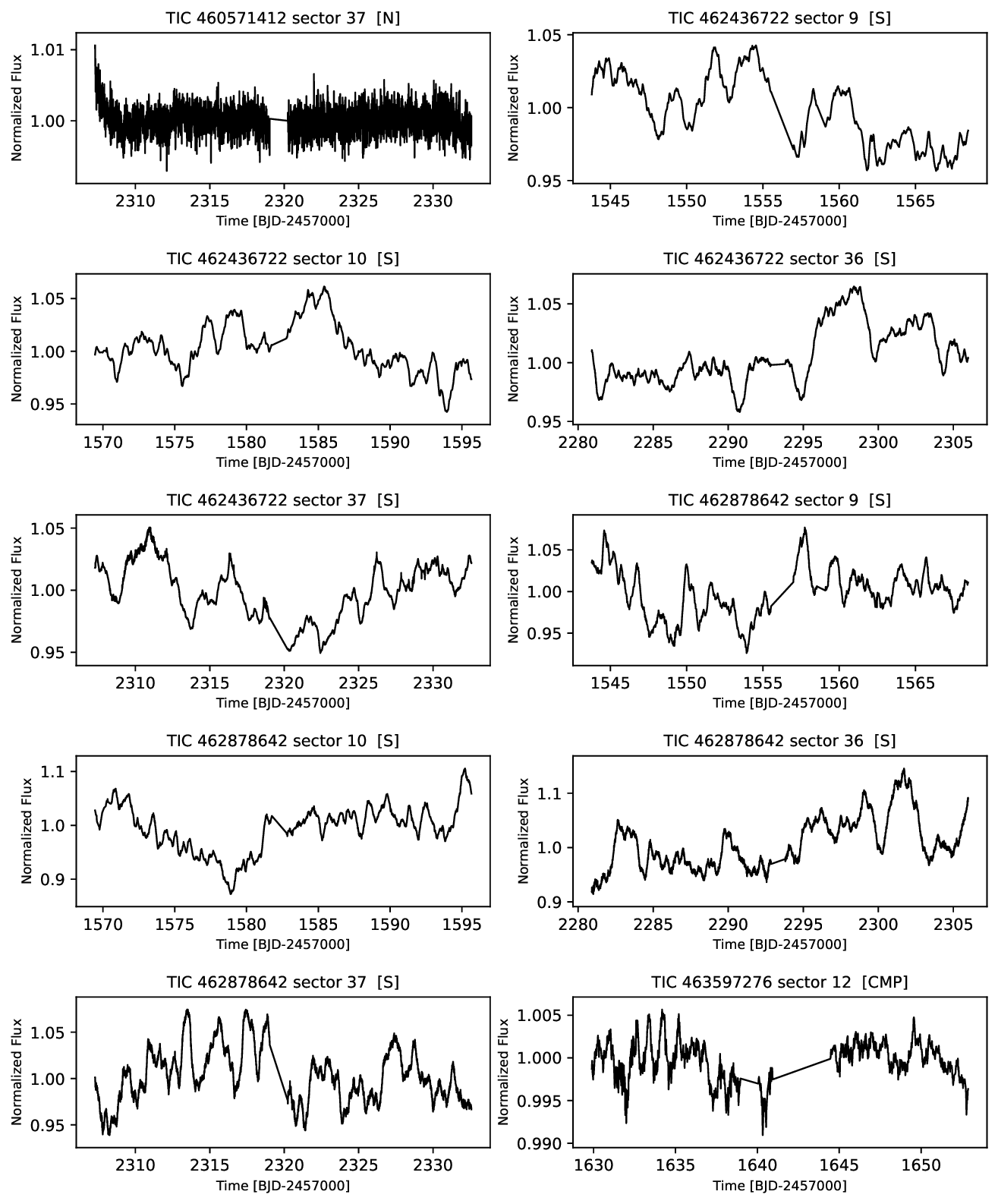}
\caption{Cont.}
\end{figure*}

\clearpage
\addtocounter{figure}{-1}
\begin{figure*}
\epsscale{0.90}
\plotone{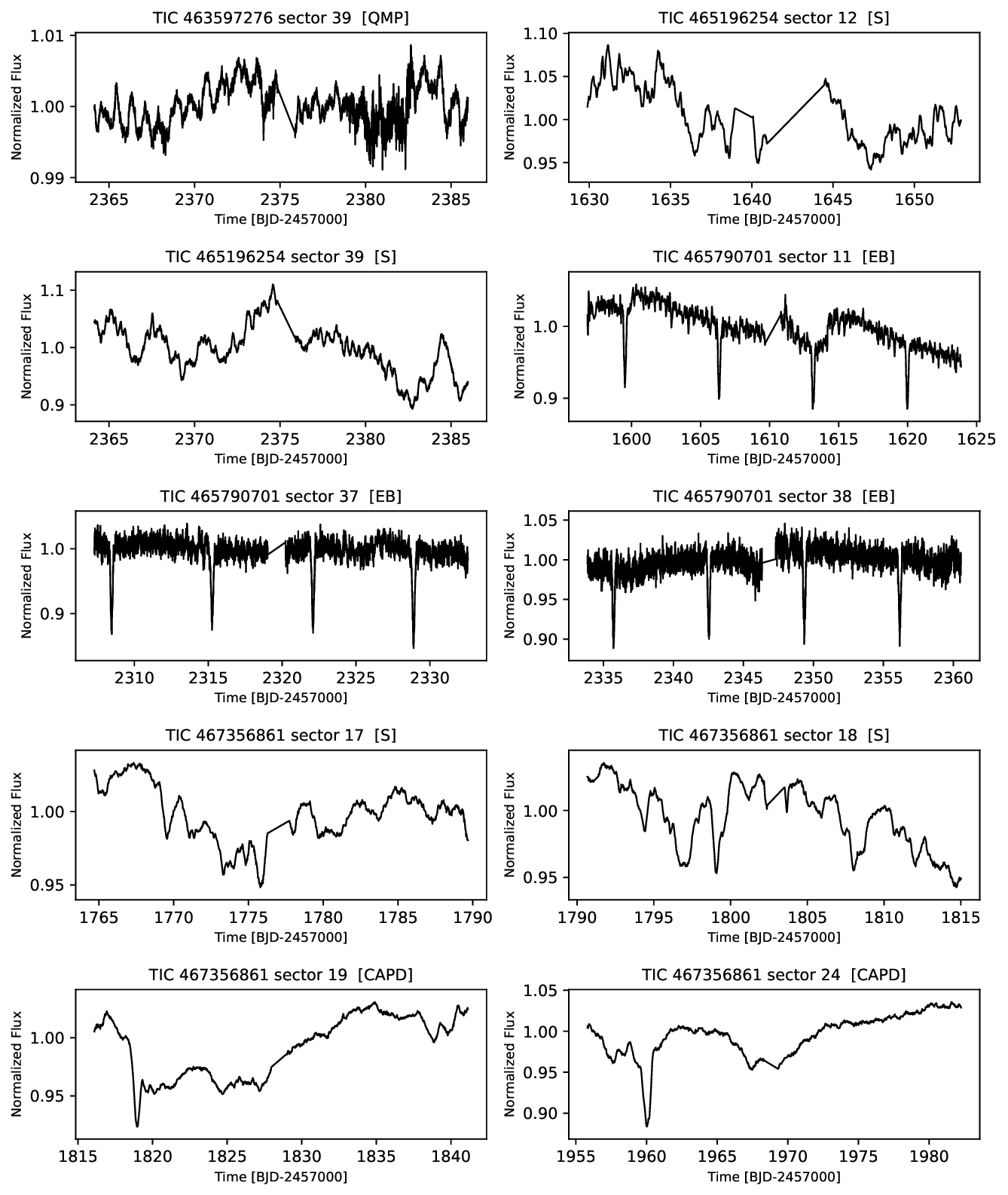}
\caption{Cont.}
\end{figure*}

\clearpage
\addtocounter{figure}{-1}
\begin{figure*}
\epsscale{0.90}
\plotone{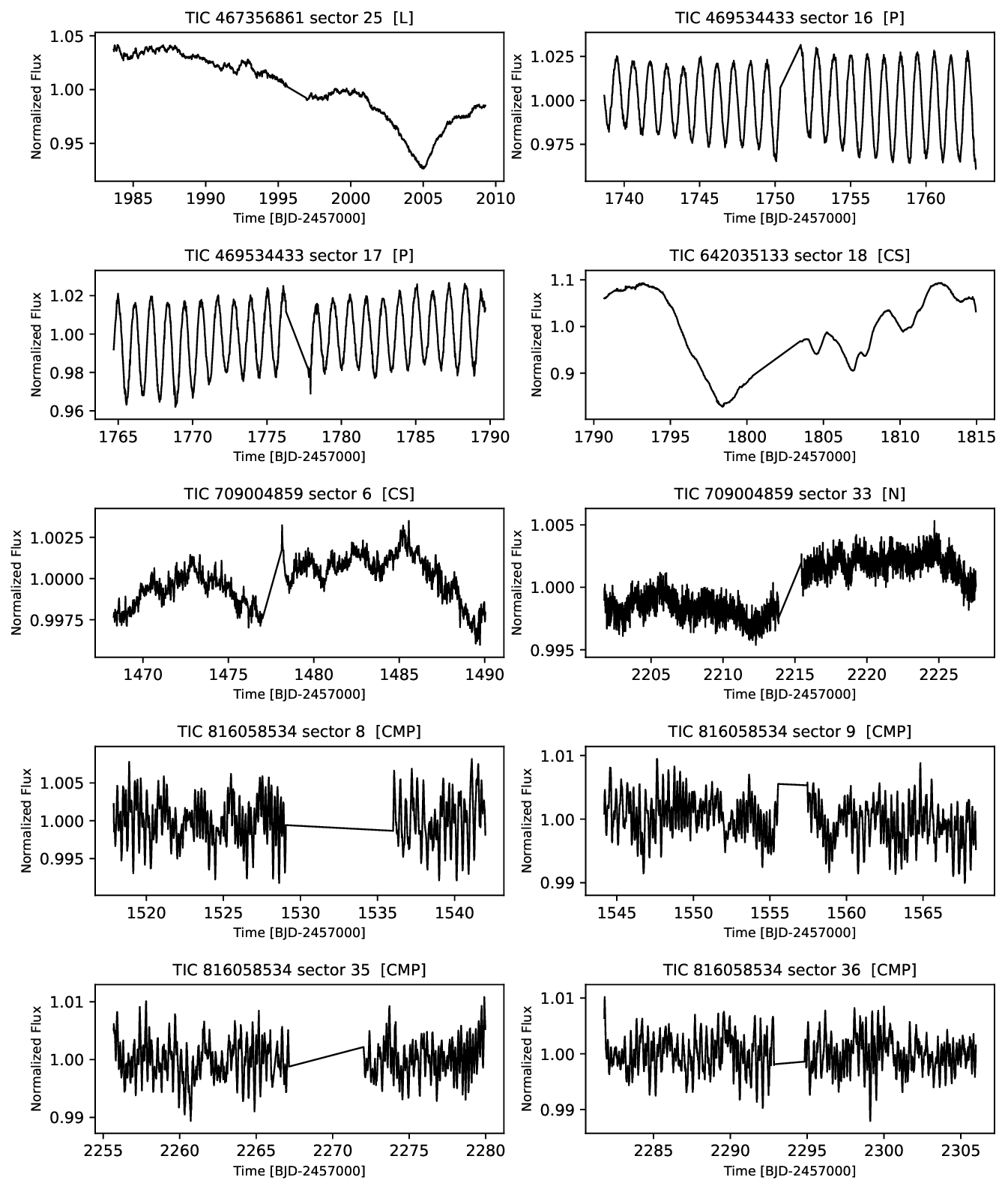}
\caption{Cont.}
\end{figure*}

\clearpage
\addtocounter{figure}{-1}
\begin{figure*}
\epsscale{0.90}
\plotone{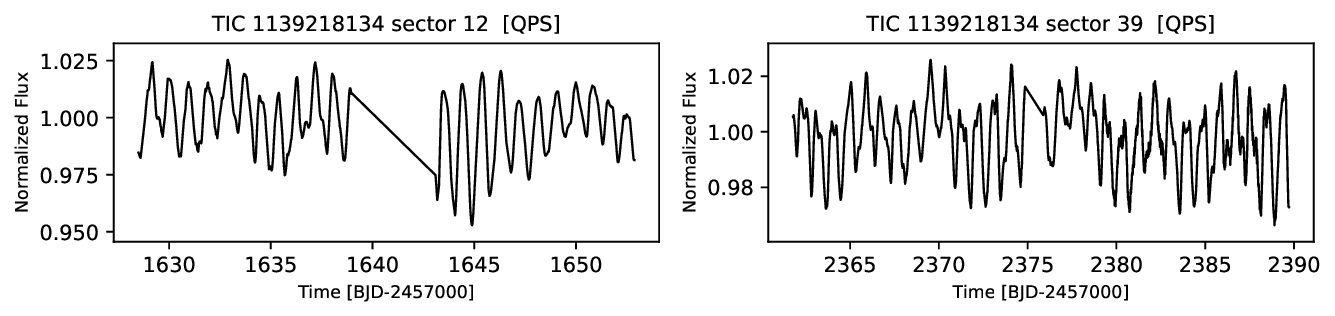}
\caption{Cont.}
\end{figure*}


\bibliography{references}{}

\begin{thebibliography}{}
\expandafter\ifx\csname natexlab\endcsname\relax\def\natexlab#1{#1}\fi
\providecommand{\url}[1]{\href{#1}{#1}}
\providecommand{\dodoi}[1]{doi:~\href{http://doi.org/#1}{\nolinkurl{#1}}}
\providecommand{\doeprint}[1]{\href{http://ascl.net/#1}{\nolinkurl{http://ascl.net/#1}}}
\providecommand{\doarXiv}[1]{\href{https://arxiv.org/abs/#1}{\nolinkurl{https://arxiv.org/abs/#1}}}

\bibitem[{{Alecian} {et~al.}(2013){Alecian}, {Wade}, {Catala}, {Grunhut}, {Landstreet}, {B{\"o}hm}, {Folsom}, \& {Marsden}}]{2013MNRAS.429.1027A}
{Alecian}, E., {Wade}, G.~A., {Catala}, C., {et~al.} 2013, \mnras, 429, 1027, \dodoi{10.1093/mnras/sts384}

\bibitem[{{Arun} {et~al.}(2019){Arun}, {Mathew}, {Manoj}, {Ujjwal}, {Kartha}, {Viswanath}, {Narang}, \& {Paul}}]{2019AJ....157..159A}
{Arun}, R., {Mathew}, B., {Manoj}, P., {et~al.} 2019, \aj, 157, 159, \dodoi{10.3847/1538-3881/ab0ca1}

\bibitem[{{Barsunova} {et~al.}(2013){Barsunova}, {Mel'nikov}, {Grinin}, {Katysheva}, \& {Shugarov}}]{2013ARep...57...89B}
{Barsunova}, O.~Y., {Mel'nikov}, S.~Y., {Grinin}, V.~P., {Katysheva}, N.~A., \& {Shugarov}, S.~Y. 2013, Astronomy Reports, 57, 89, \dodoi{10.1134/S1063772913020029}

\bibitem[{{Benisty} {et~al.}(2013){Benisty}, {Perraut}, {Mourard}, {Stee}, {Lima}, {Le Bouquin}, {Borges Fernandes}, {Chesneau}, {Nardetto}, {Tallon-Bosc}, {McAlister}, {Ten Brummelaar}, {Ridgway}, {Sturmann}, {Sturmann}, {Turner}, {Farrington}, \& {Goldfinger}}]{2013A&A...555A.113B}
{Benisty}, M., {Perraut}, K., {Mourard}, D., {et~al.} 2013, \aap, 555, A113, \dodoi{10.1051/0004-6361/201219893}

\bibitem[{{Bibo} \& {The}(1991)}]{1991A&AS...89..319B}
{Bibo}, E.~A., \& {The}, P.~S. 1991, \aaps, 89, 319

\bibitem[{{Bodman} {et~al.}(2017){Bodman}, {Quillen}, {Ansdell}, {Hippke}, {Boyajian}, {Mamajek}, {Blackman}, {Rizzuto}, \& {Kastner}}]{2017MNRAS.470..202B}
{Bodman}, E. H.~L., {Quillen}, A.~C., {Ansdell}, M., {et~al.} 2017, \mnras, 470, 202, \dodoi{10.1093/mnras/stx1034}

\bibitem[{{Boehm} \& {Catala}(1995)}]{1995AA...301..155B}
{Boehm}, T., \& {Catala}, C. 1995, \aap, 301, 155

\bibitem[{{Bouvier} {et~al.}(1993){Bouvier}, {Cabrit}, {Fernandez}, {Martin}, \& {Matthews}}]{1993A&A...272..176B}
{Bouvier}, J., {Cabrit}, S., {Fernandez}, M., {Martin}, E.~L., \& {Matthews}, J.~M. 1993, \aap, 272, 176

\bibitem[{{Bouvier} {et~al.}(2003){Bouvier}, {Grankin}, {Alencar}, {Dougados}, {Fern{\'a}ndez}, {Basri}, {Batalha}, {Guenther}, {Ibrahimov}, {Magakian}, {Melnikov}, {Petrov}, {Rud}, \& {Zapatero Osorio}}]{2003A&A...409..169B}
{Bouvier}, J., {Grankin}, K.~N., {Alencar}, S.~H.~P., {et~al.} 2003, \aap, 409, 169, \dodoi{10.1051/0004-6361:20030938}

\bibitem[{{Brasseur} {et~al.}(2019){Brasseur}, {Phillip}, {Hargis}, {Mullally}, {Fleming}, {Fox}, \& {Smith}}]{2019ASPC..523..397B}
{Brasseur}, C.~E., {Phillip}, C., {Hargis}, J., {et~al.} 2019, in Astronomical Society of the Pacific Conference Series, Vol. 523, Astronomical Data Analysis Software and Systems XXVII, ed. P.~J. {Teuben}, M.~W. {Pound}, B.~A. {Thomas}, \& E.~M. {Warner}, 397

\bibitem[{{Brittain} {et~al.}(2023){Brittain}, {Kamp}, {Meeus}, {Oudmaijer}, \& {Waters}}]{2023SSRv..219....7B}
{Brittain}, S.~D., {Kamp}, I., {Meeus}, G., {Oudmaijer}, R.~D., \& {Waters}, L.~B.~F.~M. 2023, \ssr, 219, 7, \dodoi{10.1007/s11214-023-00949-z}

\bibitem[{{Caldwell} {et~al.}(2020){Caldwell}, {Tenenbaum}, {Twicken}, {Jenkins}, {Ting}, {Smith}, {Hedges}, {Fausnaugh}, {Rose}, \& {Burke}}]{2020RNAAS...4..201C}
{Caldwell}, D.~A., {Tenenbaum}, P., {Twicken}, J.~D., {et~al.} 2020, Research Notes of the American Astronomical Society, 4, 201, \dodoi{10.3847/2515-5172/abc9b3}

\bibitem[{{Calvet} \& {Hartmann}(1992)}]{1992ApJ...386..239C}
{Calvet}, N., \& {Hartmann}, L. 1992, \apj, 386, 239, \dodoi{10.1086/171010}

\bibitem[{{Cardelli} {et~al.}(1989){Cardelli}, {Clayton}, \& {Mathis}}]{1989ApJ...345..245C}
{Cardelli}, J.~A., {Clayton}, G.~C., \& {Mathis}, J.~S. 1989, \apj, 345, 245, \dodoi{10.1086/167900}

\bibitem[{{Carmona} {et~al.}(2010){Carmona}, {van den Ancker}, {Audard}, {Henning}, {Setiawan}, \& {Rodmann}}]{2010A&A...517A..67C}
{Carmona}, A., {van den Ancker}, M.~E., {Audard}, M., {et~al.} 2010, \aap, 517, A67, \dodoi{10.1051/0004-6361/200913800}

\bibitem[{{Cauley} \& {Johns-Krull}(2015)}]{2015ApJ...810....5C}
{Cauley}, P.~W., \& {Johns-Krull}, C.~M. 2015, \apj, 810, 5, \dodoi{10.1088/0004-637X/810/1/5}

\bibitem[{{Cody} \& {Hillenbrand}(2018)}]{2018AJ....156...71C}
{Cody}, A.~M., \& {Hillenbrand}, L.~A. 2018, \aj, 156, 71, \dodoi{10.3847/1538-3881/aacead}

\bibitem[{{Cody} {et~al.}(2022){Cody}, {Hillenbrand}, \& {Rebull}}]{2022AJ....163..212C}
{Cody}, A.~M., {Hillenbrand}, L.~A., \& {Rebull}, L.~M. 2022, \aj, 163, 212, \dodoi{10.3847/1538-3881/ac5b73}

\bibitem[{{Cody} {et~al.}(2014){Cody}, {Stauffer}, {Baglin}, {Micela}, {Rebull}, {Flaccomio}, {Morales-Calder{\'o}n}, {Aigrain}, {Bouvier}, {Hillenbrand}, {Gutermuth}, {Song}, {Turner}, {Alencar}, {Zwintz}, {Plavchan}, {Carpenter}, {Findeisen}, {Carey}, {Terebey}, {Hartmann}, {Calvet}, {Teixeira}, {Vrba}, {Wolk}, {Covey}, {Poppenhaeger}, {G{\"u}nther}, {Forbrich}, {Whitney}, {Affer}, {Herbst}, {Hora}, {Barrado}, {Holtzman}, {Marchis}, {Wood}, {Medeiros Guimar{\~a}es}, {Lillo Box}, {Gillen}, {McQuillan}, {Espaillat}, {Allen}, {D'Alessio}, \& {Favata}}]{2014AJ....147...82C}
{Cody}, A.~M., {Stauffer}, J., {Baglin}, A., {et~al.} 2014, \aj, 147, 82, \dodoi{10.1088/0004-6256/147/4/82}

\bibitem[{{Costa} {et~al.}(2025){Costa}, {Shepherd}, {Bressan}, {Addari}, {Chen}, {Fu}, {Volpato}, {Nguyen}, {Girardi}, {Marigo}, {Mazzi}, {Pastorelli}, {Trabucchi}, {Bossini}, \& {Zaggia}}]{2025A&A...694A.193C}
{Costa}, G., {Shepherd}, K.~G., {Bressan}, A., {et~al.} 2025, \aap, 694, A193, \dodoi{10.1051/0004-6361/202452573}

\bibitem[{{Donati} \& {Landstreet}(2009)}]{2009ARA&A..47..333D}
{Donati}, J.~F., \& {Landstreet}, J.~D. 2009, \araa, 47, 333, \dodoi{10.1146/annurev-astro-082708-101833}

\bibitem[{{Donehew} \& {Brittain}(2011)}]{2011AJ....141...46D}
{Donehew}, B., \& {Brittain}, S. 2011, \aj, 141, 46, \dodoi{10.1088/0004-6256/141/2/46}

\bibitem[{{Dullemond} {et~al.}(2003){Dullemond}, {van den Ancker}, {Acke}, \& {van Boekel}}]{2003ApJ...594L..47D}
{Dullemond}, C.~P., {van den Ancker}, M.~E., {Acke}, B., \& {van Boekel}, R. 2003, \apjl, 594, L47, \dodoi{10.1086/378400}

\bibitem[{{Efimova} {et~al.}(2022){Efimova}, {Arkharov}, {Grinin}, {Rostopchina-Shakhovskaya}, {Shakhovskoi}, {Larionov}, {Klimanov}, \& {Gorshanov}}]{2022ARep...66..236E}
{Efimova}, N.~V., {Arkharov}, A.~A., {Grinin}, V.~P., {et~al.} 2022, Astronomy Reports, 66, 236, \dodoi{10.1134/S1063772922030027}

\bibitem[{{Eiroa} {et~al.}(2002){Eiroa}, {Oudmaijer}, {Davies}, {de Winter}, {Garz{\'o}n}, {Palacios}, {Alberdi}, {Ferlet}, {Grady}, {Collier Cameron}, {Deeg}, {Harris}, {Horne}, {Mer{\'\i}n}, {Miranda}, {Montesinos}, {Mora}, {Penny}, {Quirrenbach}, {Rauer}, {Schneider}, {Solano}, {Tsapras}, \& {Wesselius}}]{2002A&A...384.1038E}
{Eiroa}, C., {Oudmaijer}, R.~D., {Davies}, J.~K., {et~al.} 2002, \aap, 384, 1038, \dodoi{10.1051/0004-6361:20020096}

\bibitem[{{Fairlamb} {et~al.}(2015){Fairlamb}, {Oudmaijer}, {Mendigut{\'\i}a}, {Ilee}, \& {van den Ancker}}]{2015MNRAS.453..976F}
{Fairlamb}, J.~R., {Oudmaijer}, R.~D., {Mendigut{\'\i}a}, I., {Ilee}, J.~D., \& {van den Ancker}, M.~E. 2015, \mnras, 453, 976, \dodoi{10.1093/mnras/stv1576}

\bibitem[{{Feinstein} {et~al.}(2019){Feinstein}, {Montet}, {Foreman-Mackey}, {Bedell}, {Saunders}, {Bean}, {Christiansen}, {Hedges}, {Luger}, {Scolnic}, \& {Cardoso}}]{2019PASP..131i4502F}
{Feinstein}, A.~D., {Montet}, B.~T., {Foreman-Mackey}, D., {et~al.} 2019, \pasp, 131, 094502, \dodoi{10.1088/1538-3873/ab291c}

\bibitem[{{Ferrario} {et~al.}(2009){Ferrario}, {Pringle}, {Tout}, \& {Wickramasinghe}}]{2009MNRAS.400L..71F}
{Ferrario}, L., {Pringle}, J.~E., {Tout}, C.~A., \& {Wickramasinghe}, D.~T. 2009, \mnras, 400, L71, \dodoi{10.1111/j.1745-3933.2009.00765.x}

\bibitem[{{Finkenzeller} \& {Mundt}(1984)}]{1984A&AS...55..109F}
{Finkenzeller}, U., \& {Mundt}, R. 1984, \aaps, 55, 109

\bibitem[{{Folsom} {et~al.}(2012){Folsom}, {Bagnulo}, {Wade}, {Alecian}, {Landstreet}, {Marsden}, \& {Waite}}]{2012MNRAS.422.2072F}
{Folsom}, C.~P., {Bagnulo}, S., {Wade}, G.~A., {et~al.} 2012, \mnras, 422, 2072, \dodoi{10.1111/j.1365-2966.2012.20718.x}

\bibitem[{{Fossati} {et~al.}(2014){Fossati}, {Zwintz}, {Castro}, {Langer}, {Lorenz}, {Schneider}, {Kuschnig}, {Matthews}, {Alecian}, {Wade}, {Barnes}, \& {Thoul}}]{2014A&A...562A.143F}
{Fossati}, L., {Zwintz}, K., {Castro}, N., {et~al.} 2014, \aap, 562, A143, \dodoi{10.1051/0004-6361/201323214}

\bibitem[{{Gaia Collaboration} {et~al.}(2016){Gaia Collaboration}, {Prusti}, {de Bruijne}, {Brown}, {Vallenari}, {Babusiaux}, {Bailer-Jones}, {Bastian}, {Biermann}, {Evans}, {Eyer}, {Jansen}, {Jordi}, {Klioner}, {Lammers}, {Lindegren}, {Luri}, {Mignard}, {Milligan}, {Panem}, {Poinsignon}, {Pourbaix}, {Randich}, {Sarri}, {Sartoretti}, {Siddiqui}, {Soubiran}, {Valette}, {van Leeuwen}, {Walton}, {Aerts}, {Arenou}, {Cropper}, {Drimmel}, {H{\o}g}, {Katz}, {Lattanzi}, {O'Mullane}, {Grebel}, {Holland}, {Huc}, {Passot}, {Bramante}, {Cacciari}, {Casta{\~n}eda}, {Chaoul}, {Cheek}, {De Angeli}, {Fabricius}, {Guerra}, {Hern{\'a}ndez}, {Jean-Antoine-Piccolo}, {Masana}, {Messineo}, {Mowlavi}, {Nienartowicz}, {Ord{\'o}{\~n}ez-Blanco}, {Panuzzo}, {Portell}, {Richards}, {Riello}, {Seabroke}, {Tanga}, {Th{\'e}venin}, {Torra}, {Els}, {Gracia-Abril}, {Comoretto}, {Garcia-Reinaldos}, {Lock}, {Mercier}, {Altmann}, {Andrae}, {Astraatmadja}, {Bellas-Velidis}, {Benson}, {Berthier}, {Blomme}, {Busso}, {Carry}, {Cellino}, {Clementini},
  {Cowell}, {Creevey}, {Cuypers}, {Davidson}, {De Ridder}, {de Torres}, {Delchambre}, {Dell'Oro}, {Ducourant}, {Fr{\'e}mat}, {Garc{\'\i}a-Torres}, {Gosset}, {Halbwachs}, {Hambly}, {Harrison}, {Hauser}, {Hestroffer}, {Hodgkin}, {Huckle}, {Hutton}, {Jasniewicz}, {Jordan}, {Kontizas}, {Korn}, {Lanzafame}, {Manteiga}, {Moitinho}, {Muinonen}, {Osinde}, {Pancino}, {Pauwels}, {Petit}, {Recio-Blanco}, {Robin}, {Sarro}, {Siopis}, {Smith}, {Smith}, {Sozzetti}, {Thuillot}, {van Reeven}, {Viala}, {Abbas}, {Abreu Aramburu}, {Accart}, {Aguado}, {Allan}, {Allasia}, {Altavilla}, {{\'A}lvarez}, {Alves}, {Anderson}, {Andrei}, {Anglada Varela}, {Antiche}, {Antoja}, {Ant{\'o}n}, {Arcay}, {Atzei}, {Ayache}, {Bach}, {Baker}, {Balaguer-N{\'u}{\~n}ez}, {Barache}, {Barata}, {Barbier}, {Barblan}, {Baroni}, {Barrado y Navascu{\'e}s}, {Barros}, {Barstow}, {Becciani}, {Bellazzini}, {Bellei}, {Bello Garc{\'\i}a}, {Belokurov}, {Bendjoya}, {Berihuete}, {Bianchi}, {Bienaym{\'e}}, {Billebaud}, {Blagorodnova}, {Blanco-Cuaresma}, {Boch},
  {Bombrun}, {Borrachero}, {Bouquillon}, {Bourda}, {Bouy}, {Bragaglia}, {Breddels}, {Brouillet}, {Br{\"u}semeister}, {Bucciarelli}, {Budnik}, {Burgess}, {Burgon}, {Burlacu}, {Busonero}, {Buzzi}, {Caffau}, {Cambras}, {Campbell}, {Cancelliere}, {Cantat-Gaudin}, {Carlucci}, {Carrasco}, {Castellani}, {Charlot}, {Charnas}, {Charvet}, {Chassat}, {Chiavassa}, {Clotet}, {Cocozza}, {Collins}, {Collins}, {Costigan}, {Crifo}, {Cross}, {Crosta}, {Crowley}, {Dafonte}, {Damerdji}, {Dapergolas}, {David}, {David}, {De Cat}, {de Felice}, {de Laverny}, {De Luise}, {De March}, {de Martino}, {de Souza}, {Debosscher}, {del Pozo}, {Delbo}, {Delgado}, {Delgado}, {di Marco}, {Di Matteo}, {Diakite}, {Distefano}, {Dolding}, {Dos Anjos}, {Drazinos}, {Dur{\'a}n}, {Dzigan}, {Ecale}, {Edvardsson}, {Enke}, {Erdmann}, {Escolar}, {Espina}, {Evans}, {Eynard Bontemps}, {Fabre}, {Fabrizio}, {Faigler}, {Falc{\~a}o}, {Farr{\`a}s Casas}, {Faye}, {Federici}, {Fedorets}, {Fern{\'a}ndez-Hern{\'a}ndez}, {Fernique}, {Fienga}, {Figueras}, {Filippi},
  {Findeisen}, {Fonti}, {Fouesneau}, {Fraile}, {Fraser}, {Fuchs}, {Furnell}, {Gai}, {Galleti}, {Galluccio}, {Garabato}, {Garc{\'\i}a-Sedano}, {Gar{\'e}}, {Garofalo}, {Garralda}, {Gavras}, {Gerssen}, {Geyer}, {Gilmore}, {Girona}, {Giuffrida}, {Gomes}, {Gonz{\'a}lez-Marcos}, {Gonz{\'a}lez-N{\'u}{\~n}ez}, {Gonz{\'a}lez-Vidal}, {Granvik}, {Guerrier}, {Guillout}, {Guiraud}, {G{\'u}rpide}, {Guti{\'e}rrez-S{\'a}nchez}, {Guy}, {Haigron}, {Hatzidimitriou}, {Haywood}, {Heiter}, {Helmi}, {Hobbs}, {Hofmann}, {Holl}, {Holland}, {Hunt}, {Hypki}, {Icardi}, {Irwin}, {Jevardat de Fombelle}, {Jofr{\'e}}, {Jonker}, {Jorissen}, {Julbe}, {Karampelas}, {Kochoska}, {Kohley}, {Kolenberg}, {Kontizas}, {Koposov}, {Kordopatis}, {Koubsky}, {Kowalczyk}, {Krone-Martins}, {Kudryashova}, {Kull}, {Bachchan}, {Lacoste-Seris}, {Lanza}, {Lavigne}, {Le Poncin-Lafitte}, {Lebreton}, {Lebzelter}, {Leccia}, {Leclerc}, {Lecoeur-Taibi}, {Lemaitre}, {Lenhardt}, {Leroux}, {Liao}, {Licata}, {Lindstr{\o}m}, {Lister}, {Livanou}, {Lobel}, {L{\"o}ffler},
  {L{\'o}pez}, {Lopez-Lozano}, {Lorenz}, {Loureiro}, {MacDonald}, {Magalh{\~a}es Fernandes}, {Managau}, {Mann}, {Mantelet}, {Marchal}, {Marchant}, {Marconi}, {Marie}, {Marinoni}, {Marrese}, {Marschalk{\'o}}, {Marshall}, {Mart{\'\i}n-Fleitas}, {Martino}, {Mary}, {Matijevi{\v{c}}}, {Mazeh}, {McMillan}, {Messina}, {Mestre}, {Michalik}, {Millar}, {Miranda}, {Molina}, {Molinaro}, {Molinaro}, {Moln{\'a}r}, {Moniez}, {Montegriffo}, {Monteiro}, {Mor}, {Mora}, {Morbidelli}, {Morel}, {Morgenthaler}, {Morley}, {Morris}, {Mulone}, {Muraveva}, {Musella}, {Narbonne}, {Nelemans}, {Nicastro}, {Noval}, {Ord{\'e}novic}, {Ordieres-Mer{\'e}}, {Osborne}, {Pagani}, {Pagano}, {Pailler}, {Palacin}, {Palaversa}, {Parsons}, {Paulsen}, {Pecoraro}, {Pedrosa}, {Pentik{\"a}inen}, {Pereira}, {Pichon}, {Piersimoni}, {Pineau}, {Plachy}, {Plum}, {Poujoulet}, {Pr{\v{s}}a}, {Pulone}, {Ragaini}, {Rago}, {Rambaux}, {Ramos-Lerate}, {Ranalli}, {Rauw}, {Read}, {Regibo}, {Renk}, {Reyl{\'e}}, {Ribeiro}, {Rimoldini}, {Ripepi}, {Riva}, {Rixon},
  {Roelens}, {Romero-G{\'o}mez}, {Rowell}, {Royer}, {Rudolph}, {Ruiz-Dern}, {Sadowski}, {Sagrist{\`a} Sell{\'e}s}, {Sahlmann}, {Salgado}, {Salguero}, {Sarasso}, {Savietto}, {Schnorhk}, {Schultheis}, {Sciacca}, {Segol}, {Segovia}, {Segransan}, {Serpell}, {Shih}, {Smareglia}, {Smart}, {Smith}, {Solano}, {Solitro}, {Sordo}, {Soria Nieto}, {Souchay}, {Spagna}, {Spoto}, {Stampa}, {Steele}, {Steidelm{\"u}ller}, {Stephenson}, {Stoev}, {Suess}, {S{\"u}veges}, {Surdej}, {Szabados}, {Szegedi-Elek}, {Tapiador}, {Taris}, {Tauran}, {Taylor}, {Teixeira}, {Terrett}, {Tingley}, {Trager}, {Turon}, {Ulla}, {Utrilla}, {Valentini}, {van Elteren}, {Van Hemelryck}, {van Leeuwen}, {Varadi}, {Vecchiato}, {Veljanoski}, {Via}, {Vicente}, {Vogt}, {Voss}, {Votruba}, {Voutsinas}, {Walmsley}, {Weiler}, {Weingrill}, {Werner}, {Wevers}, {Whitehead}, {Wyrzykowski}, {Yoldas}, {{\v{Z}}erjal}, {Zucker}, {Zurbach}, {Zwitter}, {Alecu}, {Allen}, {Allende Prieto}, {Amorim}, {Anglada-Escud{\'e}}, {Arsenijevic}, {Azaz}, {Balm}, {Beck}, {Bernstein},
  {Bigot}, {Bijaoui}, {Blasco}, {Bonfigli}, {Bono}, {Boudreault}, {Bressan}, {Brown}, {Brunet}, {Bunclark}, {Buonanno}, {Butkevich}, {Carret}, {Carrion}, {Chemin}, {Ch{\'e}reau}, {Corcione}, {Darmigny}, {de Boer}, {de Teodoro}, {de Zeeuw}, {Delle Luche}, {Domingues}, {Dubath}, {Fodor}, {Fr{\'e}zouls}, {Fries}, {Fustes}, {Fyfe}, {Gallardo}, {Gallegos}, {Gardiol}, {Gebran}, {Gomboc}, {G{\'o}mez}, {Grux}, {Gueguen}, {Heyrovsky}, {Hoar}, {Iannicola}, {Isasi Parache}, {Janotto}, {Joliet}, {Jonckheere}, {Keil}, {Kim}, {Klagyivik}, {Klar}, {Knude}, {Kochukhov}, {Kolka}, {Kos}, {Kutka}, {Lainey}, {LeBouquin}, {Liu}, {Loreggia}, {Makarov}, {Marseille}, {Martayan}, {Martinez-Rubi}, {Massart}, {Meynadier}, {Mignot}, {Munari}, {Nguyen}, {Nordlander}, {Ocvirk}, {O'Flaherty}, {Olias Sanz}, {Ortiz}, {Osorio}, {Oszkiewicz}, {Ouzounis}, {Palmer}, {Park}, {Pasquato}, {Peltzer}, {Peralta}, {P{\'e}turaud}, {Pieniluoma}, {Pigozzi}, {Poels}, {Prat}, {Prod'homme}, {Raison}, {Rebordao}, {Risquez}, {Rocca-Volmerange}, {Rosen},
  {Ruiz-Fuertes}, {Russo}, {Sembay}, {Serraller Vizcaino}, {Short}, {Siebert}, {Silva}, {Sinachopoulos}, {Slezak}, {Soffel}, {Sosnowska}, {Strai{\v{z}}ys}, {ter Linden}, {Terrell}, {Theil}, {Tiede}, {Troisi}, {Tsalmantza}, {Tur}, {Vaccari}, {Vachier}, {Valles}, {Van Hamme}, {Veltz}, {Virtanen}, {Wallut}, {Wichmann}, {Wilkinson}, {Ziaeepour}, \& {Zschocke}}]{2016A&A...595A...1G}
{Gaia Collaboration}, {Prusti}, T., {de Bruijne}, J.~H.~J., {et~al.} 2016, \aap, 595, A1, \dodoi{10.1051/0004-6361/201629272}

\bibitem[{{GRAVITY Collaboration} {et~al.}(2019){GRAVITY Collaboration}, {Perraut}, {Labadie}, {Lazareff}, {Klarmann}, {Segura-Cox}, {Benisty}, {Bouvier}, {Brandner}, {Caratti O Garatti}, {Caselli}, {Dougados}, {Garcia}, {Garcia-Lopez}, {Kendrew}, {Koutoulaki}, {Kervella}, {Lin}, {Pineda}, {Sanchez-Bermudez}, {van Dishoeck}, {Abuter}, {Amorim}, {Berger}, {Bonnet}, {Buron}, {Cantalloube}, {Cl{\'e}net}, {Coud{\'e} Du Foresto}, {Dexter}, {de Zeeuw}, {Duvert}, {Eckart}, {Eisenhauer}, {Eupen}, {Gao}, {Gendron}, {Genzel}, {Gillessen}, {Gordo}, {Grellmann}, {Haubois}, {Haussmann}, {Henning}, {Hippler}, {Horrobin}, {Hubert}, {Jocou}, {Lacour}, {Le Bouquin}, {L{\'e}na}, {M{\'e}rand}, {Ott}, {Paumard}, {Perrin}, {Pfuhl}, {Rabien}, {Ray}, {Rau}, {Rousset}, {Scheithauer}, {Straub}, {Straubmeier}, {Sturm}, {Vincent}, {Waisberg}, {Wank}, {Widmann}, {Wieprecht}, {Wiest}, {Wiezorrek}, {Woillez}, \& {Yazici}}]{2019A&A...632A..53G}
{GRAVITY Collaboration}, {Perraut}, K., {Labadie}, L., {et~al.} 2019, \aap, 632, A53, \dodoi{10.1051/0004-6361/201936403}

\bibitem[{{GRAVITY Collaboration} {et~al.}(2021){GRAVITY Collaboration}, {Perraut}, {Labadie}, {Bouvier}, {M{\'e}nard}, {Klarmann}, {Dougados}, {Benisty}, {Berger}, {Bouarour}, {Brandner}, {Caratti O Garatti}, {Caselli}, {de Zeeuw}, {Garcia-Lopez}, {Henning}, {Sanchez-Bermudez}, {Sousa}, {van Dishoeck}, {Al{\'e}cian}, {Amorim}, {Cl{\'e}net}, {Davies}, {Drescher}, {Duvert}, {Eckart}, {Eisenhauer}, {F{\"o}rster-Schreiber}, {Garcia}, {Gendron}, {Genzel}, {Gillessen}, {Grellmann}, {Hei{\ss}el}, {Hippler}, {Horrobin}, {Hubert}, {Jocou}, {Kervella}, {Lacour}, {Lapeyr{\`e}re}, {Le Bouquin}, {L{\'e}na}, {Lutz}, {Ott}, {Paumard}, {Perrin}, {Scheithauer}, {Shangguan}, {Shimizu}, {Stadler}, {Straub}, {Straubmeier}, {Sturm}, {Tacconi}, {Vincent}, {von Fellenberg}, \& {Widmann}}]{2021A&A...655A..73G}
---. 2021, \aap, 655, A73, \dodoi{10.1051/0004-6361/202141624}

\bibitem[{{G{\"u}rtler} {et~al.}(1999){G{\"u}rtler}, {Friedemann}, {Reimann}, {Splittgerber}, \& {Rudolph}}]{1999A&AS..140..293G}
{G{\"u}rtler}, J., {Friedemann}, C., {Reimann}, H.~G., {Splittgerber}, E., \& {Rudolph}, E. 1999, \aaps, 140, 293, \dodoi{10.1051/aas:1999423}

\bibitem[{{Guzm{\'a}n-D{\'\i}az} {et~al.}(2021){Guzm{\'a}n-D{\'\i}az}, {Mendigut{\'\i}a}, {Montesinos}, {Oudmaijer}, {Vioque}, {Rodrigo}, {Solano}, {Meeus}, \& {Marcos-Arenal}}]{2021AA...650A.182G}
{Guzm{\'a}n-D{\'\i}az}, J., {Mendigut{\'\i}a}, I., {Montesinos}, B., {et~al.} 2021, \aap, 650, A182, \dodoi{10.1051/0004-6361/202039519}

\bibitem[{Harris {et~al.}(2020)Harris, Millman, van~der Walt, Gommers, Virtanen, Cournapeau, Wieser, Taylor, Berg, Smith, Kern, Picus, Hoyer, van Kerkwijk, Brett, Haldane, del R{\'{i}}o, Wiebe, Peterson, G{\'{e}}rard-Marchant, Sheppard, Reddy, Weckesser, Abbasi, Gohlke, \& Oliphant}]{numpy}
Harris, C.~R., Millman, K.~J., van~der Walt, S.~J., {et~al.} 2020, Nature, 585, 357, \dodoi{10.1038/s41586-020-2649-2}

\bibitem[{{Hartmann} {et~al.}(2016){Hartmann}, {Herczeg}, \& {Calvet}}]{2016ARA&A..54..135H}
{Hartmann}, L., {Herczeg}, G., \& {Calvet}, N. 2016, \araa, 54, 135, \dodoi{10.1146/annurev-astro-081915-023347}

\bibitem[{{Hartmann} {et~al.}(1994){Hartmann}, {Hewett}, \& {Calvet}}]{1994ApJ...426..669H}
{Hartmann}, L., {Hewett}, R., \& {Calvet}, N. 1994, \apj, 426, 669, \dodoi{10.1086/174104}

\bibitem[{{Herbig}(1960)}]{1960ApJS....4..337H}
{Herbig}, G.~H. 1960, \apjs, 4, 337, \dodoi{10.1086/190050}

\bibitem[{{Herbst} {et~al.}(1994){Herbst}, {Herbst}, {Grossman}, \& {Weinstein}}]{1994AJ....108.1906H}
{Herbst}, W., {Herbst}, D.~K., {Grossman}, E.~J., \& {Weinstein}, D. 1994, \aj, 108, 1906, \dodoi{10.1086/117204}

\bibitem[{{Herbst} \& {Shevchenko}(1999)}]{1999AJ....118.1043H}
{Herbst}, W., \& {Shevchenko}, V.~S. 1999, \aj, 118, 1043, \dodoi{10.1086/300966}

\bibitem[{{Hern{\'a}ndez} {et~al.}(2004){Hern{\'a}ndez}, {Calvet}, {Brice{\~n}o}, {Hartmann}, \& {Berlind}}]{2004AJ....127.1682H}
{Hern{\'a}ndez}, J., {Calvet}, N., {Brice{\~n}o}, C., {Hartmann}, L., \& {Berlind}, P. 2004, \aj, 127, 1682, \dodoi{10.1086/381908}

\bibitem[{{Hern{\'a}ndez} {et~al.}(2005){Hern{\'a}ndez}, {Calvet}, {Hartmann}, {Brice{\~n}o}, {Sicilia-Aguilar}, \& {Berlind}}]{2005AJ....129..856H}
{Hern{\'a}ndez}, J., {Calvet}, N., {Hartmann}, L., {et~al.} 2005, \aj, 129, 856, \dodoi{10.1086/426918}

\bibitem[{{Hillenbrand} {et~al.}(2022){Hillenbrand}, {Kiker}, {Gee}, {Lester}, {Braunfeld}, {Rebull}, \& {Kuhn}}]{2022AJ....163..263H}
{Hillenbrand}, L.~A., {Kiker}, T.~J., {Gee}, M., {et~al.} 2022, \aj, 163, 263, \dodoi{10.3847/1538-3881/ac62d8}

\bibitem[{{Hillenbrand} {et~al.}(1992){Hillenbrand}, {Strom}, {Vrba}, \& {Keene}}]{1992ApJ...397..613H}
{Hillenbrand}, L.~A., {Strom}, S.~E., {Vrba}, F.~J., \& {Keene}, J. 1992, \apj, 397, 613, \dodoi{10.1086/171819}

\bibitem[{{Hubrig} {et~al.}(2019){Hubrig}, {J{\"a}rvinen}, {Sch{\"o}ller}, {Carroll}, {Ilyin}, \& {Pogodin}}]{2019ASPC..518...18H}
{Hubrig}, S., {J{\"a}rvinen}, S.~P., {Sch{\"o}ller}, M., {et~al.} 2019, in Astronomical Society of the Pacific Conference Series, Vol. 518, Physics of Magnetic Stars, ed. D.~O. {Kudryavtsev}, I.~I. {Romanyuk}, \& I.~A. {Yakunin}, 18, \dodoi{10.48550/arXiv.1812.04482}

\bibitem[{{Hubrig} \& {Sch{\"o}ller}(2021)}]{2021mfob.book.....H}
{Hubrig}, S., \& {Sch{\"o}ller}, M. 2021, {Magnetic Fields in O, B, and A Stars}, \dodoi{10.1088/2514-3433/abefcc}

\bibitem[{{Hubrig} {et~al.}(2010){Hubrig}, {Sch{\"o}ller}, {Savanov}, {Gonz{\'a}lez}, {Cowley}, {Sch{\"u}tz}, {Arlt}, \& {R{\"u}diger}}]{2010AN....331..361H}
{Hubrig}, S., {Sch{\"o}ller}, M., {Savanov}, I., {et~al.} 2010, Astronomische Nachrichten, 331, 361, \dodoi{10.1002/asna.201011346}

\bibitem[{{Hubrig} {et~al.}(2004){Hubrig}, {Sch{\"o}ller}, \& {Yudin}}]{2004A&A...428L...1H}
{Hubrig}, S., {Sch{\"o}ller}, M., \& {Yudin}, R.~V. 2004, \aap, 428, L1, \dodoi{10.1051/0004-6361:200400091}

\bibitem[{Hunter(2007)}]{matplotlib}
Hunter, J.~D. 2007, Computing in Science \& Engineering, 9, 90, \dodoi{10.1109/MCSE.2007.55}

\bibitem[{{J{\"a}rvinen} {et~al.}(2019){J{\"a}rvinen}, {Carroll}, {Hubrig}, {Ilyin}, \& {Sch{\"o}ller}}]{2019MNRAS.489..886J}
{J{\"a}rvinen}, S.~P., {Carroll}, T.~A., {Hubrig}, S., {Ilyin}, I., \& {Sch{\"o}ller}, M. 2019, \mnras, 489, 886, \dodoi{10.1093/mnras/stz2190}

\bibitem[{{Lightkurve Collaboration} {et~al.}(2018){Lightkurve Collaboration}, {Cardoso}, {Hedges}, {Gully-Santiago}, {Saunders}, {Cody}, {Barclay}, {Hall}, {Sagear}, {Turtelboom}, {Zhang}, {Tzanidakis}, {Mighell}, {Coughlin}, {Bell}, {Berta-Thompson}, {Williams}, {Dotson}, \& {Barentsen}}]{2018ascl.soft12013L}
{Lightkurve Collaboration}, {Cardoso}, J.~V.~d.~M., {Hedges}, C., {et~al.} 2018, {Lightkurve: Kepler and TESS time series analysis in Python}, Astrophysics Source Code Library.
\newblock \doeprint{1812.013}

\bibitem[{{Luhman}(2022)}]{2022AJ....163...25L}
{Luhman}, K.~L. 2022, \aj, 163, 25, \dodoi{10.3847/1538-3881/ac35e3}

\bibitem[{{Luhman}(2023)}]{2023AJ....165...37L}
---. 2023, \aj, 165, 37, \dodoi{10.3847/1538-3881/ac9da3}

\bibitem[{{Marconi} \& {Palla}(1998)}]{1998ApJ...507L.141M}
{Marconi}, M., \& {Palla}, F. 1998, \apjl, 507, L141, \dodoi{10.1086/311704}

\bibitem[{{Marconi} {et~al.}(2000){Marconi}, {Ripepi}, {Alcal{\'a}}, {Covino}, {Palla}, \& {Terranegra}}]{2000AA...355L..35M}
{Marconi}, M., {Ripepi}, V., {Alcal{\'a}}, J.~M., {et~al.} 2000, \aap, 355, L35, \dodoi{10.48550/arXiv.astro-ph/0002466}

\bibitem[{{Meeus} {et~al.}(2001){Meeus}, {Waters}, {Bouwman}, {van den Ancker}, {Waelkens}, \& {Malfait}}]{2001A&A...365..476M}
{Meeus}, G., {Waters}, L.~B.~F.~M., {Bouwman}, J., {et~al.} 2001, \aap, 365, 476, \dodoi{10.1051/0004-6361:20000144}

\bibitem[{{Mendigut{\'\i}a} {et~al.}(2011){Mendigut{\'\i}a}, {Eiroa}, {Montesinos}, {Mora}, {Oudmaijer}, {Mer{\'\i}n}, \& {Meeus}}]{2011A&A...529A..34M}
{Mendigut{\'\i}a}, I., {Eiroa}, C., {Montesinos}, B., {et~al.} 2011, \aap, 529, A34, \dodoi{10.1051/0004-6361/201015821}

\bibitem[{{Meyer} {et~al.}(1997){Meyer}, {Calvet}, \& {Hillenbrand}}]{1997AJ....114..288M}
{Meyer}, M.~R., {Calvet}, N., \& {Hillenbrand}, L.~A. 1997, \aj, 114, 288, \dodoi{10.1086/118474}

\bibitem[{{Monnier} \& {Millan-Gabet}(2002)}]{2002ApJ...579..694M}
{Monnier}, J.~D., \& {Millan-Gabet}, R. 2002, \apj, 579, 694, \dodoi{10.1086/342917}

\bibitem[{{Monnier} {et~al.}(2005){Monnier}, {Millan-Gabet}, {Billmeier}, {Akeson}, {Wallace}, {Berger}, {Calvet}, {D'Alessio}, {Danchi}, {Hartmann}, {Hillenbrand}, {Kuchner}, {Rajagopal}, {Traub}, {Tuthill}, {Boden}, {Booth}, {Colavita}, {Gathright}, {Hrynevych}, {Le Mignant}, {Ligon}, {Neyman}, {Swain}, {Thompson}, {Vasisht}, {Wizinowich}, {Beichman}, {Beletic}, {Creech-Eakman}, {Koresko}, {Sargent}, {Shao}, \& {van Belle}}]{2005ApJ...624..832M}
{Monnier}, J.~D., {Millan-Gabet}, R., {Billmeier}, R., {et~al.} 2005, \apj, 624, 832, \dodoi{10.1086/429266}

\bibitem[{{Nguyen} {et~al.}(2022){Nguyen}, {Costa}, {Girardi}, {Volpato}, {Bressan}, {Chen}, {Marigo}, {Fu}, \& {Goudfrooij}}]{2022A&A...665A.126N}
{Nguyen}, C.~T., {Costa}, G., {Girardi}, L., {et~al.} 2022, \aap, 665, A126, \dodoi{10.1051/0004-6361/202244166}

\bibitem[{{Paegert} {et~al.}(2021){Paegert}, {Stassun}, {Collins}, {Pepper}, {Torres}, {Jenkins}, {Twicken}, \& {Latham}}]{2021arXiv210804778P}
{Paegert}, M., {Stassun}, K.~G., {Collins}, K.~A., {et~al.} 2021, arXiv e-prints, arXiv:2108.04778, \dodoi{10.48550/arXiv.2108.04778}

\bibitem[{{Pecaut} \& {Mamajek}(2013)}]{2013ApJS..208....9P}
{Pecaut}, M.~J., \& {Mamajek}, E.~E. 2013, \apjs, 208, 9, \dodoi{10.1088/0067-0049/208/1/9}

\bibitem[{{Poxon}(2015)}]{2015JAVSO..43...35P}
{Poxon}, M. 2015, \jaavso, 43, 35

\bibitem[{{Ricker}(2019)}]{2019ESS.....410001R}
{Ricker}, G. 2019, in AAS/Division for Extreme Solar Systems Abstracts, Vol.~51, AAS/Division for Extreme Solar Systems Abstracts, 100.01

\bibitem[{{Ricker} {et~al.}(2015){Ricker}, {Winn}, {Vanderspek}, {Latham}, {Bakos}, {Bean}, {Berta-Thompson}, {Brown}, {Buchhave}, {Butler}, {Butler}, {Chaplin}, {Charbonneau}, {Christensen-Dalsgaard}, {Clampin}, {Deming}, {Doty}, {De Lee}, {Dressing}, {Dunham}, {Endl}, {Fressin}, {Ge}, {Henning}, {Holman}, {Howard}, {Ida}, {Jenkins}, {Jernigan}, {Johnson}, {Kaltenegger}, {Kawai}, {Kjeldsen}, {Laughlin}, {Levine}, {Lin}, {Lissauer}, {MacQueen}, {Marcy}, {McCullough}, {Morton}, {Narita}, {Paegert}, {Palle}, {Pepe}, {Pepper}, {Quirrenbach}, {Rinehart}, {Sasselov}, {Sato}, {Seager}, {Sozzetti}, {Stassun}, {Sullivan}, {Szentgyorgyi}, {Torres}, {Udry}, \& {Villasenor}}]{2015JATIS...1a4003R}
{Ricker}, G.~R., {Winn}, J.~N., {Vanderspek}, R., {et~al.} 2015, Journal of Astronomical Telescopes, Instruments, and Systems, 1, 014003, \dodoi{10.1117/1.JATIS.1.1.014003}

\bibitem[{{Rivinius} {et~al.}(2013){Rivinius}, {Carciofi}, \& {Martayan}}]{2013A&ARv..21...69R}
{Rivinius}, T., {Carciofi}, A.~C., \& {Martayan}, C. 2013, \aapr, 21, 69, \dodoi{10.1007/s00159-013-0069-0}

\bibitem[{{Strom} {et~al.}(1972){Strom}, {Strom}, {Yost}, {Carrasco}, \& {Grasdalen}}]{1972ApJ...173..353S}
{Strom}, S.~E., {Strom}, K.~M., {Yost}, J., {Carrasco}, L., \& {Grasdalen}, G. 1972, \apj, 173, 353, \dodoi{10.1086/151425}

\bibitem[{{Suh}(2016)}]{2016JASS...33..119S}
{Suh}, K.-W. 2016, Journal of Astronomy and Space Sciences, 33, 119, \dodoi{10.5140/JASS.2016.33.2.119}

\bibitem[{{The}(1994)}]{1994ASPC...62...23T}
{The}, P.~S. 1994, in Astronomical Society of the Pacific Conference Series, Vol.~62, The Nature and Evolutionary Status of Herbig Ae/Be Stars, ed. P.~S. {The}, M.~R. {Perez}, \& E.~P.~J. {van den Heuvel}, 23

\bibitem[{{Th\'{e}} {et~al.}(1994){Th\'{e}}, {de Winter}, \& {Perez}}]{1994A&AS..104..315T}
{Th\'{e}}, P.~S., {de Winter}, D., \& {Perez}, M.~R. 1994, \aaps, 104, 315

\bibitem[{{Thomas} {et~al.}(2023){Thomas}, {Rodgers}, {van der Bliek}, {Doppmann}, {Bouvier}, {Salvo}, {Beuzit}, \& {Rigaut}}]{2023AJ....165..135T}
{Thomas}, S.~J., {Rodgers}, B., {van der Bliek}, N.~S., {et~al.} 2023, \aj, 165, 135, \dodoi{10.3847/1538-3881/aca803}

\bibitem[{{Torres} {et~al.}(1995){Torres}, {Quast}, {de La Reza}, {Gregorio-Hetem}, \& {Lepine}}]{1995AJ....109.2146T}
{Torres}, C.~A.~O., {Quast}, G., {de La Reza}, R., {Gregorio-Hetem}, J., \& {Lepine}, J.~R.~D. 1995, \aj, 109, 2146, \dodoi{10.1086/117440}

\bibitem[{{Uemura} {et~al.}(2004){Uemura}, {Kato}, {Ishioka}, {Yoshida}, {Kadota}, {Ohkura}, {Henden}, {Pejcha}, {Kinugasa}, {Fujii}, {Simonsen}, {Greaves}, {Dubovsky}, {Poyner}, {West}, {Stine}, {Taylor}, {Poxon}, {Muyllaert}, {Ripero}, {Reszelski}, \& {Jones}}]{2004PASJ...56S.183U}
{Uemura}, M., {Kato}, T., {Ishioka}, R., {et~al.} 2004, \pasj, 56, S183, \dodoi{10.1093/pasj/56.sp1.S183}

\bibitem[{{van den Ancker} {et~al.}(1998){van den Ancker}, {de Winter}, \& {Tjin A Djie}}]{1998A&A...330..145V}
{van den Ancker}, M.~E., {de Winter}, D., \& {Tjin A Djie}, H.~R.~E. 1998, \aap, 330, 145

\bibitem[{{Venuti} {et~al.}(2021){Venuti}, {Cody}, {Rebull}, {Beccari}, {Irwin}, {Thanvantri}, {Howell}, \& {Barentsen}}]{2021AJ....162..101V}
{Venuti}, L., {Cody}, A.~M., {Rebull}, L.~M., {et~al.} 2021, \aj, 162, 101, \dodoi{10.3847/1538-3881/ac0536}

\bibitem[{{Vieira} {et~al.}(2003){Vieira}, {Corradi}, {Alencar}, {Mendes}, {Torres}, {Quast}, {Guimar{\~a}es}, \& {da Silva}}]{2003AJ....126.2971V}
{Vieira}, S.~L.~A., {Corradi}, W.~J.~B., {Alencar}, S.~H.~P., {et~al.} 2003, \aj, 126, 2971, \dodoi{10.1086/379553}

\bibitem[{{Vioque} {et~al.}(2018){Vioque}, {Oudmaijer}, {Baines}, {Mendigut{\'\i}a}, \& {P{\'e}rez-Mart{\'\i}nez}}]{2018A&A...620A.128V}
{Vioque}, M., {Oudmaijer}, R.~D., {Baines}, D., {Mendigut{\'\i}a}, I., \& {P{\'e}rez-Mart{\'\i}nez}, R. 2018, \aap, 620, A128, \dodoi{10.1051/0004-6361/201832870}

\bibitem[{{Vioque} {et~al.}(2020){Vioque}, {Oudmaijer}, {Schreiner}, {Mendigut{\'\i}a}, {Baines}, {Mowlavi}, \& {P{\'e}rez-Mart{\'\i}nez}}]{2020A&A...638A..21V}
{Vioque}, M., {Oudmaijer}, R.~D., {Schreiner}, M., {et~al.} 2020, \aap, 638, A21, \dodoi{10.1051/0004-6361/202037731}

\bibitem[{{Waters} \& {Waelkens}(1998)}]{1998ARA&A..36..233W}
{Waters}, L.~B.~F.~M., \& {Waelkens}, C. 1998, \araa, 36, 233, \dodoi{10.1146/annurev.astro.36.1.233}

\bibitem[{{Wichittanakom} {et~al.}(2020){Wichittanakom}, {Oudmaijer}, {Fairlamb}, {Mendigut{\'\i}a}, {Vioque}, \& {Ababakr}}]{2020MNRAS.493..234W}
{Wichittanakom}, C., {Oudmaijer}, R.~D., {Fairlamb}, J.~R., {et~al.} 2020, \mnras, 493, 234, \dodoi{10.1093/mnras/staa169}

\bibitem[{{Zhang} {et~al.}(2006){Zhang}, {Yang}, \& {Liu}}]{2006Ap&SS.305...11Z}
{Zhang}, P., {Yang}, H.~T., \& {Liu}, J. 2006, \apss, 305, 11, \dodoi{10.1007/s10509-005-9031-6}

\bibitem[{{Zhang} {et~al.}(2022){Zhang}, {Luo}, {Jiang}, {Hou}, {Zuo}, {Du}, {Li}, \& {Zhao}}]{2022ApJ...936..151Z}
{Zhang}, Y.-J., {Luo}, A.~L., {Jiang}, B., {et~al.} 2022, \apj, 936, 151, \dodoi{10.3847/1538-4357/ac84da}

\end{thebibliography}
\bibliographystyle{aasjournal}



\end{document}